\documentclass[graybox]{svmult}

\usepackage{helvet}         
\usepackage{courier}        
                            
\usepackage{makeidx}         
\usepackage{graphicx}                    
\usepackage{multicol}     
\smartqed  

\usepackage[tablename=Box]{caption}

\usepackage{amsmath}
\usepackage{amsfonts}
\usepackage{amssymb,amsbsy}
\usepackage{bm}
 \usepackage{mathtools}

\usepackage[]{subfigure}
\usepackage{url}
\usepackage{soul}
\usepackage{verbatim}
\usepackage{csquotes}
\usepackage{breqn}
\usepackage{mdframed}

\renewcommand{\contentsname}{Table of content}

\makeindex             

\begin{document}

\title*{Advanced statistical methods for eye movement  analysis and modelling: a gentle introduction}
\titlerunning{Advanced statistical methods}

\author{Giuseppe Boccignone} 


\institute{G. Boccignone \at Department of Computer Science, 
 Universit\'a di Milano\\
via Comelico 39/41, 10135 Milano, Italy\\
\url{http://boccignone.di.unimi.it}
\email{giuseppe.boccignone@unimi.it}
 }

\maketitle
%

\tableofcontents

\section{Summary}
\label{sec:sum}

In this Chapter we  consider eye movements and, in particular, the resulting sequence of gaze shifts to be  the observable outcome of a stochastic process.  Crucially, we show that, under such assumption, a wide variety of tools become available for analyses and modelling beyond conventional statistical methods. Such tools encompass random walk analyses and  more complex techniques borrowed from the Pattern Recognition and  Machine Learning fields.

After a brief,  though critical,  probabilistic tour of current computational models of eye
movements and visual attention, we lay down  the basis for gaze shift pattern analysis. 
To this end, the concepts of  Markov Processes,  the Wiener process and related  random walks within the Gaussian framework of the Central Limit Theorem will be introduced. Then, we will deliberately violate  fundamental assumptions of the Central Limit Theorem to  elicit a larger perspective, rooted in statistical physics, for analysing and modelling eye movements in terms of anomalous, non-Gaussian, random walks and modern foraging theory.

Eventually, by resorting to Statistical Machine Learning techniques, we discuss how  the analyses of movement patterns   can develop into the inference of hidden patterns of the mind: inferring the observer's task, assessing cognitive impairments, classifying  expertise.

\section{Introduction}
\label{sec:intro}

\begin{figure}[b]
\sidecaption[t]
\includegraphics[scale=0.2,keepaspectratio=true]{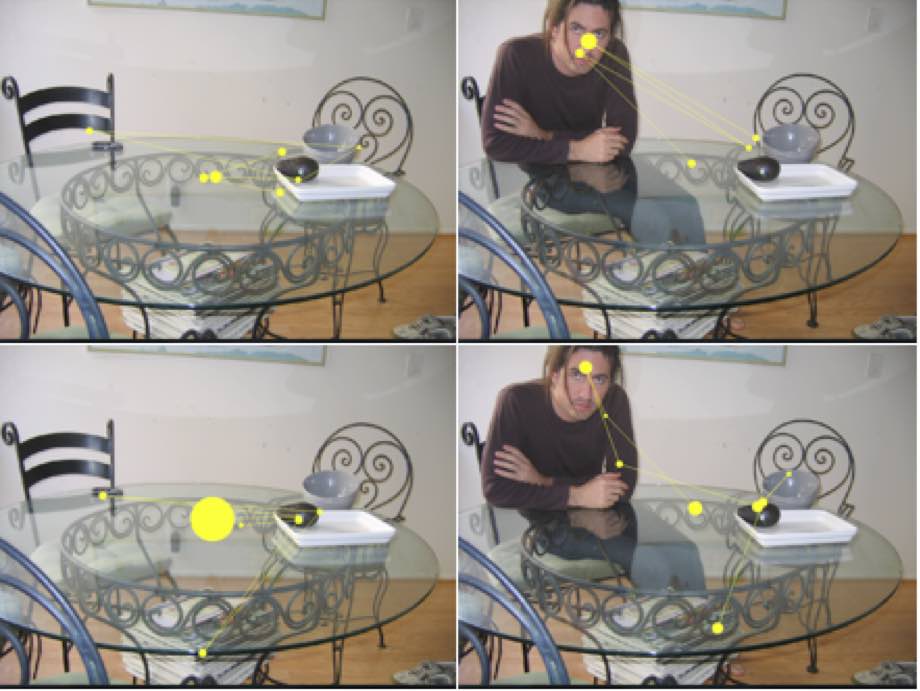}
\caption{Different scan paths on a pair of images  eye-tracked from different human observers. Left, free viewing of a natural scene; right, natural scene embedding a face. The area of yellow disks marking fixations between saccades is proportional to fixation time (images and eye tracking data from  the \textsf{Fixations in FAces} dataset}
\label{Fig:topdown}
\end{figure}

Consider Fig. \ref{Fig:topdown}: it shows  typical \emph{scan paths} (in this case a succession of saccades and fixations) produced by two human observers on a natural image: circular spots  and
lines joining spots graphically represent
fixations and gaze shifts between subsequent fixations, respectively. 

When looking at scan paths, the first question arising is: How can we characterise the shape  and the statistical properties of such trajectories? Answering this question entails a \emph{data analysis} issue. The second question is:  What factors determine the shape and the statistical properties? and it relates to the \emph{modelling} issue.

From a mere research practice standpoint these two issues need not  be related (yet, from a more general theoretical standpoint  such attitude is at least debatable). A great deal of research can be conducted by performing an eye tracking experiment based on  a specific paradigm, and then analysing data by running standard statistical tools (e.g., ANOVA)  on scan path ``features'' such as fixation frequency, mean fixation time, mean saccadic amplitudes, scan path length, etc. The ``data-driven'' attitude can be preserved even in the case where standard tools are abandoned in favour  of more complex techniques borrowed from the Pattern Recognition and  Machine Learning fields;  for instance,  in the endeavour of inferring or classifying the observer's mental task  or the expertise behind his gaze shifts (e.g., \cite{henderson2013predicting,boccignone_jemr2014}). 

In the same vein, it is possible to set up a gaze shift model and successively  assess its performance against  eye tracking data in terms of classic statistical analyses. For instance, one might set up a probabilistic dynamic model of gaze shifting; then ``synthetic'' shifts can be generated from the model-based simulation. The distribution of their features can so be compared against the feature distribution of  human gaze shifts - on the same stimuli -  by exploiting a suitable goodness-of-fit test (e.g., \cite{BocFerSMCB2013}, \cite{LiberatiASD2017}).    

Clearly, the program of following the data  lies at the heart of scientific methodology. When trying to understand a complex process in nature, the empirical evidence is essential. Hypotheses must be compared with the actual data, but the empirical evidence itself may have limitations; that is, it may not be sufficiently large or accurate either to confirm or rule out hypotheses, models, explanations, or assumptions, even when the most sophisticated analytical tools  are used. 

\begin{figure}[b]
\sidecaption[b]
\includegraphics[scale=0.12,keepaspectratio=true]{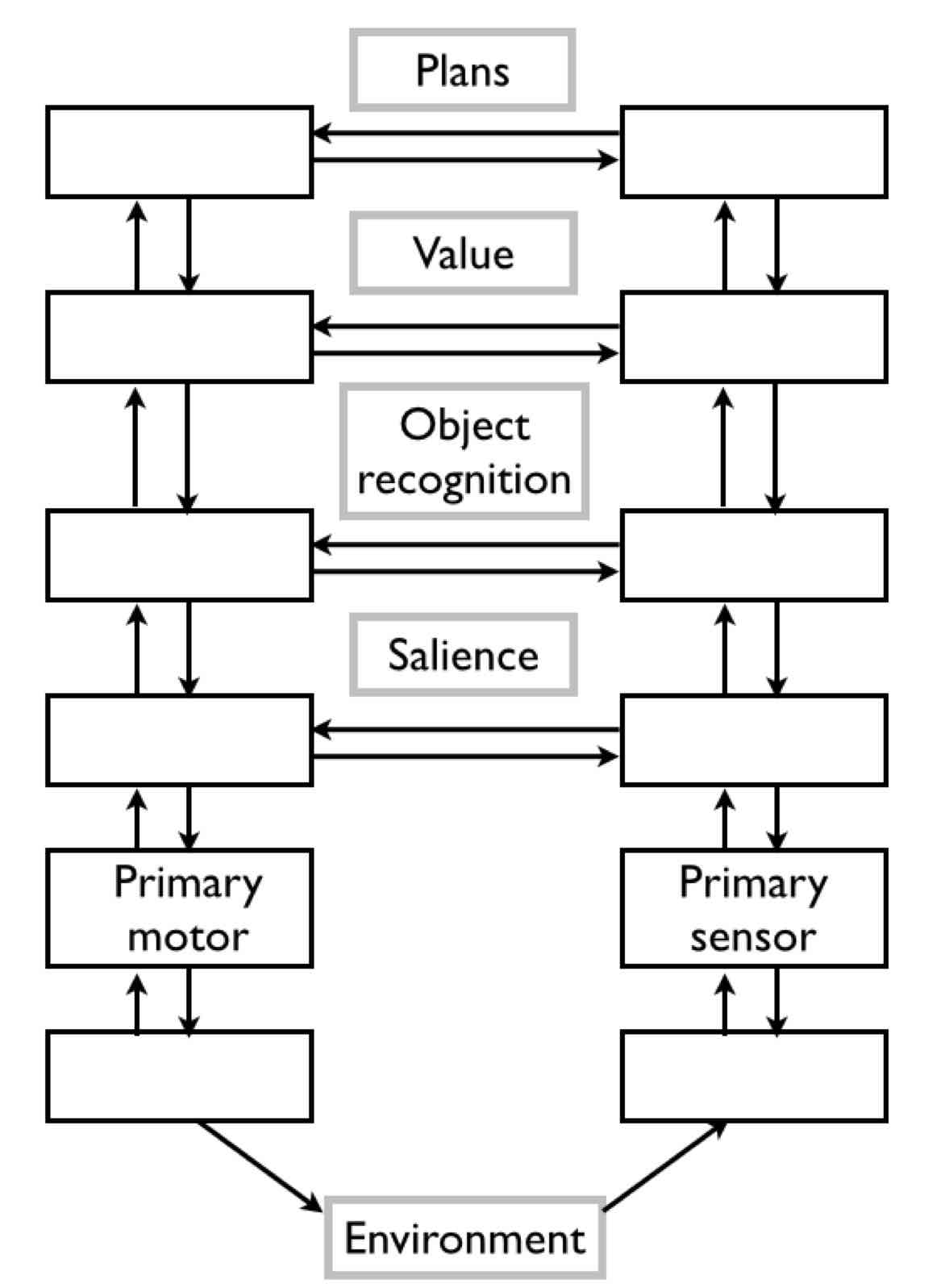}
\caption{Framework for the control of  eye movements. There are several interacting layers of control that influence  target selection: the scheme highlights, top to bottom,  the contributions of plans, value,   object recognition and salience to  target selection. The left hand route summarises the motor components, the right hand one, the perceptual components. Figure modified after Sch{\"u}tz \emph{et al} \cite{schutz2011eye}}
\label{Fig:levels}
\end{figure}

For eye movement patterns, this issue may be in some cases particularly delicate. Such patterns are, in some sense, a summary of all the motor and perceptual activities in which the observer has been involved during data collection. As sketched in Fig. \ref{Fig:levels}, from a functional standpoint,  there are several interacting action / perception loops that drive eye movements.  These factors act on different levels of representation and processing: salience, for instance, is a typical bottom-up process, while plans are typical top-down processes \cite{schutz2011eye}.

In principle, all such activities should be taken into account when analysing and modelling actual eye movements in visual attention behaviour. Clearly, this is a mind-blowing endeavour. 

This raises  the question of what is a computational model  and how it can support more advanced  analyses of experimental data.   In this Chapter we  discuss a minimal phenomenological model.  

\begin{figure}[t]
\includegraphics[scale=0.22,keepaspectratio=true]{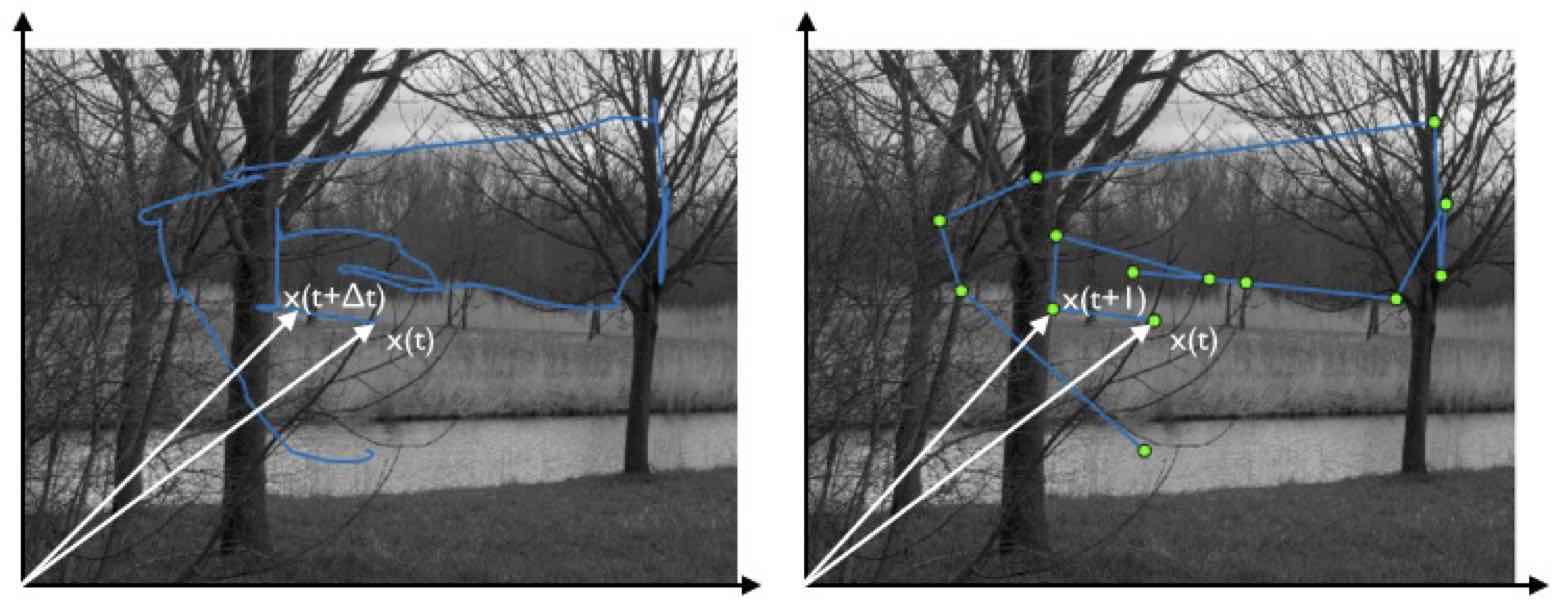}
\caption{Scan path of  an eye-tracked human observer rendered as a temporal sequence  of gaze position represented by time-varying location vectors $\mathbf{x}(t)$. The left image shows the continuous raw-data trajectory; the right  image, the discretized sequence of fixations and saccades. Images and data are from the Doves dataset~\cite{van2009doves}, which is freely available on the Web 
}
\label{Fig:rawfix}
\end{figure}

At the most general level, the aim of a computational model of visual attention is to answer the question \emph{Where to Look Next?} by providing: 
\begin{enumerate}
\item at the \emph{computational theory level} (in the sense of Marr, \cite{Marr}; defining the input/output computation at time $t$), an account of the mapping  from visual data of a complex natural scene, say $\mathcal{D}$ (raw image data, or more usefully, features), to  a sequence of gaze locations $\mathbf{x}_{F}(1), \mathbf{x}_{F}(2),\cdots$, under a given task $\mathbf{T}$, namely 
\begin{equation}
\mathcal{D} \xmapsto[\mathbf{T}]{}  \{\mathbf{x}_{F}(1), \mathbf{x}_{F}(2),\cdots \}, 
\label{eq:mapping}
\end{equation}
where the sequence  $\{\mathbf{x}_{F}(1), \mathbf{x}_{F}(2),\cdots \}$ can be used to define a scan path (as illustrated in Fig. \ref{Fig:rawfix});

\item at  the \emph{algorithmic level}, \cite{Marr}, a procedure that simulates such mapping (we will not specifically address here the third level of neural realisation \cite{Marr}).
\end{enumerate}

Under this conceptualisation, when considering for instance the input $\mathcal{D}$ in the form of a static scene (a picture), either the raw time series $\{\mathbf{x}_{F}(1), \mathbf{x}_{F}(2),\cdots \}$ or fixation duration and saccade  (length and  direction) are the only two observable behaviours of the underlying control mechanism. When, $\mathcal{D}$ is a dynamic or time varying scene (e.g. a video), then pursuit needs to be taken also into account.  Thus, it is convenient to adopt  the generic terms of gaze shifts (either pursuit or saccades) and gaze shift amplitudes. Fixation duration and shift amplitude vary greatly during visual scanning of the scene. As previously discussed, such variation reflects moment-to-moment changes in the visual input, processes occurring at different levels of representation, the state of the oculomotor system and stochastic variability in neuromotor force pulses. 

We can summarize this state of affairs by stating that fixation duration and  the time series $\{\mathbf{x}_{F}(1), \mathbf{x}_{F}(2),\cdots \}$ (or equivalently,  gaze shift lengths and directions) are random variables (RVs) that are generated by an underlying random process. In other terms,  the sequence $\{\mathbf{x}_{F}(1), \mathbf{x}_{F}(2),\cdots \}$ is the realisation of a stochastic process, and the goal of a computational theory  is to develop a mathematical model that describes statistical properties of  eye movements as closely as possible.

Is this minimalist approach to computational modelling of gaze shifts a reasonable one? The answer can be positive  if ``systematic tendencies''  between fixation durations, gaze shift amplitudes and  directions of successive eye movements exist and  such sequential dependencies can be captured by the stochastic process model.
Systematic tendencies in oculomotor behaviour can be thought of as regularities that are common across all instances of, and manipulations to, behavioural tasks.  In that  case useful information about how the observers will move their eyes can be found.

%

\begin{figure}[t]
\sidecaption[t]
\includegraphics[scale=0.2,keepaspectratio=true]{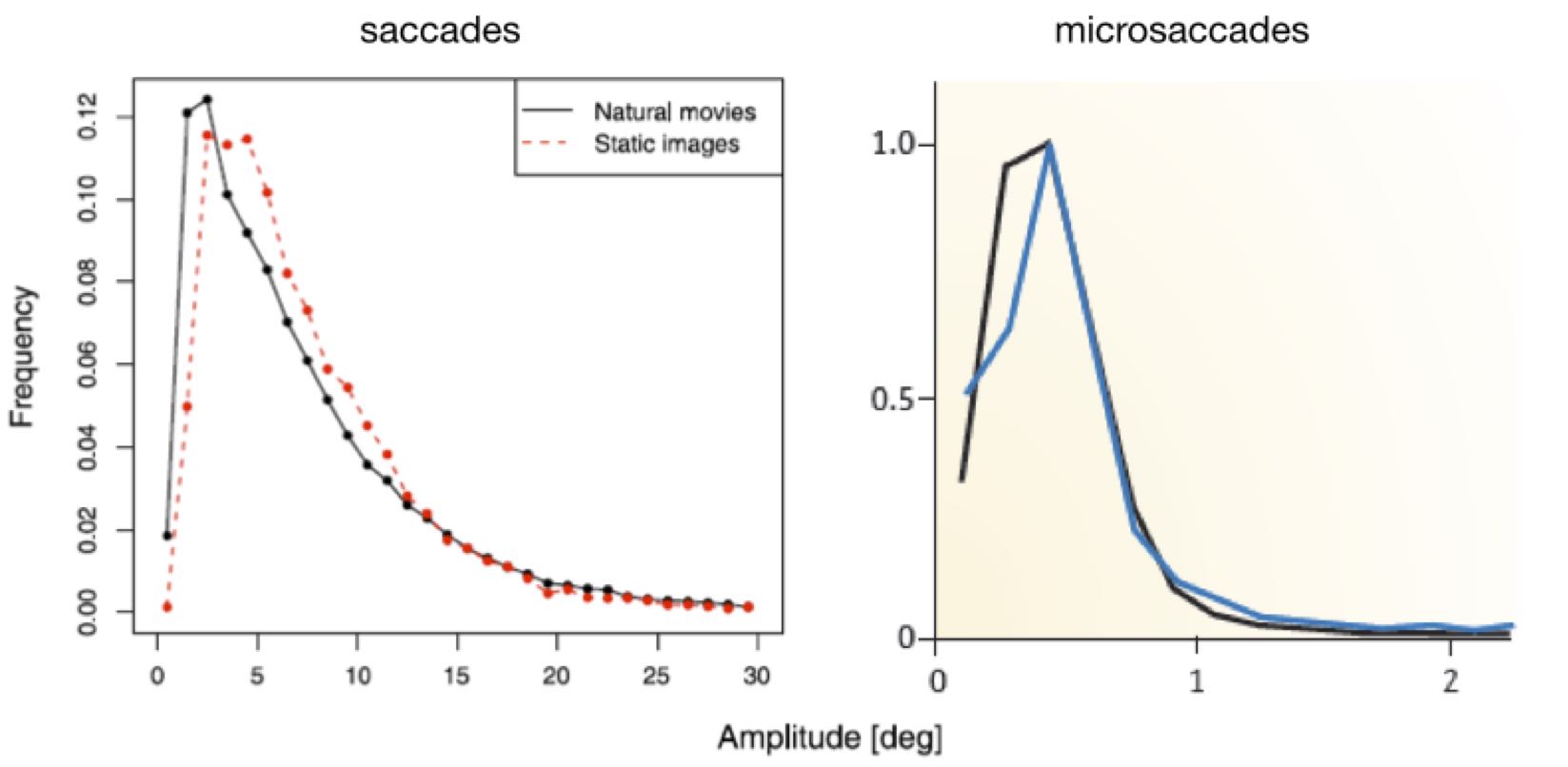}
\caption{Amplitudes distribution of saccades (on natural movies and static images, left) and microsaccades (right, recordings via video - black - and search coil - blue). In both cases amplitudes follow a positively skewed, long-tailed distribution. Figure modified after Dorr \emph{et al} \cite{dorr2010variability} and Martinez-Conde \emph{et al} \cite{martinez2013impact}}
\label{Fig:ampl}
\end{figure}

Indeed,  such systematic tendencies  or ``biases'' in the manner in which we explore scenes with our eyes are well known in the literature.  One example is provided in Fig. \ref{Fig:ampl} showing the amplitude distribution of saccades and microsaccades that typically exhibit a positively skewed, long-tailed shape.
Other paradigmatic examples of systematic tendencies in scene viewing are \cite{tatler2008systematic,tatler2009prominence}: initiating  saccades in the horizontal and vertical directions more frequently than in oblique directions;  small amplitude saccades tending to be followed by long amplitude ones and vice versa.

Such biases  may arise from a number of sources. Tatler and Vincent~\cite{tatler2009prominence} have suggested the following:
%
%
%
 biomechanical factors, saccade flight time and landing accuracy, uncertainty, distribution of objects of interest in the environment, task parameters.

Understanding biases in how we move the eyes can provide powerful new insights into the decision about where to look in complex scenes. In a remarkable study \cite{tatler2009prominence}, Tatler and Vincent  have shown that a model based solely on these biases and therefore blind to current visual information can outperform  salience-based approaches (in particular, they compared against the well known model  proposed by Itti \emph{et al}~\cite{IttiKoch98,walther2006} -  see Tom Foulsham's Chapter in this book for an introduction, and the following Section \ref{sec:tour} for a probabilistic framing of saliency models).

Summing up, the adoption of an approach based on stochastic processes  bring about  significant advantages.  First,  the analysis and modelling of eye movements  can benefit of all the ``tools'' that have been developed in the  field of stochastic processes and  time series. For example,  the approach opens the possibility of treating visual exploration strategies in terms of \emph{random walks}, e.g., \cite{engbert2006microsaccades,engbert2011integrated,carpenter1995neural}.  Indeed, this kind of conceptual shift  happened to the modern developments of econophysics \cite{mantegna2000} and  finance \cite{paul2013stochastic}.
Further, by following this path,  visual exploration can be reframed in terms of   \emph{foraging} strategies an intriguing perspective that has recently gained currency \cite{wolfe2013time,cain2012bayesian,bfpha04,BocFerSMCB2013,BocCOGN2014}. Eventually, by embracing the stochastic perspective leads  to the possibility of exploiting all the results so far achieved in the ``hot'' field of Statistical Machine Learning.

Thus, in this Chapter, we pursue the following learning objectives 
\begin{enumerate}
\item Casting eye movement analysis and modelling in probabilistic terms (Section \ref{sec:tour});
\item Understanding the essential concepts of  stochastic process, such as Markov processes, and microscopic/macroscopic levels  of description (Sections \ref{sec:stoch}, \ref{sec:markov});
\item Setting the basics of random walk analyses and modelling of eye movements either within the scope of the Central Limit Theorem or beyond, towards anomalous walks and diffusions (Sections \ref{sec:wild});
\item Moving from the analyses of scan path patterns to the inference of mental patterns by introducing the basic tools of modern probabilistic Machine learning  (Section \ref{sec:ML}). 
\end{enumerate}

As to the eye movements concepts exploited in the modelling review of  Section~\ref{sec:tour}, it is worth referring to  the related Chapters of this book. 

For all the topics covered hereafter we assume a basic calculus level or at least a familiarity with the concepts of differentiation  and integration. Box~\ref{tab:int} provides a brief introductory note. However, find an A-level text book with some diagrams if you have not seen this before. Similarly, we surmise reader's conversance with elementary notions of probability and statistics.

\begin{table}
\begin{svgraybox}
\caption{\textbf{Interlude: differential and integral calculus with no pain}}
\label{tab:int}
Differential calculus deals with the concept of \textbf{rate of change}. The rate of change of a function $f(x)$ is defined as the ratio of the change in $f$ to the change in $x$. Consider Fig.\ref{Fig:deriv} showing a plot of $f$ as a function of $x$. There are intervals during which $f$ increases and other intervals where $f$ decreases. We can quantify the ups and downs of the changes in the values of $f$ by estimating the slope, i.e., the change in the variable $f$ over a given interval $\Delta x$, say between $x_1$ and $x_2$. Denote the interval or average slope by 
$$  \frac{\Delta f}{\Delta x} = \frac{f(x_2)- f(x_2)}{x_2 -x_1} =\frac{f(x + \Delta x)- f(x)}{\Delta x} = \frac{\textsf{rise}}{\textsf{run}},$$
with $\Delta x = x_2 -x_1$. What happens as the  interval $\Delta x$ becomes smaller and smaller and approaches zero, formally, $\Delta x \rightarrow 0$?

In that case the interval or average rate of change shrinks to the \emph{instantaneous rate of change}. This is exactly what is computed by the \textbf{derivative} of $f$ with respect to $x$:
$$ \frac{df}{dx} = \lim_{\Delta x \rightarrow 0} \frac{f(x + \Delta x)- f(x)}{\Delta x}.$$
If you prefer thinking in a geometric way, the derivative at a point $x$ provides the slope of the tangent of the curve at $x$.

As an example, we calculate the derivative of the function $f(x)=x^2$. First, write the term $f(x + \Delta x)$:
$$f(x + \Delta x) = (x + \Delta x)^2= x^2+ 2x\Delta x + \Delta x^2 $$
Then, subtract $f(x)$ and divide by $\Delta x$:
$$\frac{f(x + \Delta x)- f(x)}{\Delta x} =  \frac{x^2+ 2x\Delta x + \Delta x^2 - x^2}{\Delta x} = 2x+ \Delta x $$
Now in the limit $\Delta x \rightarrow 0$ we shrink $ \Delta x$ to zero,  i.e., 
$$\lim_{\Delta x \rightarrow 0} 2x+ \Delta x = 2x.$$
Eventually,
$$ \frac{d(x^2)}{dx} = 2x.$$

If differential calculus has to do with rates of change, \textbf{integral calculus} deals with sums of many tiny incremental quantities. For instance, consider a continuous function $f$ such as the one plotted in Fig. \ref{Fig:integral} and the following sum
$$ \sum_{i=1}^{n}f(x_{i}) \Delta x = f(x_{1}) \Delta x + f(x_{2}) \Delta x + \cdots + f(x_{n}) \Delta x.$$
Here  the uppercase greek letter $\sum$  indicates a sum of successive values defined by $i$ and where $\Delta x = \frac{b-a}{n}$ and $x_{i} = a + i \Delta x$. Note that  the term
$$f(x_{i}) \Delta x = \textsf{height} \times \textsf{width} = \delta A_i$$
computes the area $\delta A_i$ of the $i$-th rectangle (see  Fig. \ref{Fig:integral}). Thus, the (Riemann) sum written above approximates the area defined by the continuous function $f$ within the left and right limits $a$ and $b$, as a the sum of tiny rectangles  covering the  area under $f$.
The sum transforms into the  \textbf{integral}
$$ \int_{a}^{b} f(x) dx = \lim_{\Delta x \rightarrow 0}  \sum_{i=1} f(x_{i}) \Delta x$$
when $\Delta x$ shrinks to $0$ (i.e. in the limit $\Delta x \rightarrow 0$) and the number $n$ of intervals grows very large ($ n \rightarrow \infty$). 

There is a deep connection between integration and differentiation, which is stated by the \textbf{fundamental theorem of calculus}: the processes of integration and differentiation are reciprocal, namely, the derivative of an integral is the original integrand.


\end{svgraybox} 
\end{table}

\begin{figure}[t]
\sidecaption[t]
\includegraphics[scale=0.15,keepaspectratio=true]{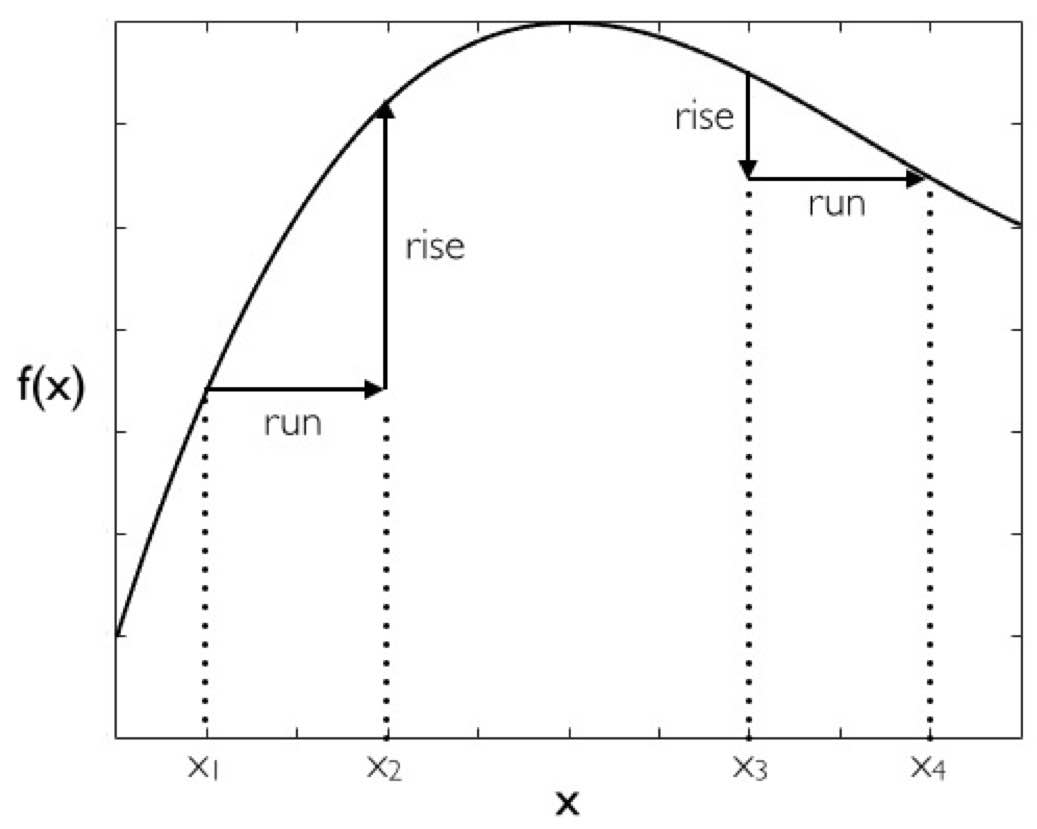}
\caption{A plot of the values of a function $f$ as a function of $x$, showing a region of negative rate of change or slope (between $x_1$ and $x_2$) and a region of positive change (between $x_3$ and $x_4$).}.
\label{Fig:deriv}
\end{figure}

\begin{figure}[t]
\sidecaption[t]
\includegraphics[scale=0.15,keepaspectratio=true]{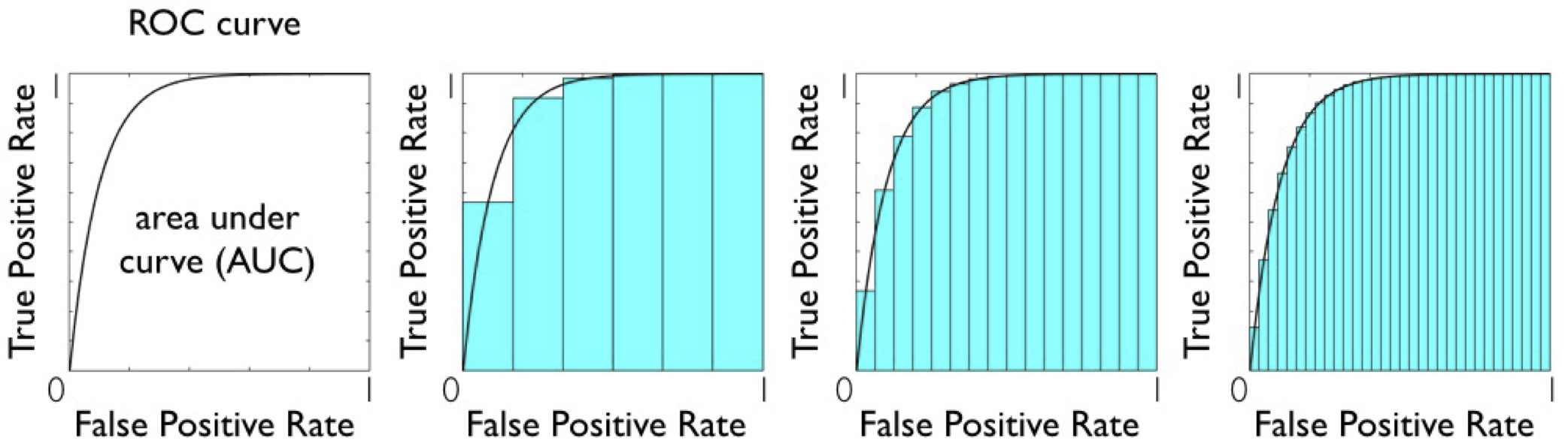}
\caption{An illustration of the integral concept by using Receiver Operating Characteristic (ROC) analysis, one of the methods  to evaluate salience map algorithms. Continuous saliency maps are processed as a binary classifier applied on every pixel:  the image pixels of the ground truth, as well as those of the prediction, are classified as fixated (or salient) or as not fixated (not salient). A simple threshold operation is used for this purpose. The ROC curve is  exploited to display the classification result for a varying  threshold used, each threshold value originating a number of False Positives and True Positives. An ROC curve is shown in the leftmost graph, and it has been obtained by plotting the False Positive Rate (FPR) as a function of the True Positive  Rate (TPR). The ROC area, or the \textbf{area under curve} (AUC), provides a measure indicating the overall performance of the classification. The second graph shows the approximate calculus of the AUC as the (Riemann) sum of approximating rectangles as discussed in Box \ref{tab:int}. Third and fourth graphs demonstrate how the computed AUC becomes more and more precise for increasing number of rectangles ($ n \rightarrow \infty$) and diminishing rectangle widths ($\Delta x \rightarrow 0$). In such limit the sum $ \sum $ becomes the integral $ \int $.}
\label{Fig:integral}
\end{figure}

\section{Historical annotations}
\label{sec:hist}
Stochastic modelling has a long and wide history encompassing different fields. The notion of stochastic trajectories possibly goes back to the scientific poem \emph{``De Rerum Natura''} (``On the Nature of Things'', circa 58 BC) by Titus Lucretius Carus:
\begin{quote}
All things keep on in everlasting motion, / Out of the infinite come the particles, / Speeding above, below, in endless dance.
\end{quote}
 Yet, it is towards the end of the nineteenth century that a major breakthrough occurred.
As Gardiner put it \cite{gardiner2009stochastic}: 
\begin{quote}
Theoretical science up to the end of the nineteenth century can be viewed as the study of solutions of differential equations and the modelling of natural phenomena by deterministic solutions of these differential equations. It was at that time commonly thought that if all initial data could only be collected, one would be able to predict the future with certainty.
\end{quote}

Quantum theory, on the one hand, and the concept of chaos (a simple differential equation, due to any error in the initial conditions that is rapidly magnified,  can give rise to essentially unpredictable behaviour) on the other, have undermined such a Laplacian conception.
However, even without  dealing with quantum and chaotic phenomena, there are limits to deterministic predictability. Indeed, the rationale behind this Chapter is that of ``limited predictability'' \cite{gardiner2009stochastic} mostly arising when \emph{fluctuating} phenomena are taken into account. As a matter of fact, stochastic processes are much closer to observations than deterministic descriptions in modern science and everyday life.  Indeed, it is the existence of fluctuations that calls out for a statistical account. Statistics had already been used by Maxwell and Boltzmann in their  gas theories. But it is Einstein's explanation~\cite{einstein1905motion} of the nature of Brownian motion (after  the  Scottish botanist Robert Brown who observed under microscope, in 1827,  the random highly erratic motion of small pollen grains suspended in water), which can be regarded as the beginning of stochastic modelling of natural phenomena\footnote{Actually, the first  who noted the Brownian motion was the Dutch physician, Jan Ingen-Housz in 1794,  in the Austrian court of Empress Maria Theresa. He observed that finely powdered charcoal floating on an alcohol surface executed a highly random motion}. Indeed, Einstein's elegant paper is worth a look, even by the non specialist, since containing  all the basic concepts which will make up the subject matter of this Chapter: the Markov assumption, the Chapman-Kolmogorov equation,  the random or stochastic differential equation for a particle path, the diffusion equation describing the behaviour of an ensemble of particles, and so forth.
Since then, research in the field has quickly progressed. For an historically and technically detailed account  the reader might refer to Nelson's \emph{``Dynamical Theories of Brownian Motion''}\footnote{Freely available at \url{https://web.math.princeton.edu/~nelson/books/bmotion.pdf}}, \cite{nelson1967dynamical}.

To make a long story short, Einstein's seminal paper has provided inspiration for subsequent works, in particular that by Langevin~\cite{langevin1908theorie} who, relying upon the analysis of a single particle random trajectory, achieved a different derivation of Einstein's results. Langevin's equation was the first example of the stochastic differential equation, namely a differential equation with a random term  and  whose solution is, in some sense, a random function. Langevin initiated a train of thought that, in 1930, culminated in the work  by Ornstein and Uhlenbeck~\cite{uhlenbeck1930theory}, representing a truly dynamical theory  of Brownian motion.
Although the approach of Langevin was improved and expanded upon by Ornstein and Uhlenbeck, some more fundamental problems remained, markedly related to the differentiability and integrability of a stochastic process. The major contribution to the mathematical theory of Brownian motion has been brought by Wiener~\cite{wiener1930generalized}, who  proved that the trajectories of a  Brownian process  are continuous almost everywhere but are not differentiable anywhere. These problems were addressed by Doob (who came to probability from complex analysis) in  his famous paper of 1942 \cite{doob1942brownian}.  Doob aimed  at applying the methods and results of modern probability theory to the analysis of the Ornstein-Uhlenbeck distribution. His efforts, together with those of   It\^o, Markov,  Kac, Feller, Bernstein, L\'evy, Kolmogorov, Stratonovich and others  lead the  \emph{theory of random processes}  to become an important branch of mathematics. A nice historical account of  stochastic processes from 1950 to the present is provided by Meyer~\cite{meyer2009stochastic}.

Along with theoretical achievements, many more cases of random phenomena materialised in science and engineering. The major developments  came in the 1950's and 1960's through the analysis of electrical circuits and radio wave propagation.  A great deal of highly irregular electrical signals were given the collective name of ``noise'': uncontrollable fluctuations in electric circuits (e.g, thermal noise, namely the distribution of voltages and currents in a network due to thermal electron agitation); scattering of electromagnetic waves caused by inhomogeneities in the refractive index of the atmosphere (fading). 
The fundamental theorem of Nyquist is based on the principle of thermal equilibrium, the same used by Einstein and Langevin \cite{papoulis2002probability}. Beyond those early days the theory of random processes has become a central topic in the basic training of engineers, and lays the foundation for spectral representation and estimation of signals in noise, filtering and prediction, entropy and information theory~\cite{papoulis2002probability}. Clearly, electrical noise, albeit very important, is far from a unique case. As other examples, one might consider the pressure, temperature and velocity vector of a fluid particle in a turbulent flow. A substantial overlap between the topics of neuroscience and stochastic systems has been acknowledged \cite{laing2010stochastic}.

Interestingly, beyond the realm of  the natural sciences and engineering, analyses of the random character of stock market prices started to gain currency in the 1950's. Osborne ``rediscovered'' the Brownian motion of stock markets in 1959 \cite{osborne1959brownian}. Computer simulations of ``microscopic'' interacting-agent models of financial markets have been performed  as early as 1964~\cite{stigler1964public}. Brownian motion became an important model for the financial market: Paul  Samuelson for his contributions on such topic  received the 1970 Nobel Prize in Economics; in 1973,   Merton and  Scholes, in collaboration with the late Fischer Black,  have used the geometric Brownian motion to construct a theory for determining the price of stock options; their achievements were also honoured by the Nobel Prize  (Scholes and Merton, 1997). The theory represents a milestone in the development of mathematical finance and today's daily capital market practice. Interestingly enough,  such body of work builds on the early dissertation of a PhD student of Henri Poincar\'e, named Louis Bachelier. In 1900 Bachelier defended his thesis entitled \emph{``Th\'eorie de la Sp\'eculation''} at the Sorbonne University of Paris~\cite{bachelier1900theorie}. He had developed,  five years before Einstein,  the theory of the random walk as a suitable probabilistic description for price fluctuations on the financial market. Unfortunately,  such humble application   was not acknowledged by the scientific community at that time; hence, Bachelier's work fell into complete oblivion until the early 1940's (when  It\^o used it as a motivation to introduce his calculus).  Osborne himself made no mention of it~\cite{osborne1959brownian}. For an historical account of the role played by stochastic processes in the development of mathematical finance theory, see \cite{jarrow2004short}.

A relevant step, which is of major importance for this Chapter, was taken by Richardson's work~\cite{richardson1926atmospheric}. Twenty years later Einstein  and Langevin works,  he presented empirical data related to the ``superdiffusion'' of an admixture cloud in a turbulent atmosphere being  in contradiction with the normal diffusion.  Such anomalous diffusion can be explained as a deviation of the real statistics of fluctuations from the Gaussian law. Subsequently, anomalous diffusion in the form of  L\'evy flights has been discovered in many other physical, chemical, biological, and financial systems (L\'evy flights, as we will see, are stochastic processes characterised by the occurrence of extremely long jumps, so that their trajectories are not continuous anymore). The first studies on the subject were those of Kolmogorov~\cite{kolmogorov1941dissipation} on the scale invariance of turbulence in the 1940's. This topic was later on addressed by many physicists and mathematicians, particularly by Mandelbrot (the father of fractal mathematics). In the 1960s  he  applied it not only to the phenomenon of turbulence but also to the behaviour of financial markets~\cite{mandelbrot1963variation}. As Mandelbrot lucidly summarised~\cite{mandelbrot1963variation}:
\begin{quote}
Despite the fundamental importance of Bachelier's  process, which has come to be called ``Brownian motion,'' it is now obvious that it does not account for the abundant data accumulated since 1900 by empirical economists, simply because the empirical distributions of price changes are usually too ``picked'' to be relative to samples from Gaussian populations.
\end{quote}
  An historical  but rather technical perspective on anomalous diffusion and L\'evy flights is detailed by Dubkov \emph{et al.}~\cite{dubkov2008levy}; a more affordable presentation is outlined by Schinckus~\cite{schinckus2013physicists}. Today, these kinds of processes are important to characterise a multitude of systems (e.g., microfluidics, nanoscale devices, genetic circuits that underlie cellular behaviour). L\'evy flights are recognised to underlie many aspects of human dynamics and behaviour~\cite{baronchelli2013levy}. Eye movement processes make no exception, as we will see.
  
  Nowadays, the effective application of the theory of random processes and, more generally, of probabilistic models in the real world is gaining pace. The advent of cheap computing power and the developments in Markov chain Monte Carlo simulation produced a revolution within the field of Bayesian statistics around the beginning of the 1990's. This allowed a true ``model liberation''.  Computational tools and the latest developments in approximate inference, including both deterministic and stochastic approximations, facilitate coping  with complex stochastic process based
models that previously we could only dream of dealing with~\cite{insua2012bayesian}.  We are witnessing an impressive cross-fertilisation between random process theory and the more recently established areas of Statistical Machine Learning and Pattern Recognition,  where  the commonalities in  models and techniques emerge, with Probabilistic Graphical Models playing an important role in  guiding intuition~\cite{barber2011bayesian}. 


\begin{table}
\begin{svgraybox}
\caption{\textbf{Visual attention models: a brief critical review}}
\label{tab:models}

 In the field of psychology, there exists a wide variety of theories and models on visual attention (see, e.g., the review by Heinke and Humphreys \cite{heinke2005computational}). Among the most influential for computational attention systems: the well known Treisman's  \textbf{Feature Integration Theory} (FIT) \cite{treisman1980feature,treisman1998feature};  Wolfe's \textbf{Guided Search Model} \cite{wolfe1994guided}, aiming at explaining and predicting the results of visual search experiments; Desimone and Duncan's \textbf{Biased Competition Model} (BCM, \cite{desimone1995neural}), Rensink's triadic architecture \cite{rensink2000dynamic}, the \textbf{Koch and Ullman's model} \cite{koch1985shifts}, and Tsotsos' \textbf{Selective Tuning} (ST) model \cite{tsotsos1995modelling}. 

Other psychophysical models  have  addressed  attention modelling in a more formal framework. One notable example is Bundensen's \textbf{Theory of Visual Attention} (TVA, \cite{bundesen1998}), further developed by Logan into the \textbf{CODE theory} of visual attention (CTVA, \cite{logan1996code}). Also,  theoretical approaches  to visual search have been devised by exploiting Signal Detection Theory  \cite{palmer2000psychophysics}.
  
At a different level of explanation, other proposals have  been conceived in terms  of  connectionist models, such as 
\textbf{MORSEL} (Multiple Object Recognition and attentional SELection, \cite{mozer1987}), \textbf{SLAM} (SeLective Attention Model) \cite{phaf1990slam}, \textbf{SERR} (SEarch via Recursive Rejection) \cite{humphreys1993search}, and  \textbf{SAIM} (Selective Attention for Identification Model by Heinke and Humphreys  \cite{heinke2003attention}) subsequently refined in the Visual Search SAIM (VS-SAIM) \cite{heinke2011modelling}.

To a large extent, the psychological literature was conceived and fed on simple stimuli, nevertheless  the key role that the above models continue to play in understanding attentive behaviour should not be overlooked. For example,  many current computational approaches, by and large,  build upon the bottom-up salience based model by  Itti \emph{et al.} \cite{IttiKoch98}, which in turn is the computational counterpart of Koch and Ullman and  Treisman's FIT models.
The seminal work of Torralba et al.  \cite{Torralba}, draws on an important component of Rensink's triadic architecture \cite{rensink2000dynamic},  in that it considers contextual information  such as gist - the abstract meaning of a scene, e.g., a city scene, etc. - and  layout - the spatial arrangement of the objects in a scene. 
More recently, Wischnewski \emph{et al.} \cite{anna} have presented a computational model that integrates Bundensen's TVA \cite{bundesen1998}.

However, in the last  three decades, psychological models  have been adapted and extended in many respects, within  the \textbf{computational vision} field where the goal is to deal with  attention models and systems that are able to cope  with natural complex scenes rather than simple  stimuli and synthetical images (e.g., see \cite{frintrop2010} and the most recent review by Borji  and Itti \cite{BorItti2012}). 
The adoption of complex stimuli has sustained a new brand of computational theories, though this theoretical development is still at an early stage: up to this date, nobody has really succeeded in predicting the sequence of fixations of a human observer looking at an arbitrary scene \cite{frintrop2010}. This is not surprising given the complexity of the problem. One might think that issues of generalisation from simple to complex  contexts are nothing more than a minor theoretical inconvenience; but, indeed, the generalisation from simple to complex  patterns might not be straightforward. As it has been noted in the case of attentive search,  a model  that exploits handpicked features may fail utterly when dealing with realistic objects or scenes \cite{zelinsky2008theory}.

Current approaches within this field suffer from a number of limitations:  they mostly rely on a low-level salience based representation of the visual input, they  seldom take into account the task's role, and eventually they overlook the eye guidance problem, in particular the actual generation of gaze-shifts (but see Tatler \emph{et al} \cite{TatlerBallard2011eye} for a lucid critical review of current methods). We will  discuss such limitations in some detail in Section~\ref{sec:tour}.
\end{svgraybox} 
\end{table}

\section{A probabilistic tour of  current computational models of eye movements and visual attention (with some criticism)}
\label{sec:tour}

Many models in psychology and in the computational vision literature have investigated limited aspects of the problem of  eye movements in visual attention behaviour  (see Box \ref{tab:models}, for a quick review).  And, up to now, no model has really succeeded in predicting the sequence of fixations of a human observer looking at an arbitrary scene \cite{frintrop2010}.

\begin{table}
\begin{svgraybox}
\caption{\textbf{Dangerous relationships: A rendezvous with Bayesian Probabilities}}
\label{tab:prob}
We assume  the readers  to be  already familiar with the elementary notions (say, undergrad level)  of probability and \textbf{random variables} (RVs).  Thus, a warning. Sometimes we talk about probabilities of events that are ``out there'' in the world. The face of a flipped coin is one such event. But sometimes we talk about probabilities of events that  are just possible \textbf{beliefs} ``inside the head.'' Our belief about the fairness of a coin is an example of such an event.  Clearly, it might be bizarre to say that we randomly sample from our beliefs, like we  sample from a sack of coins. To cope with such embarrassing situation, we shall use probabilities  to express our information and beliefs about unknown quantities. $P(A)$ denotes the probability that the event $A$ is true. But event $A$ could stand for   logical expressions such as ``there is a red car in the bottom of the scene'' or ``an elephant will enter the pub''. In this perspective, probability is used to quantify our uncertainty about something; hence, it is fundamentally related to information rather than repeated trials. Stated more clearly:   we are adopting the Bayesian interpretation of probability in this Chapter. 

Fortunately, the basic rules of probability theory are the same, no matter which interpretation is adopted (but not that smooth, if we truly addressed  inferential statistics). For what follows,  we just need to refresh a few.

Let $X$ and $Y$ be RVs, that is numbers associate to events. For example, the quantitative outcome of a survey, experiment or study is a RV; the amplitudes of saccades or the fixation duration times recorded in a trial are RVs.  In Bayesian inference a RV (either discrete or continuous) is defined as an unknown numerical quantity about which we make probability statements. 
 Call $P(X,Y)$ their joint probability. The \textbf{conditional probability} of $X$ given $Y$ is:
\begin{equation}
\label{eq:cond}
P(X \mid Y) \equiv \frac{P(X, Y)}{P(Y)} \text{   if   } P(Y)\neq 0.
\end{equation}

In Bayesian probability we always deal with conditional probabilities: at least we condition on the assumptions or set of \textbf{hypotheses} $\mathcal{H}$ on which the probabilities are based. In data modelling and Machine Learning, the following holds~\cite{Mackay}:  
\begin{center}
\emph{You cannot do inference without making assumptions}
\end{center} 
Then, the  rules below will be useful:
\begin{description}
  \item[\textbf{Product rule} (or chain rule)] 
\begin{equation}
\label{eq:prod}
P(X, Y \mid \mathcal{H}) = P(X \mid Y,  \mathcal{H}) P(Y \mid  \mathcal{H})
\end{equation}

  \item[\textbf{Sum rule} (marginalisation)]
\begin{eqnarray}
\label{eq:sum }
P(Y \mid  \mathcal{H}) &=& \sum_{X} P(X, Y \mid \mathcal{H})     \text{    (discrete RVs)}  \\
P(Y \mid  \mathcal{H}) &=& \int_{X} P(X, Y \mid \mathcal{H}) dX    \text{    (continuous RVs)}
\end{eqnarray}
\item[\textbf{Bayes' rule} (see Fig.~\ref{Fig:Bayes} for a simple example)] 
  \begin{equation}
\label{eq:Bayes}
P(X \mid Y, \mathcal{H}) =  \frac{P(Y \mid X,  \mathcal{H}) P(X \mid  \mathcal{H})}{P(Y \mid  \mathcal{H})} \leftrightarrow 
\text{ posterior} = \frac{ \text{ likelihood}  \times  \text{prior}}{\text{evidence}}
\end{equation}
\end{description}
To avoid burying the reader under notations, we have used $P(\cdot)$ to denote both the probability of a discrete outcome (\textbf{probability mass function}, PMF) and the probability of a continuous outcome (\textbf{probability density function}, pdf). We  let context make things clear. Also,  we may adopt the form $X=x$ for a specific choice of value (or outcome) of the RV $X$. Briefer notation will sometimes be used: for example, $P(X= x)$ may be written as $P(x)$.  A bold $\mathbf{X}$ might denote a set of RVs or a random vector/matrix.

The ``bible'' of the Bayesian approach is the treatise of Jaynes~\cite{jaynes2003probability}. A succinct introduction with an eye to \textbf{inference and learning} problems can be found in Chapter 2 of the beautiful book by MacKay~\cite{Mackay}, which is also available for free online, \url{http://www.inference.phy.cam.ac.uk/mackay/itila/}. 

\end{svgraybox} 
\end{table}

\begin{figure}[t]
\sidecaption[t]
\includegraphics[scale=0.15,keepaspectratio=true]{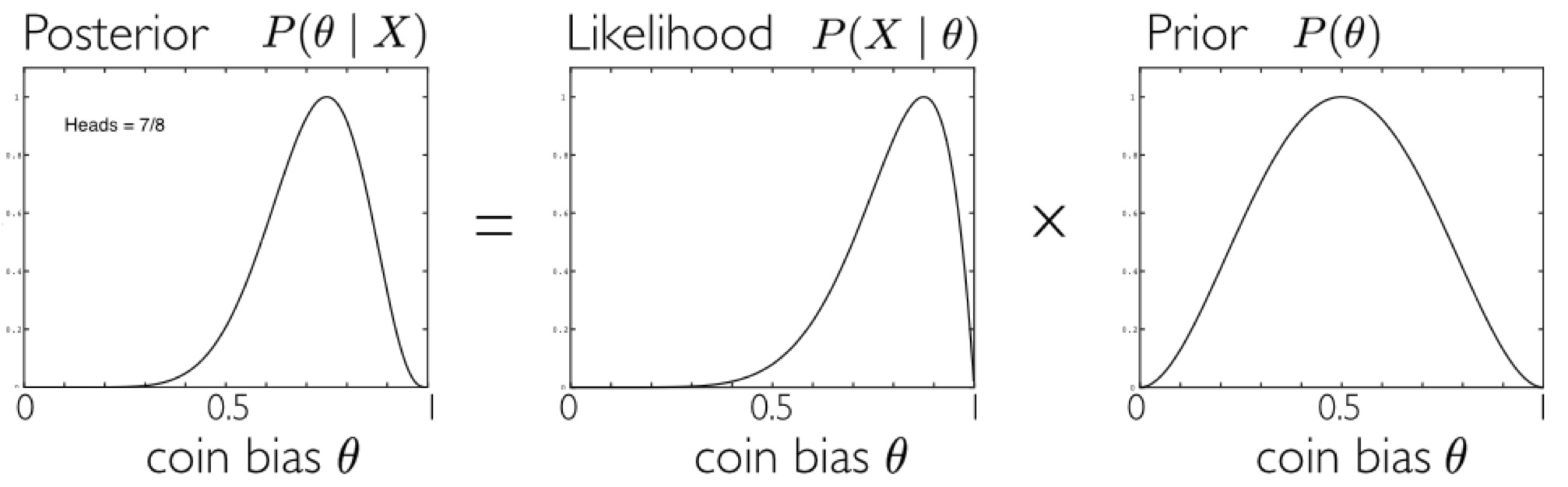}
\caption{An illustration of the use of the Bayes' rule for inferring the bias of a coin  on the basis of coin tossing results. The prior probability $P(\theta)$ for the coin bias $\theta$ captures the assumption that the coin is likely to be a fair one (the pdf is ``peaked'' on $\theta = 0.5$). However, $7$ heads occur after $8$ tosses. Such experimental result  is captured by the shape of the likelihood $P(X \mid \theta)$ strongly biased to the right. Bayes' rule computes the posterior pdf $P(\theta \mid X)$ by ``updating'' the initial prior through the ``observed'' likelihood (the evidence term is not shown in the figure and it has been treated as a normalisation factor to constrain probabilities between $0$ and $1$)}.
\label{Fig:Bayes}
\end{figure}

The issue of devising a computational model of eye guidance as related to visual attention -  i.e. answering the question \emph{Where to Look Next?} in a formal way - can be set in a probabilistic Bayesian framework (see Box~\ref{tab:prob} for a brief introduction). Tatler and Vincent~\cite{tatler2009prominence} have re-phrased this question in terms of Bayes' rule:
\begin{equation}
\overbrace{P(\mathbf{x} \mid \mathcal{D})}^{\text{posterior prob. of gaze shift }} =  \frac{ \overbrace{P(\mathcal{D} \mid \mathbf{x} )}^{\text{data likelihood under the shift}}} {P(\mathcal{D})}   \overbrace{P(\mathbf{x})}^{\text{gaze shift prior}}  ,
\label{eq:BayesTatler}
\end{equation}
\noindent where $\mathbf{x} = \mathbf{x}_{F}(t) - \mathbf{x}_{F}(t-1)$ is the random vector representing the gaze shift (in \cite{tatler2009prominence}, saccades), and $\mathcal{D}$ generically stands for the input data. 
As Tatler and Vincent put it, ``The beauty of this approach is that the data could come from a variety of data sources such as simple feature cues, derivations such as Itti's definition of salience, object-or other high-level sources''.

In Eq. \ref{eq:BayesTatler}, the first term on the right hand side accounts for the likelihood of particular visual data (e.g., features, such as edges or colors)  occurring at a gaze shift target location normalized by  $P(\mathcal{D})$ the pdf of these visual data occurring in the environment. 
As we will see in brief, this first term bears a close resemblance to approaches previously employed to evaluate the possible involvement of visual features in eye guidance.

Most interesting, and related to issues raised in the introductory Section, is the Bayesian prior $P(\mathbf{x})$, i.e., the probability of shifting the gaze to a location \emph{irrespective of the visual information} at that location. Indeed,  this term will encapsulate any systematic tendencies in the manner in which we explore scenes with our eyes. The striking result obtained by Tatler and Vincent~\cite{tatler2009prominence} is that if we  learn $P(\mathbf{x})$ from actual observer's behaviour, then we can \textbf{sample} gaze shifts (cfr. Box \ref{tab:Sample}), i.e.,
\begin{equation}
\mathbf{x}(t) \sim P(\mathbf{x}), \; \;\; \; t=1,2,\cdots
\label{eq:priorTatlersamp}
\end{equation}
\noindent so to obtain  scan paths that, blind to visual information, out-perform feature-based accounts of eye guidance~\cite{tatler2009prominence}: $0.648$ area under the receiver operator curve (AUC, which has been illustrated in Fig. \ref{Fig:integral}) as opposed to $0.593$ for edge information (namely, an orientation map computed from edge maps constructed over a range of spatial scales, by convolving the image with four oriented odd-phase Gabor  filters)  and  $0.565$ for salience information as derived through the Itti \emph{et al} model ~\cite{IttiKoch98}\footnote{More precisely, they used  the latest version of Itti's salience algorithm, available at \url{http://www.saliencytoolbox.net}~\cite{walther2006}, with defaults parameters setting. One may argue that  since then the methods of saliency computation have developed and improved significantly so far. However, if one compares the predictive power results obtained by  salience maps obtained within the very complex computational framework of deep networks, e.g., via the PDP system (with fine tuning) \cite{jetley2016end},  against a simple central bias map (saliency inversely proportional to distance from centre, blind to image information), one can read an AUC performance of  $0.875$ against  $0.780$ on the large VOCA dataset \cite{jetley2016end} (on the same dataset, the Itti \emph{et al} model achieves $0.533$ AUC).  Note that a central bias map  can be computed in a few Matlab lines \cite{mathe2013action}.}

\begin{table}
\begin{svgraybox}
\caption{\textbf{When God plays dice: the art (and magic) of sampling}}
\label{tab:Sample}
Eye movements can be considered a natural form of sampling. Another example of actual physical sampling is tossing a coin, as in the example illustrated in Fig. \ref{Fig:FigSample}, or throwing dice. Nevertheless, we can (and need to) \textbf{simulate}  sampling that occurs in nature (and thus the underlying process).  Indeed, for both  computational modelling and analysis we assume of being capable of the fundamental operation of generating a sample $\mathbf{X}=\mathbf{x}$ from a probability distribution $P(\mathbf{X})$. We  denote the sampling action via the $\sim$ symbol:
\begin{equation}
\mathbf{x}\sim P(\mathbf{X}).
\label{eq:samp}
\end{equation}
For instance, tossing a coin like we did in the example of Fig. \ref{Fig:Bayes} can be simulated by sampling $\mathbf{x}$ from a \textbf{Bernoulli distribution}, $\mathbf{x}\sim Bern(\mathbf{X}; \theta)$, where $\theta$ is the parameter standing for the coin bias ($\theta= \frac{1}{2}=0.5$  denotes a fair coin).

Surprisingly, to simulate nature,  we  need a minimal  capability: that of generating realisations of RVs uniformly distributed on the interval $\left[ 0,1\right]$. In practical terms, we just need a programming language or a toolbox in which a \texttt{rand()} function is available implementing the $\mathbf{u} \sim Uniform(0 ,1)$ operation.  Indeed, given the RVs $\mathbf{u}$, we can generate the realisations of any other RV with appropriate ``transformations'' of $\mathbf{u}$.

There is a wide variety of ``transformations'' for generating samples, from simple ones (e.g. inverse transform sampling  and rejection sampling)   to more sophisticated, like those relying on \textbf{Markov Chain Monte Carlo} methods (e.g., \textbf{Gibbs sampling} and \textbf{Metropolis sampling}). Again,  MacKay's book~\cite{Mackay} provides a very clear introduction to the art of random sampling.

You can qualitatively assess the results of your computational sampling procedure using sample \textbf{histograms}.  Recall from your basic statistic courses that  an histogram  is an empirical estimate of the probability distribution of a continuous variable. It is obtained by "binning" the range of values -- that is, by dividing the entire range of values into a series of small intervals --,  and then counting how many values fall into each interval.
Intuitively, if we look at the empirical distribution of the set of samples $\{\mathbf{x}(t) \}_{t=1}^{T}$ obtained for a large number $T$ of sampling trials  $\mathbf{x}(t) \sim P(\mathbf{X})$, $t = 1,2, \cdots , T$,   we expect  the shape of the histogram to approximate the originating theoretical density. Examples are provided in Fig. \ref{Fig:FigSample} where $1000$ samples have been generated experimenting with the \textbf{Uniform distribution}, the \textbf{Gaussian distribution} and the \textbf{Cauchy distribution}, respectively. 
\end{svgraybox} 
\end{table}

\begin{figure}[h!]
\sidecaption[t]
\includegraphics[scale=0.17,keepaspectratio=true]{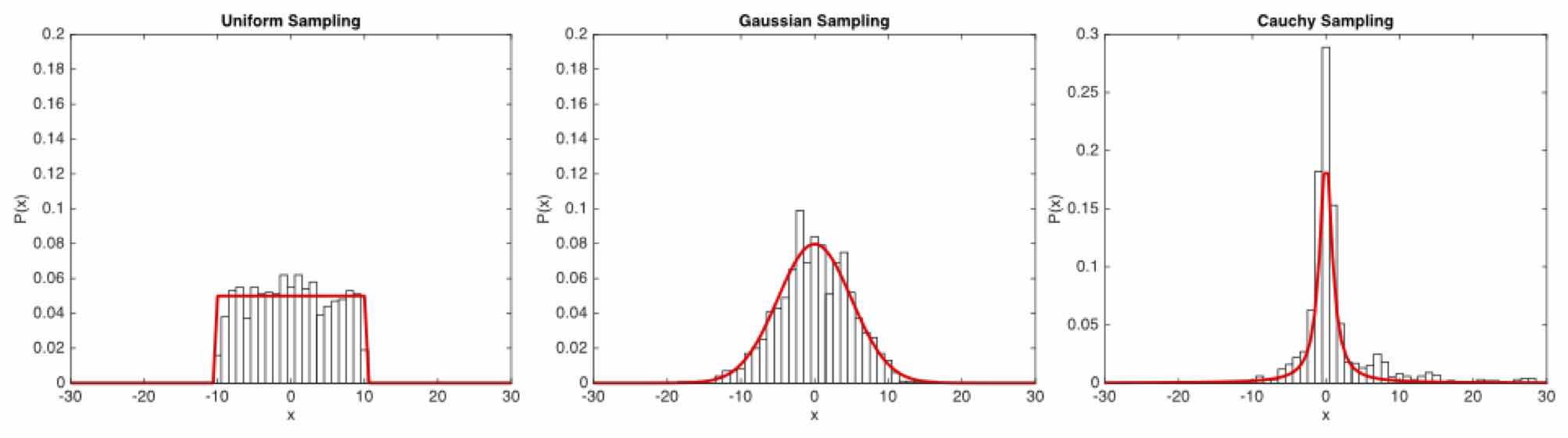}
\caption{From left to right, the empirical distributions (histograms)   for $n = 1000$ samples drawn from  Uniform,  Gaussian and  and Cauchy pdfs.  Uniform and Gaussian sampling have been performed via the Matlab functions \texttt{rand()}  and \texttt{randn()}, respectively; samples from the Cauchy pdf have been generated resorting to Metropolis sampling.   Each histogram is overlaid with the generating theoretical density  depicted as  a continuous red curve}.
\label{Fig:FigSample}
\end{figure}

Learning is basically obtained by empirically collecting through eye tracking the observer's  behaviour on an image data set (formally, the joint pdf $P(\mathbf{x}, \mathcal{D}$)  and then factoring out the informative content of the specific images, briefly, via marginalisation,i.e., $P(\mathbf{x}) = \sum_{\mathcal{D}}P(\mathbf{x}, \mathcal{D})$.

Note that the apparent simplicity of the prior term $P(\mathbf{x})$ hides a number of subtleties. For instance, Tatler and Vincent expand the  random vector  $\mathbf{x}$ in terms of its components, amplitude $l$ and direction $\theta$. Thus, $P(\mathbf{x})= P(l, \theta)$. This simple statement paves the way to different options. First easy option: such  RVs are marginally independent, thus,  $P(l, \theta) = P(l) P(\theta)$. In this case, gaze guidance, solely relying on biases, could be simulated by expanding Eq. (\ref{eq:priorTatlersamp}) via independent sampling of both components, i.e. at each time $t$, $l(t) \sim P(l(t)), \theta(t) \sim P(\theta(t))$. Alternative option: conjecture some kind of dependency, e.g. amplitude on direction, so that $P(l, \theta) = P(l \mid \theta) P(\theta)$. In this case, the gaze shift sampling procedure would turn into the sequence $\widehat{\theta}(t) \sim P(\theta(t)), l(t) \sim P(l(t) \mid \widehat{\theta}(t) )$. Further:  assume that there is some persistence in the direction of the shift. This gives rise to a stochastic process in which  subsequent directions are correlated, i.e., $\theta(t) \sim P(\theta(t) \mid \theta(t-1))$, and so on.

\begin{figure}[t]
\sidecaption[t]
\includegraphics[scale=0.15,keepaspectratio=true]{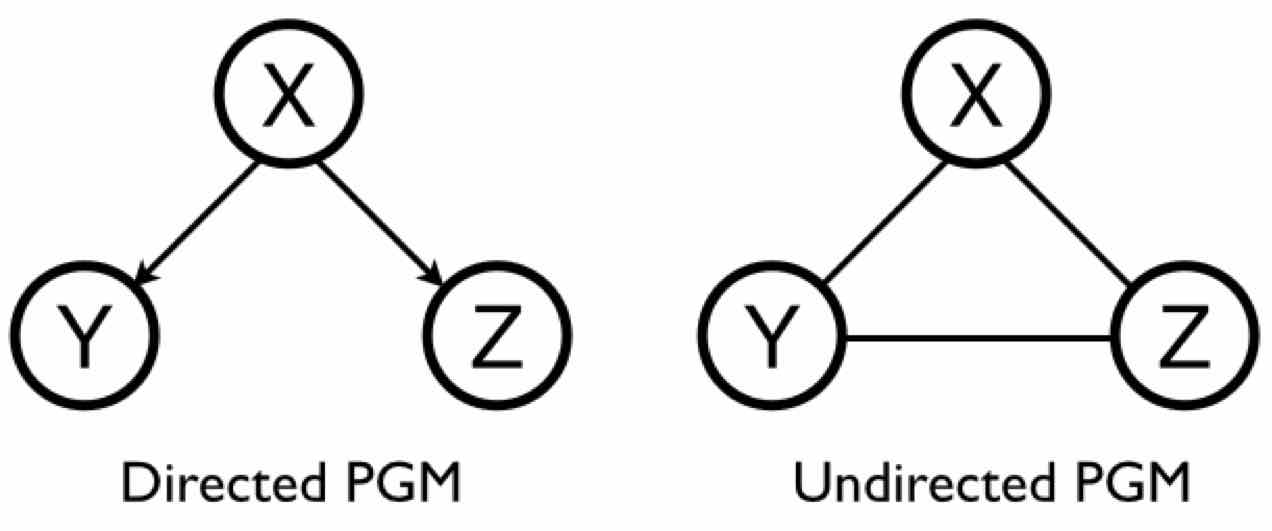}
\caption{Probabilistic Graphical Models: a directed PGM (left, a.k.a. Bayesian Network) and an undirected PGM (right, a.k.a. Markov Random Field). Nodes represents  RVs and arcs express probabilistic relationships between RVs}
\label{Fig:PGM}
\end{figure}
\begin{table}
\begin{svgraybox}
\caption{\textbf{Probabilistic Graphical Models (PGM)}}
\label{tab:PGM}
A PGM \cite{koller2009probabilistic} is a graph-based representation (see Fig. \ref{Fig:PGM}) where \textbf{nodes} (also called vertices)  are connected by \textbf{arcs} (or \textbf{edges}). In a PGM, each node represents a RV (or group of RVs), and the arcs express probabilistic relationships between these variables. 
Graphs where arcs are   arrows are \textbf{directed} PGM, a generalisation of \textbf{Bayesian Networks} (BN), well known in the Artificial Intelligence community. The other major class of PGMs are  \textbf{undirected} PGM (Fig. \ref{Fig:PGM}, right), in which the links have no directional significance, but are suitable to express soft constraints between RVs.  The latter are also known as \textbf{Markov Random Fields} (MRF), largely exploited in Computer Vision. 

We shall focus on directed PGM representations where arrows represent  conditional dependencies (Fig. \ref{Fig:PGM}, left). 
For instance   the arrow $X\rightarrow Y$  encodes the probabilistic dependency of RV $Y$ on $X$  quantified through the conditional probability  $P(Y \mid X)$.  Note that arrows do not generally represent causal relations, though  in some circumstances it could be the case. 
We will mainly exploit PGMs as a descriptive tool: indeed, 1) they provide a simple way to visualise the structure of a probabilistic model and can be used to design and motivate new models; 2) they offer  insights into the properties of the model, including conditional independence properties, which can be obtained by inspection of the graph. 

PGMs capture the way in which the joint distribution over all of the RVs can be decomposed into a product of factors each depending only on a subset of the variables.  Assume that we want to describe a simple object-based attention model (namely, the one presented at the centre of Figure~\ref{Fig:models}), so to deal with: (i) objects (e.g., red triangles vs. blue squares), (ii) their possible locations, and  (iii) the visual features  sensed from the observed scene. Such ``world' can be described by the  joint pdf $P(Objects, Location, Features)$ which we denote, more formally, through the RVs $O,L,F$:  $P(Objects, Location, Features) \equiv P(O, L, F)$.
 Recall that - via the product rule - the joint pdf could be factorised in a combinatorial variety of ways, all equivalent and admissible:
\begin{eqnarray}
P(O, L, F) &=& P(O \mid L,F) P(L \mid F) P(F)  \\
&=& P(L \mid O,F) P(O\mid F) P(F) \nonumber \\
&=& P(F \mid O,L) P(O\mid L) P(L) \nonumber \\
&=& \cdots \nonumber
\label{eq:OLFpdfdec}
\end{eqnarray} 
The third factorisation is, actually,   the  meaningful one: the likelihood of observing certain features (e.g, color) in the visual scene depends on \emph{what} kind of objects are present and on \emph{where} they are located; thus, the factor $P(F \mid O,L)$ makes sense.  $P(L)$ represents the prior probability of choosing certain locations within the scene (e.g., it could code the \emph{center bias} effect \cite{tatler2007central}). Eventually, the  $P(O\mid L)$ factor might code the prior probability of certain kinds of objects (e.g., we may live in a world where red triangles are more frequent than blue squares). As to $P(O\mid L)$ we can  assume that  the object location and object identity are independent, formally, $P(O\mid L)= P(O)$,   finally leading to 
\begin{eqnarray}
P(O, L, F) = P(F \mid O,L) P(O) P(L). 
\label{eq:OLFpdffin}
\end{eqnarray} 
This factorisation is exactly that captured by the structure of the directed PGM presented at the centre of  Figure~\ref{Fig:models}. Indeed, the graph renders the most suitable factorisation of the unconstrained joint pdf, under  the assumptions and the constraints we are adopting to build our model. We can ``query'' the PGM for making any kind of \textbf{probabilistic inference}. For instance, we could ask what is the posterior probability $P(O, L \mid F)$ of observing certain objects at certain locations given the observed features. By using the definition of conditional probability  and  Eq. \ref{eq:OLFpdffin}:  
\begin{eqnarray}
P(O, L \mid F) = \frac{P(O, L, F)}{P(F)} = \frac{P(F \mid O,L) P(O) P(L)}{P(F)} 
\label{eq:OLFquery}
\end{eqnarray}   
Complex computations for \textbf{ inference and learning} in sophisticated probabilistic models  can be expressed in terms of   graph-based algorithms. PGMs are a formidable tool to such end, and nowadays are widely adopted in modern probabilistic \textbf{Machine Learning} and \textbf{Pattern Recognition}. An affordable introduction can be found in Bishop~\cite{BishopPRML}. The PGM ``bible'' is the textbook by Koeller~\cite{koller2009probabilistic}.
\end{svgraybox} 
\end{table}


To summarise, by simply  taking into account the prior  $P(\mathbf{x})$, a richness of possible behaviours and analyses  are brought into the game. To further explore  this perspective, we recommend the  thorough and  up-to-date review by Le Meur  and Coutrot~\cite{le2016introducing}.

 Unfortunately, most computational accounts of eye movements and visual attention have overlooked this issue.  We noticed before, by inspecting Eq. (\ref{eq:BayesTatler})  that the term  $\frac{ P(\mathcal{D} \mid \mathbf{x} )} {P(\mathcal{D})} $ bears a close resemblance to many approaches proposed in the literature. This is an optimistic view. Most of the approaches actually discard the dynamics of gaze shifts, say $\mathbf{x}_{F}(t) \rightarrow \mathbf{x}_{F}(t+1)$, implicitly captured through the shift vector $\mathbf{x}(t)$. In practice, most models are more likely to be described  by a simplified version of Eq. (\ref{eq:BayesTatler}): 
\begin{equation}
\overbrace{P(\mathbf{x}_{F} \mid \mathcal{D})}^{\text{posterior prob. of gazing at}} =  \frac{\overbrace{ P(\mathcal{D} \mid \mathbf{x}_{F})}^{\text{data likelihood under gaze at}}} {P(\mathcal{D})}   \overbrace{P(\mathbf{x}_{F})}^{\text{prior prob. of gazing at}}  ,
\label{eq:BayesTatler2}
\end{equation}
By careful inspection, it can be noted that the posterior $P(\mathbf{x}_{F} \mid \mathcal{D}) $ answers the query ``What is the probability of \emph{fixating} location $\mathbf{x}_{F}$ given visual data $\mathcal{D}$?''.  Further, the prior  $P(\mathbf{x}_{F})$ accounts for the probability of \emph{fixating} location $\mathbf{x}_{F}$ irrespective of the visual information at that location. The difference between Eq. \ref{eq:BayesTatler} and Eq. \ref{eq:BayesTatler2} is subtle. But, as a matter of fact, Eq. \ref{eq:BayesTatler2} bears no dynamics. In probabilistic terms we may re-phrase this result as the outcome of an assumption of independence: 
\begin{equation}
P(\mathbf{x})   = P(\mathbf{x}_{F}(t) - \mathbf{x}_{F}(t-1)) \nonumber
  \simeq P(\mathbf{x}_{F}(t) \mid \mathbf{x}_{F}(t-1)) = P(\mathbf{x}_{F}(t)).
\end{equation}

To make things even clearer, let us explicitly substitute $\mathbf{x}_{F}$ with a RV $\mathbf{L}$ denoting locations in the scene, and $\mathcal{D}$ with RV $\mathbf{F}$ denoting features (whatever they may be);  then Eq.~\ref{eq:BayesTatler2} boils down to  the following
\begin{equation}
\overbrace{P(\mathbf{L} \mid \mathbf{F})}^{\text{posterior prob. of selecting location L}} =  \frac{ \overbrace{P(\mathbf{F} \mid \mathbf{L})}^{\text{feature likelihood under location L}}} {P(\mathbf{F})}   \overbrace{P(\mathbf{L})}^{\text{prior prob. of location L}}
\label{eq:models}
\end{equation}

\begin{figure}[t]
\sidecaption[t]
\includegraphics[scale=0.15,keepaspectratio=true]{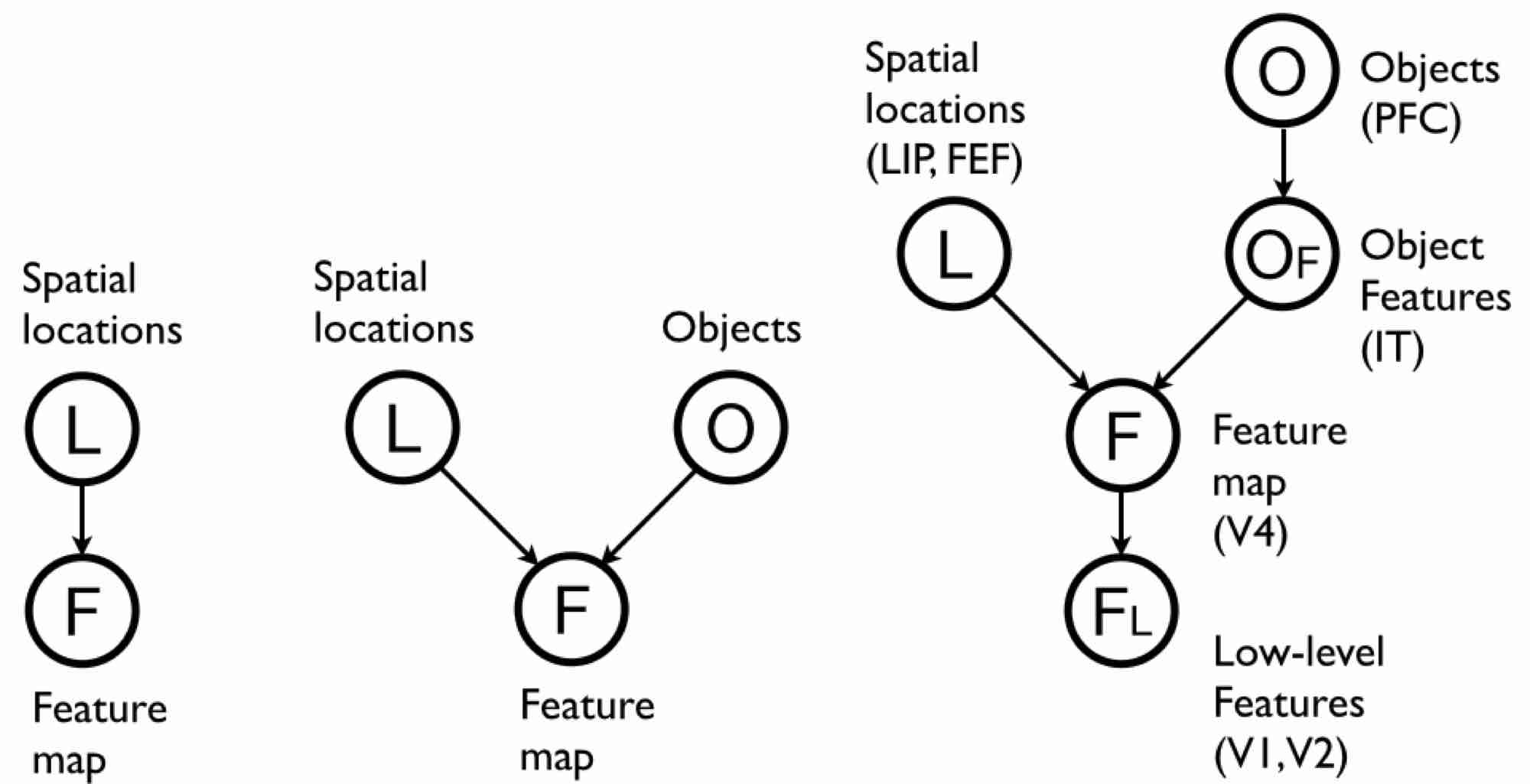}
\caption{PGMs of increasing level of representational complexity (left to right) that can account for most models proposed in the computational vision field. Left: feature-based model. Centre: object-based model. Right: the Bayesian model by Chikkerur  \emph{et al.} \cite{Poggio2010},  which extends the object-based model in the centre and  maps the resulting PGM structure  to brain areas underpinning visual attention: early visual areas V1 and V2,  V4,  lateral intraparietal (LIP), frontal eye fields (FEF), inferotemporal (IT), prefrontal cortex (PFC).}
\label{Fig:models}
\end{figure}

The feature-based \textbf{Probabilistic Graphical Model} underlying  this query (see Box~\ref{tab:PGM} for a brief PGM overview) is a very simple one and is represented on the left of Fig.~\ref{Fig:models}. 
As it can be seen, it is a subgraph of the  object-based model PGM (Fig.~\ref{Fig:models}, centre), which is the one previously discussed in Box~\ref{tab:PGM}.

Surprisingly enough, this simple model is sufficiently powerful  to account for a large number of visual attention models that have been proposed in computational vision. This can be easily appreciated by setting $P(\mathbf{F} \mid \mathbf{L}) = const.,  P(\mathbf{L})=const.$  so that Eq. \ref{eq:models} reduces to
\begin{equation}
\overbrace{P(\mathbf{L} \mid \mathbf{F})}^{\text{posterior prob. of selecting location L }} \propto  \overbrace{\frac{1}  {P(\mathbf{F})}}^{\text{salience at location L}} .
\label{eq:modelItti}
\end{equation}  
Eq.~\ref{eq:modelItti} tells that the probability of fixating a spatial location $\mathbf{L}= (x,y)$ is higher when ``unlikely'' features ($\frac{1}  {P(\mathbf{F})}$)  occur at that location. In a natural scene, it is typically the case of high contrast regions (with respect to either luminance, color, texture or motion) and clearly relates to entropy and information theory concepts \cite{BocICIAP2001}. This is nothing but the  most prominent salience-based model in the literature proposed by Itti \emph{et al} \cite{IttiKoch98}, which Eq.~\ref{eq:modelItti} only re-phrases in probabilistic terms.

A thorough reading of the recent review by Borji and Itti \cite{BorItti2012} is sufficient to gain the understanding that a great deal computational models so far proposed  are more or less variations  of this leitmotif (experimenting with different features, different weights for combining them, etc.).  The weakness of such a pure bottom-up approach has been largely discussed (see, e.g. \cite{TatlerBallard2011eye,foulsham2008,EinhauserSpainPerona2008}). Indeed, the effect of early saliency  on attention is likely to be    a correlational effect rather than  an actual causal one \cite{foulsham2008,schutz2011eye},  though  salience may be still more predictive than chance  while preparing for a memory test as discussed by Foulsham and Underwood \cite{foulsham2008}. 

Thus, recent efforts have tried to go beyond this simple stage with the aim of climbing the representational hierarchy shown in Fig. \ref{Fig:levels}. This entails a first shift from  Eq.~\ref{eq:modelItti}  (based on an oversimplified  representation)  back to Eq.~\ref{eq:models}.  Torralba \emph{et al.} \cite{Torralba} have shown that using prior knowledge on the typical spatial location of the search target, as well as  contextual information (the ``gist'' of a scene) to  modulate early saliency  improves its fixation prediction. 

Next shift is  exploiting object knowledge  for top-down ``tuning'' early salience; thus, moving to the PGM representation at the centre of Figure~\ref{Fig:models}. 
Indeed, objects and their semantic value  have been  deemed as  fundamental for visual attention and eye guidance (e.g., \cite{mozer1987,bundesen1998,rensink2000dynamic,heinke2011modelling}, but see Scholl \cite{scholl} for a review). 
For instance, when  dealing with faces within the scene,  a face detection step can provide a reliable  cue to complement early conspicuity maps, as it has been shown by Cerf \emph{et al} \cite{cerf2008predicting}, deCroon \emph{et al} \cite{postma2011}, Marat \emph{et al} \cite{marat2013improving}, or a useful prior for Bayesian integration with low level cues \cite{bocc08tcsvt}. This is indeed an important issue since faces may drive attention in a direct fashion \cite{cerf2009faces}. The same holds for text regions \cite{cerf2008predicting,BocCOGN2014}
Other notable exceptions are  those provided by Rao \emph{et al.} \cite{Rao2002},  Sun \emph{et al.} \cite{Sun2008}, the Bayesian models discussed by   Borji  \emph{et al.}  \cite{borji2012object} and  Chikkerur  \emph{et al.} \cite{Poggio2010}. In particular the model by Chikkerur  \emph{et al.}, which is shown at right of Fig. \ref{Fig:models} is the most complete to the best of our knowledge (though it does not consider contextual scene information \cite{Torralba}, but the latter could be easily incorporated). Interestingly enough, the authors  have the merit of making the effort of providing links between the structure of the PGM and the brain areas that could support  computations.

Further, again in the effort of climbing the representational hierarchy (Fig. \ref{Fig:levels}), attempts have been made for incorporating task and value information (see \cite{BocCOGN2014,schutz2011eye} for a brief review, and \cite{TatlerBallard2011eye} for a discussion).

Now, a simple question arises: where have the eye movements gone?

To answer such question is useful to summarise the brief overview above.  The common practice of computational approaches is to conceive the mapping (\ref{eq:mapping}), as a two step procedure: 
\begin{enumerate}
\item obtain a suitable representation $\mathcal{R}$, i.e., $\mathcal{D} \xmapsto[\mathbf{T}]{}  \mathcal{R}$; 
\item use $\mathcal{R}$ to generate the scanpath, $\mathcal{R} \xmapsto[\mathbf{T}]{}  \{\mathbf{x}_{F}(1), \mathbf{x}_{F}(2),\cdots \}$.
\end{enumerate}


Computational modelling has been mainly concerned with the first step: deriving a representation $\mathcal{R}$ (either probabilistic or not).
The second step, that is $\mathcal{R} \mapsto \{\mathbf{x}_{F}(1), \mathbf{x}_{F}(2),\cdots \}$, which actually brings in  the question of \emph{how} we look rather than \emph{where},  is seldom taken into account.

In spite of the fact that the  most cited work in the field,  that by Itti \emph{et al} \cite{IttiKoch98}, clearly addressed the \emph{how} issue  (gaze shifts as the result of  a Winner-Take-All, WTA, sequential selection of most salient locations), most  models simply overlook the eye movement problem.
The computed representation   $\mathcal{R}$ is usually evaluated in terms of its capacity for predicting the image regions that will be explored by covert and overt attentional shifts according to some evaluation measure  \cite{BorItti2012}. In other cases,  if needed for practical purposes, e.g. for robotic applications, the  problem  of oculomotor action selection is solved by adopting some deterministic choice procedure.   These  usually rely  on  selecting  the gaze position  $\mathbf{x}$ as the argument that  maximises a measure on the given representation $\mathcal{R}$ (in brief, see \cite{walther2006} for using the  $\arg\max_{\mathbf{x}} \mathcal{R}$ operation\footnote{$\arg\max_{x} f(x)$ is the mathematical shorthand for ``find the value of the argument $x$ that maximizes $f(\cdot)$"}  and \cite {BocFerSMCB2013,TatlerBallard2011eye}, for an in-depth discussion).

Yet, another issue arises: the variability of visual scan paths. When looking  at natural movies under a free-viewing or a general-purpose task, the  relocation of gaze can be different among observers even though the same locations are taken into account.  In practice, there is a small probability  that two observers will fixate exactly the same location at exactly the same time.  Such  variations in individual scan paths (as regards chosen fixations, spatial scanning order, and fixation duration)  still hold when the scene contains  semantically rich "objects" (e.g., faces, see Fig.~\ref{Fig:topdown}).  Variability  is  even exhibited by the same subject along different trials on equal stimuli.  Further, the consistency in fixation locations between observers decreases with prolonged viewing \cite{dorr2010variability}. 
This effect is remarkable when free-viewing static images: consistency in fixation locations selected by observers decreases over the course of the first few fixations after stimulus onset \cite{TatlerBallard2011eye} and can become idiosyncratic.

Note that, the WTA scheme  \cite{IttiKoch98,walther2006}, or the selection of  the proto-object with the highest attentional weight  \cite{anna} are deterministic procedures.
Even when  probabilistic frameworks are used to infer  where to look next, the final decision is often taken via  the maximum a posteriori (MAP) criterion\footnote{Given a  posterior distribution $P(X \mid Y)$ the MAP rule  is just about choosing  the argument $X=x$ for which $P(X \mid Y)$ reaches its maximum value (the $\arg\max$) ; thus,   if $P(X \mid Y)$ is a Gaussian distribution, then the $\arg\max$ corresponds to the mode, which for the Gaussian is also the mean value.},  which again is an $\arg\max$ operation   (e.g., \cite{elazary2010bayesian,bocc08tcsvt,geisler2005,ChernyakStark}), or  variants such as the robust mean (arithmetic mean with maximum value) over candidate positions \cite{begum2010probabilistic}. As a result, for  a chosen visual data input $\mathcal{D}$
the mapping $\mathcal{R} \xmapsto[\mathbf{T}]{}  \{\mathbf{x}_{F}(1), \mathbf{x}_{F}(2),\cdots \}$ will always generate the same scanpath across different trials.

There are few notable exceptions to this current state of affairs (see \cite{BocFerSMCB2013} for a discussion). In \cite{kimura2008dynamic}  simple eye-movement patterns, in the vein of  \cite{tatler2009prominence}, are straightforwardly incorporated as a prior of a dynamic Bayesian network to guide the sequence of eye focusing positions on videos. The model presented in \cite{ho2009computational} embeds at least one parameter suitable to be tuned to obtain different saccade length distributions on static images, although statistics obtained by varying such parameter are still far from those of human data. The model by Keech and Resca \cite{keech2010eye1}  mimics phenomenologically the observed eye movement trajectories and where  randomness is captured  through a Monte Carlo selection of a particular eye movement based on its probability;  probabilistic modelling of eye movement data has been also discussed in \cite{rutishauser2007probabilistic}. However, both models address the specific task of conjunctive visual search and are limited to static scenes. Other exceptions are given, but in the very peculiar field of eye-movements in reading \cite{feng2006eye}. Other works have addressed the variability issue in the framework of foraging random walks~\cite{bfpha04,BocFerSMCB2013,BocFerAnnals2012,BocFerSPIC2012,BocCOGN2014,napboc_TIP2015}. 

What we need at least is to bring stochasticity back into the game. As Canosa put it \cite{canosa2009real}:
\begin{quote}
Where we choose to look next at any given moment in time is not completely deterministic, yet neither is it completely random.
\end{quote}
\section{Stochastic processes and eye movements}
\label{sec:stoch}
\begin{figure}[t]
\sidecaption[t]
\includegraphics[scale=0.18,keepaspectratio=true]{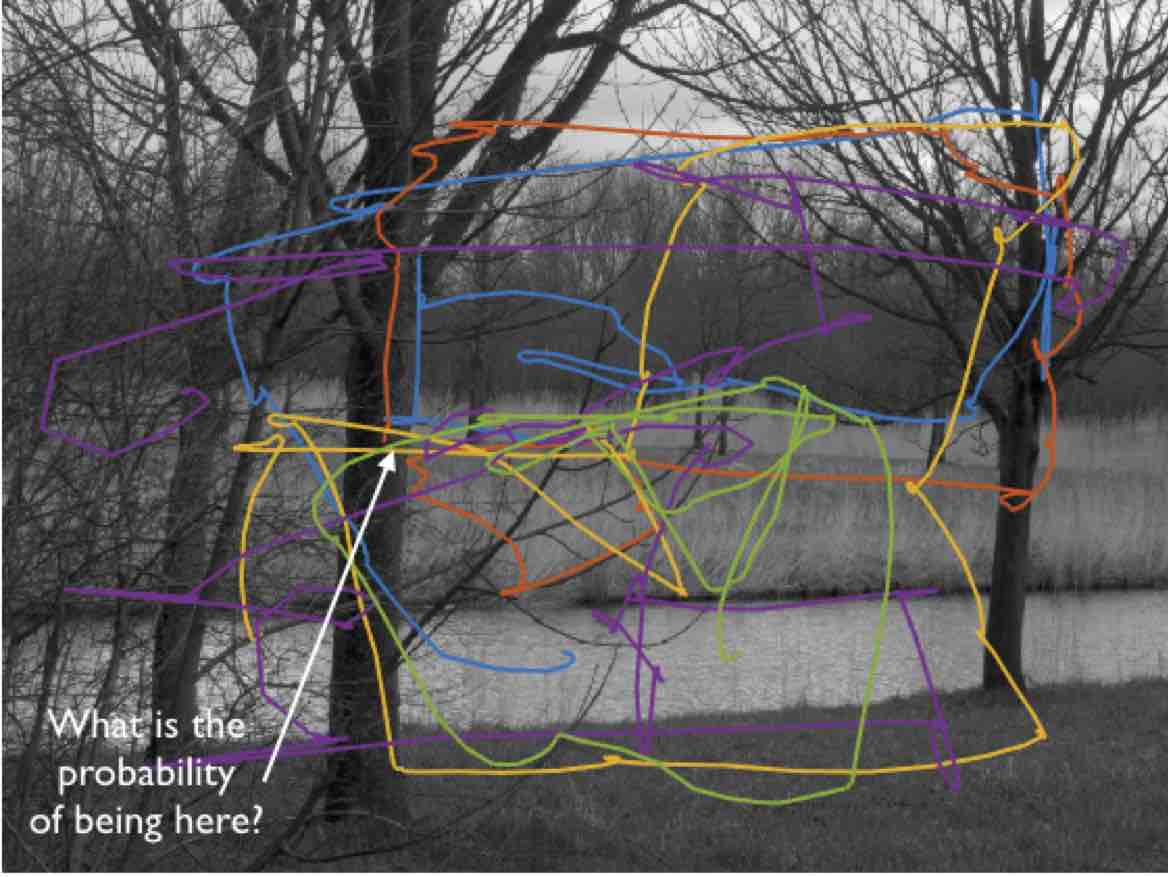}
\caption{An  ensemble of  scan paths recorded from different observers while viewing  the same image. For visualisation purposes, only five trajectories are shown,   different colors coding individual trajectories. If such ensemble is considered to represent the outcome of a stochastic process, the fundamental question that should be answered is: What is the probability $P(\mathbf{x}, t)$ of gazing at location $\mathbf{x}$ at time $t$? Images and data are from the Doves dataset~\cite{van2009doves}} 
\label{Fig:ensemble}
\end{figure}

When we randomly sample a sequence $\{\mathbf{x}(t=1), \mathbf{x}(t=2), \mathbf{x}(t=3),\cdots \}$ of gaze shifts from the pdf $P(\mathbf{x})$ (cfr., Eq.\ref{eq:priorTatlersamp}), we  set up a stochastic process.  For example, the \emph{ensemble} of different scan paths  on the same viewed image can be conceived as the record of a stochastic process (Fig. \ref{Fig:ensemble})

Stochastic processes are systems that evolve probabilistically in time or more precisely, systems in which a certain time-dependent random variable $\mathbf{X}(t)$ exists (as to notation, we may sometimes write $\mathbf{X}_t$ instead of $\mathbf{X}(t)$) The variable $t$  usually denotes time and it can be integer or real valued: in the first case,  
 $\mathbf{X}(t)$ is a discrete time stochastic process; in the second case,  it  is a continuous time stochastic process. 
We can observe realisations of the process, that is we can measure values $$\mathbf{X}(t_{1})=\mathbf{x}_{1}, \; \;\mathbf{X}(t_{2})=\mathbf{X}_{2}, \; \;\mathbf{X}(t_{3})=\mathbf{x}_{3}, \cdots,$$ at times $t_{1}, t_{2}, t_{3}, \cdots$. The set  $\mathcal{S}$  whose elements are the values of the process is called \textbf{state space}.  

\begin{figure}[t]
\sidecaption[t]
\includegraphics[scale=0.2,keepaspectratio=true]{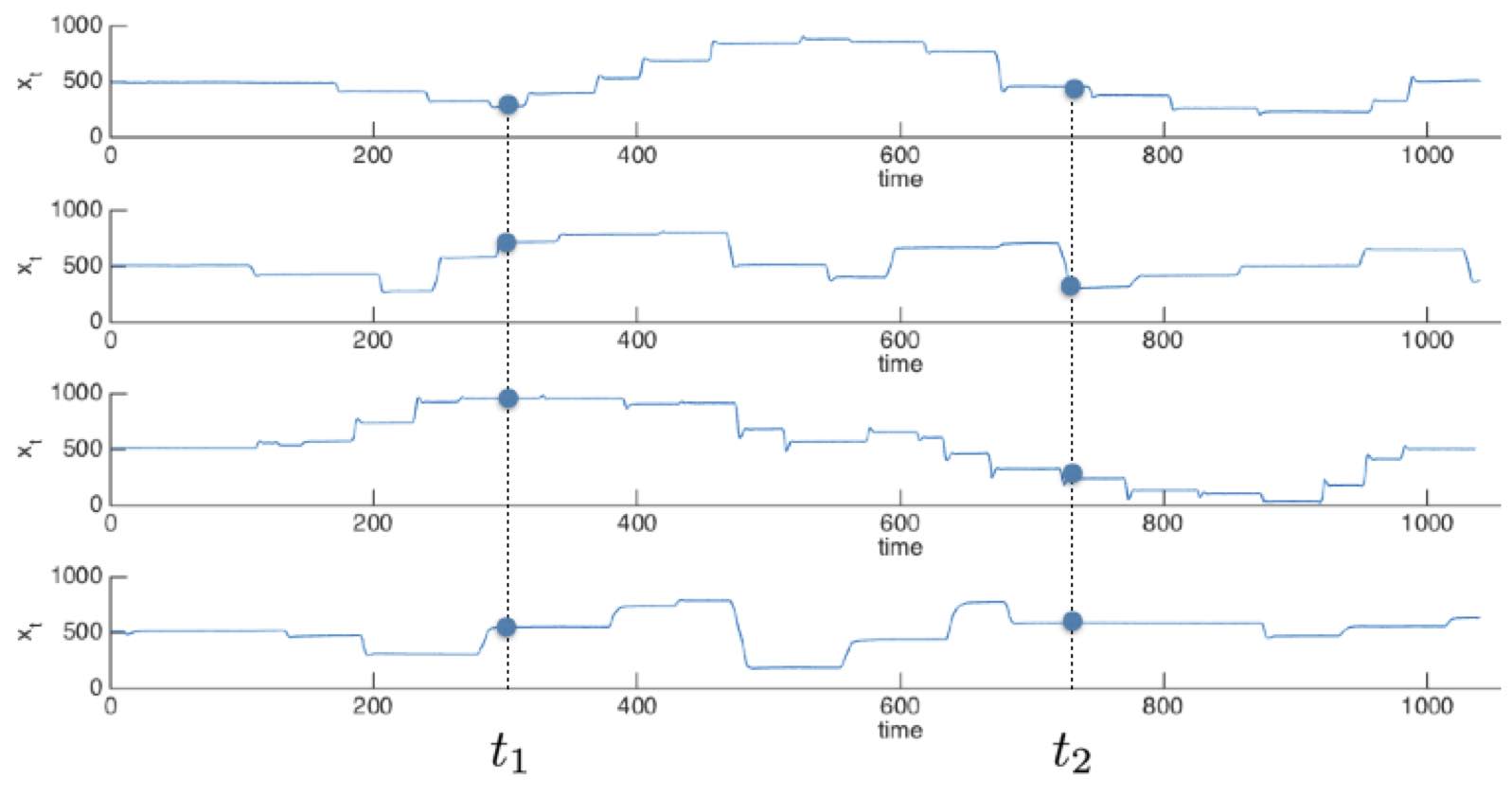}
\caption{An ensemble of paths representing a stochastic process. Each path represents the sequence in time of raw $x$ coordinates from different scan paths recorded  on the same picture (cfr. Fig. \ref{Fig:ensemble}). We can conceive   the trajectories (namely, time series) of such ensemble as realisations of a stochastic process.}
\label{Fig:stoch}
\end{figure}

Thus, we can conceive the stochastic process $\mathbf{X}(t)$  as an ensemble of paths as shown in Fig. \ref{Fig:rawfix} or, more simply, as illustrated  in Fig. \ref{Fig:stoch}: here,  for concreteness, we show four series of   the raw $x$ coordinates of different eye-tracked subjects gazing at  picture shown in Fig. \ref{Fig:rawfix}. Note that if we fix the time, e.g., $t=t_1$, then $\mathbf{X}(t_1)$ boils down to a RV (vertical values); the same holds if we choose one path $\mathbf{x}$ and we (horizontally) consider the set of values $\mathbf{x}_{1}, \; \;\mathbf{x}_{2}, \; \;\mathbf{x}_{3}, \cdots,$ at times $t_{1}, t_{2}, t_{3}, \cdots$.

To sum up, a stochastic process can be regarded as either a family of realisations of a random variable in time, or as a family of random variables at fixed time. Interestingly enough, referring back to  Section~\ref{sec:hist}, notice that Einstein's point of view was to treat Brownian motion as a distribution of a random variable describing position, while Langevin took the point of view that Newton's law's of motion apply to an individual realisation.

In order to be more compact with notation, we will  use Huang's  abbreviation \cite{huang2001introduction}
$$k \leftrightarrow \{\mathbf{x}_{k}, t_{k} \}, $$   
\noindent where, e.g., $P(1)$ succintly stands for $P(\mathbf{x}_{1}, t_{1} )$.

To describe the process completely we need to know the correlations in time, that is the hierarchy of pdfs (but see Box \ref{tab:quant}, for a discussion of correlation):
\begin{eqnarray}
P(1): \text{   the 1 point pdf}\\ \nonumber
P(1,2):  \text{    the 2 points  pdf}\\\ \nonumber
P(1,2,3): \text{    the 3 points  pdf}\\\ \nonumber
\cdots
\label{eq:hier}
\end{eqnarray}
\noindent up to the $n$ point joint pdf. The $n$ point joint pdf must imply all the lower $k$ point pdfs, $k<n$:
\begin{equation}
P(1, \cdots, k) = \int P(1, \cdots, n) d \mathbf{x}_{k+1} d \mathbf{x}_{k+2} \cdots d \mathbf{x}_{n}
\label{eq:marg}
\end{equation}
\noindent where $P(1, \cdots, n) d \mathbf{x}_{k+1} d \mathbf{x}_{k+2} \cdots d \mathbf{x}_{n}$ stands for the joint probability of  finding that $\mathbf{x}$ has a certain value
\begin{description}
\centering
\item[ ]$ \mathbf{x}_{k+1} < \mathbf{x} \leq \mathbf{x}_{k+1} + d \mathbf{x}_{k+1}$ at time $t_{k+1}$
\item[ ]$ \mathbf{x}_{k+2} < \mathbf{x} \leq \mathbf{x}_{k+2} + d \mathbf{x}_{k+2}$ at time $t_{k+2}$
\item[ ]$\cdots$
\end{description}

\begin{table}
\begin{svgraybox}
\caption{\textbf{How to observe a stochastic process}}
\label{tab:quant}

Consider a series of time signals. The signal fluctuates up and down in a seemingly erratic way.
The measurements that are in practice available at one time of a measurable quantity $x(t)$   are the mean  and the variance. However, the latter do not tell a great deal about the underlying dynamics of what is happening. A fundamental question in time series analysis is: to what extent  the value of a RV variable measured at one time can be predicted from knowledge of its value measured at some earlier time? Does the signal at $t_0$ influence what is measured at a later time $t_0 + t$? We are not interested in any specific time instant $t_0$ but rather in the typical (i.e., the statistical) properties of the fluctuating signal. The amount of dependence, or history in the signal can be characterised by  the \textbf{autocorrelation} function
 \begin{equation}
C_{xx}(\tau)=\lim_{T \to \infty}  \frac{1}{T} \int_{0}^{T} x(t)x(t+ \tau) dt.
\end{equation}
\noindent This is the time average of a two-time product over an arbitrary large time $T$, which is then allowed to become infinite.
Put simply, is the integral of the product of  the time series with the series simply displaced with respect to itself by an amount $\tau$.  An autocorrelated time series is predictable, probabilistically, because future values depend on current and past values. In practice, collected time series  are of finite length, say $N$. Thus, the estimated autocorrelation function is best described as the \textbf{sample autocorrelation}
\begin{equation}
c_{xx}(\Delta)= \frac{1}{N}\sum_{n=0}^{N - |\Delta|-1} x(n)x(n+ \Delta) 
\end{equation}
Measurements of $C_{xx}(\tau)$  are used to estimate the time-dependence of the changes in the joint probability distribution, where the \textbf{lag} is $\tau = t - t_0$. If there is no statistical correlation $C_{xx}(\tau)=0$. The rate at which $C_{xx}(\tau)$ approaches $0$ as $\tau$ approaches $\infty$ is a measure of the \textbf{memory} for the stochastic process, which can also be defined in terms of \textbf{correlation time}:
\begin{equation}
t_{corr}=\frac{1}{C_{xx}(0)} \int_{0}^{+\infty} C_{xx}(\tau) d \tau.
\label{eq:tcorr}
\end{equation} 
The autocorrelation function has been defined so far as a \textbf{time average} of a signal, but we may also consider the \textbf{ensemble average}, in which we repeat the same measurement many times, and compute averages, denoted by symbol $\langle \rangle$. Namely, the correlation function between $x(t)$ at two different times $t_1$ and $t_2$ is given by
\begin{equation}
\langle x(t_1), x(t_2) \rangle= \int_{-\infty}^{+\infty}\int_{-\infty}^{+\infty} x_1x_2 P(x_1,t_1;x_2,t_2)  d x_1 d x_2.
\end{equation}
For many systems the ensemble average is equal to the time average, $\langle x \rangle =\int_{-\infty}^{+\infty}x_1 P(x_1,t)  d x_1  \approx \lim_{T \to \infty}  \frac{1}{T} \int_{0}^{T} x(t) dt = \overline{x(t)}$. Such systems are termed \textbf{ergodic}.  Ergodic ensembles for which the probability distributions are invariant under time translation and only depend on the relative times $t_2 - t_1$ are \textbf{stationary} processes. If we have a stationary process, it is reasonable to expect that average measurements could be constructed by taking values of the variable $x$ at successive times, and averaging various functions of these. Correlation and memory properties of a stochastic process are typically investigated by analysing the autocorrelation function or the spectral density (\textbf{power spectrum}) $S(\omega)$, which describes how the power of a  time series is distributed over the different frequencies. These two statistical properties are equivalent for stationary stochastic processes. In this case the \emph{Wiener-Kintchine theorem} holds
 \begin{equation}
S(\omega)= \frac{1}{2\pi}  \int_{-\infty}^{-\infty} \exp{(-i \omega \tau)}C_{xx}(\tau) d\tau
\end{equation}
 \begin{equation}
C_{xx}(\tau)=   \int_{-\infty}^{-\infty} \exp{(i \omega \tau)} S(\omega) d\omega
\end{equation}
It means that one may either directly measure the autocorrelation function of a signal, or the spectrum, and convert back and forth, which by means of the Fast Fourier Transform (FFT)  is relatively straightforward. The sample power spectral density function is computed via  the FFT of $c_{xx}$, i.e. $s(\omega)= FFT(c_{xx}(\Delta))$, or viceversa by the inverse transform,   $c_{xx}(\Delta = IFFT(s(\omega))$.
\end{svgraybox} 
\end{table}


For instance, referring to  Fig. \ref{Fig:stoch}, we can calculate the joint probability $P(1,2) d \mathbf{x}_{1}d \mathbf{x}_{2}$ by following the vertical line at $t_1$ and $t_2$ and find the fraction of paths for which  $\mathbf{x}(t_1) =\mathbf{x}_{1} $ within tolerance  $d \mathbf{x}_{1}$ and $\mathbf{x}(t_2) =\mathbf{x}_{2} $ within tolerance  $d \mathbf{x}_{2}$, respectively\footnote{This gives an intuitive insight into the notion of $P(1,2)$ as a \emph{density}.}

Summing up,   the  joint probability density function, written in full notation  as 
$$P( \mathbf{x}_{1}, t_{1};   \mathbf{x}_{2}, t_{2}; \cdots ; \mathbf{x}_n, t_n), $$
\noindent  is all  we need  to fully characterise the statistical properties of a stochastic process and to calculate the quantities of interest characterising the process (see Box \ref{tab:quant}).

The \textbf{dynamics}, or \emph{evolution} of a stochastic process can be represented  through the specification of \textbf{transition probabilities}:
\begin{description}
\centering
\item[ ] $P(2 \mid 1)$ : probability  of finding $2$, when $1$ is given;
\item[ ] $P(3 \mid 1,2)$ : probability of finding $3$, when $1$ and $2$  are given;
\item[ ] $P(4 \mid 1,2,3)$ : probability of finding $4$, when $1,2$ and $3$  are given;
\item[ ] $\cdots$
\end{description}

Transition probabilities for a stochastic process are nothing but  the conditional probabilities suitable to predict the future values of $\mathbf{X}(t)$ (i.e., $\mathbf{x}_{k+1}, \mathbf{x}_{k+2}, \cdots \mathbf{x}_{k+l}$, at $t_{k+1}, t_{k+2}, \cdots t_{k+l}$), given the knowledge of the past ($\mathbf{x}_{1}, \mathbf{x}_{2}, \cdots \mathbf{x}_{k}$, at $t_{1}, t_{2}, \cdots t_{k}$). The conditional pdf  explicitly defined in terms of the joint pdf can be written:
\begin{equation}
P( \overbrace{ \mathbf{x}_{k+1}, t_{k+1}; \cdots;  \mathbf{x}_{k+l}, t_{k+l} }^{\text{future}} \mid \underbrace{\mathbf{x}_{1}, t_{1};    \cdots; \mathbf{x}_{k}, t_{k}}_{\text{past}} ) = \frac{P( \mathbf{x}_{1}, t_{1};   \cdots ; \mathbf{x}_{k+l}, t_{k+l})}{P( \mathbf{x}_{1}, t_{1};    \cdots; \mathbf{x}_{k}, t_{k} )}.
\label{eq:condpdf}
\end{equation}
\noindent assuming the time ordering $t_{1} < t_{2} < \cdots < t_{k} < t_{k+1} < \cdots < t_{k+l}$.

By using  transition probabilities and the product rule,  the following update equations can be written:
\begin{eqnarray}
\label{eq:update}
P( 1, 2) &=& P(2 \mid 1) P( 1) \\ \nonumber
P( 1, 2, 3) &=& P(3 \mid 1,2) P( 1, 2)  \\ \nonumber
P( 1, 2, 3,4) &=& P(4 \mid 1,2,3) P( 1, 2,3)\\  \nonumber
&\cdots \nonumber
\end{eqnarray}

The transition probabilities must satisfy the normalisation condition $\int P(2 \mid 1) d \mathbf{x}_{2}=1$. 
Since $P(2 )= \int P(1,2) d \mathbf{x}_{1}$ and by using the update Eqs. (\ref{eq:update}), the following evolution (integral) equation holds
 \begin{equation}
 P(2 )= \int \overbrace{P(2 \mid 1)}^{\text{propagator}}P(1) d \mathbf{x}_{1}
 \label{eq:prop}
 \end{equation}
 \noindent where $P(2 \mid 1)$ serves as the \emph{evolution  kernel} or \emph{propagator} 
 from state $1 $ to state $2$, i.e., in full notation, from state $(\mathbf{x}_{1}, t_{1})$ to state $(\mathbf{x}_{2}, t_{2})$.
 
A stochastic process whose joint pdf does not change when shifted in time is called a (strict sense) \textbf{stationary process}:
\begin{equation}
P( \mathbf{x}_{1}, t_{1};  \mathbf{x}_{2}, t_{2}; \cdots ; \mathbf{x}_n, t_n)= P( \mathbf{x}_{1}, t_{1}+\tau;   \mathbf{x}_{2}, t_{2}+\tau; \cdots ; \mathbf{x}_n, t_n+\tau)
\end{equation}
\noindent $\tau > 0$ being a time shift.
Analysis of a stationary process is frequently much simpler than for a similar process that is time-dependent:   varying $t$, all the random variables $\mathbf{X}_t$ have the same law; all the moments, if they exist, are constant in time; the distribution of $\mathbf{X}(t_{1})$ and $\mathbf{X}(t_{2})$ depends only on the difference $\tau = t_{2} -  t_{1}$ (time lag), i.e, $$P( \mathbf{x}_{1}, t_{1};   \mathbf{x}_{2}, t_{2}) = P( \mathbf{x}_{1},   \mathbf{x}_{2}; \tau).$$

\begin{figure}[t]
\sidecaption[t]
\includegraphics[scale=0.11,keepaspectratio=true]{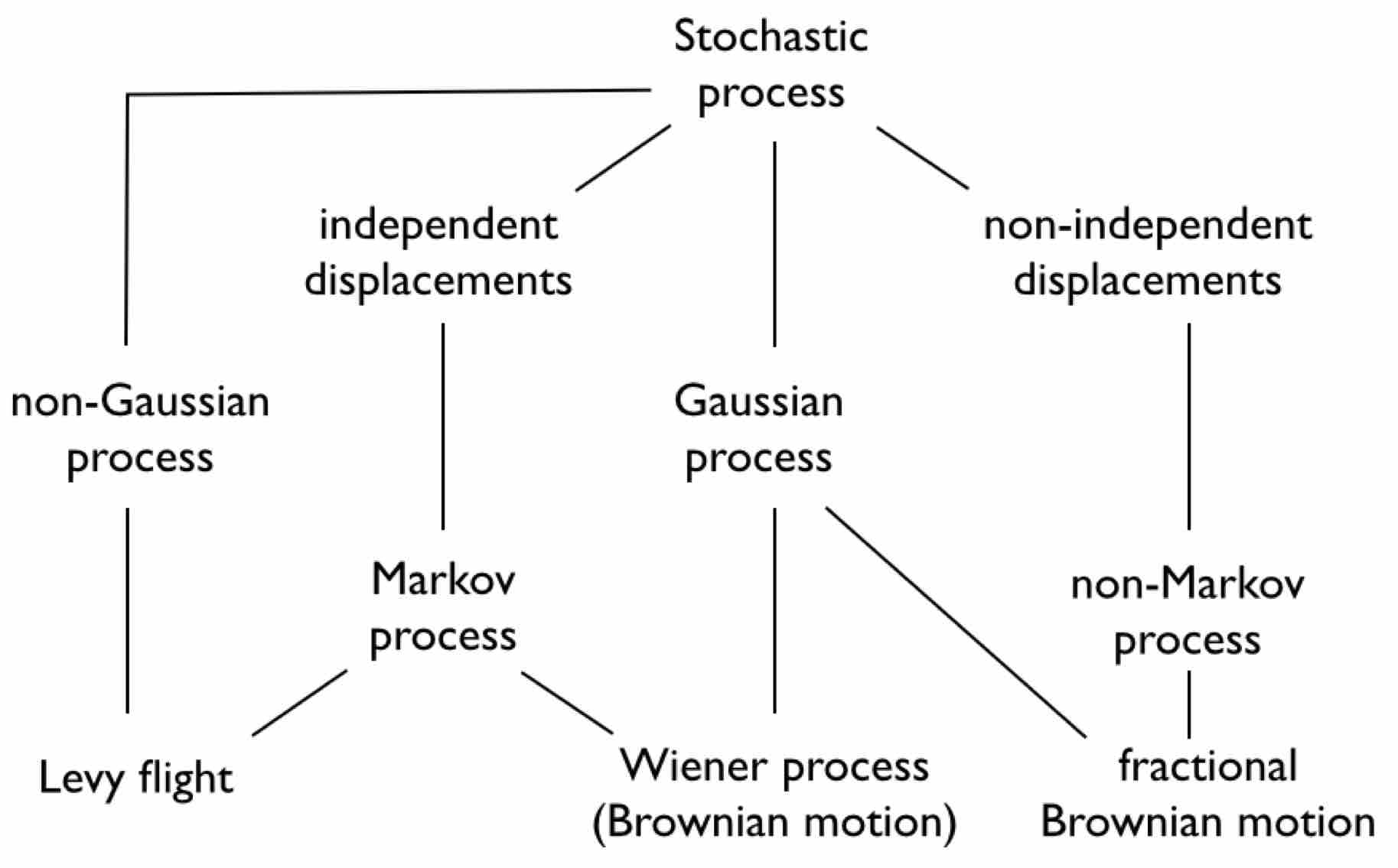}
\caption{A conceptual map of stochastic processes that are likely to play a role in eye movement modelling and analyses}.
\label{Fig:Concept}
\end{figure}

A conceptual map of main kinds of stochastic processes that we will discuss in the remainder of this Chapter is presented in Fig. \ref{Fig:Concept}.

\section{How to leave the past behind: Markov Processes}
\label{sec:markov}

 The most simple kind of stochastic process is the Purely Random Process in which there are no correlations. From Eq. (\ref{eq:update}):
 \begin{eqnarray}
P( 1, 2) &=& P( 1)  P( 2)  \\ \nonumber
P( 1, 2, 3) &=& P( 1)  P( 2)  P( 3)  \\ \nonumber
P( 1, 2, 3,4) &=& P( 1)  P( 2)  P( 3)P_1( 3)\\ \nonumber
&\cdots
\label{eq:updatePRP}
\end{eqnarray}
 
One such process can be obtained for example by repeated  coin tossing. The  complete independence property can be written explicitly as:
\begin{equation}
P( \mathbf{x}_{1}, t_{1};   \mathbf{x}_{2}, t_{2}; \cdots  ) = \prod_i P( \mathbf{x}_{i}, t_{i}),
\label{eq:PRP}
\end{equation}
\noindent  the uppercase greek letter $\prod$  indicates a product of factors, e.g., for $i=1, 2, 3$,  $P( \mathbf{x}_{1}, t_{1};   \mathbf{x}_{2}, t_{2}; \mathbf{x}_{3}, t_{3} ) = P( \mathbf{x}_{1}, t_{1})  P( \mathbf{x}_{2}, t_{2}) P( \mathbf{x}_{3}, t_{3})$.

Equation \ref{eq:PRP} means that the value of $\mathbf{X}$ at time $t$ is completely independent of its values in the past (or future). A special case occurs when the $P( \mathbf{x}_{i}, t_{i})$  are independent of $t$, so that the same probability law governs the process at all times. Thus, a completely memoryless stochastic process is composed by  a set of \textbf{independent and identically distributed} (\textbf{i.i.d}) RVs.
Put simply, a series of  i.i.d. RVs is a series of samples where individual samples  are ``independent'' of each other  and are generated from the same probability distribution (``identically distributed''). 


More realistically, we know that most processes in nature, present some correlations between consecutive values. For example, the direction of the following gaze shift is likely to be positively correlated with the direction of current gaze shift. A step towards a more realistic description consists then of assuming that the next value of each RV in the process depends explicitly on the current one (but not explicitly on any other previous to that). An intuitive example is the \emph{simple random walk}, which is briefly discussed in Box \ref{tab:simple}, and can be modelled by the simple equation
\begin{equation}
x_{t} = x_{t-1} + \xi_t ,   
\label{eq:simpleRW2}
\end{equation}
\noindent    where  the noise term $\xi_t$ is sampled, at each step $t$, from a suitable distribution $P(\xi)$.

Note that if we iterate Eq. (\ref{eq:simpleRW2}) for a number of steps $t=1,2,3,\cdots$  and collect the output sequence of the equation, that is  $x_{1}, x_{2},x_{3}, \cdots$, we  obtain a single trajectory/path of the walk,  which is one possible realisation  of the underlying stochastic process. This corresponds to consider one horizontal slice of Fig. \ref{Fig:stoch}, that is a (discrete)  \textbf{time series}. It is worth mentioning that there exist a vast literature on time series analysis, which can be  exploited in neuroscience data analysis and more generally in other fields. Indeed, the term ``time series'' more generally refers to data that  can be represented as a sequence. This includes for example financial data in which the sequence index indicates time, as in our case, but also genetic data (e.g. ACATGC . . .) where the sequence index has no temporal meaning.
Some of the methods that have been developed in this research area might be useful for  eye movement modelling and analysis. We have no place here to further discuss this issue, however Box \ref{tab:times} provides some ``pointers'' for the reader.
\begin{table}
\begin{svgraybox}
\caption{\textbf{Random walks}}
\label{tab:simple}
Random walks (RW) are a special kind of stochastic process and can be used, as we will see, to model the dynamics of many complex systems. A particle moving in a field, an animal foraging, and indeed the ``wandering'' eye can be conceived as examples of random walkers.

 In general, RWs exhibit what is called serial correlation, conditional independence for fairly small values of correlation length $t_{corr}$ (cfr., Box \ref{tab:quant}, Eq.~\ref{eq:tcorr}), and a simple stochastic historical dependence. For instance, a simple additive $1-$dimensional random walk has the form:
\begin{equation}
x_{t} = x_{t-1} + \xi_t ,   \text{     where       } \xi_t \sim P(\xi)
\label{eq:simpleRW}
\end{equation}

In the above formulation, time $t$ proceeds in discrete steps. 
 $\xi_t$ is a RV drawn i.i.d. from a distribution $P(\xi)$, called the \emph{noise} or \emph{fluctuation} distribution. Thus, the differences in sequential observations $x_{t} - x_{t-1} = \xi_t \sim P(\xi) $ are i.i.d. We have here  \textbf{independent displacements}. 
 
 However, the observations themselves are not independent, since (\ref{eq:simpleRW}) encodes the generative process, or evolution law,  $x_{t-1} \rightarrow x_{t}$ where $x_{t}$ explicitly depends on  $x_{t-1}$,  but not on earlier $x_{t-2}, x_{t-3}, x_{t-4}, \cdots$. Thus Eq. (\ref{eq:simpleRW}) represents a Markov process.

 
Conventionally, fluctuations are Gaussian distributed with mean $\mu$ and variance $\sigma^2$, that is, $\xi \sim \mathcal{N}(\mu,\sigma^2)$, as this makes mathematical analysis considerably simpler. In this case by simply extending to two dimensions  Eq. \ref{eq:simpleRW},
\begin{eqnarray}
x_{t} = x_{t-1} + \xi_{x,t } \\ \nonumber
y_{t} = y_{t-1} + \xi_{y,t },
\label{eq:simple2DRW}
\end{eqnarray}
the simulation of a simple Brownian RW can be obtained (see Fig. \ref{Fig:brown2D}).

However, any probability distribution, for instance, a Laplace (exponential tails) or double-Pareto distribution (power-law tails), also works. 

\end{svgraybox} 
\end{table}
%
%

\begin{table}
\begin{svgraybox}
\caption{\textbf{As time goes by: time series analysis}}
\label{tab:times}

The random walk model summarised by Eq. \ref{eq:simpleRW2}, can be seen as a special case of the  model
\begin{equation}
x_{t} = \alpha_{1} x_{t-1}+ \xi_{t},  
\label{eq:AR1}
\end{equation}
\noindent with model parameter $\alpha_1 = 1$.

In \textbf{time series} analysis, under the assumption that $\xi_t$ is sampled from a Gaussian distribution mean zero and variance $\sigma^2$, the model specified by Eq. \ref{eq:AR1} is known as an autoregressive or \textbf{AR model}  of order $1$, abbreviated to AR($1$).

The AR($1$) model, in turn is a special case of an autoregressive process of order $p$, denoted as AR($p$):
\begin{equation}
x_{t} = \alpha_1 x_{t-1} + \alpha_2 x_{t-2} + \cdots + \alpha_p  x_{t-p}  +  \xi_t ,  
\label{eq:ARp}
\end{equation}
\noindent with model parameters $ \alpha_p \neq 0$.
Note that such model is  a regression of $x_{t}$ on past terms from the same series; hence the use of the term ``autoregressive''.

Considering again the RW of Eq. \ref{eq:simpleRW2}. One can substitute the term $x_{t-1}$, that by using the same equation can be calculated as $x_{t-1} = x_{t-2} + \xi_{t-1}$;  thus,
\begin{equation}
x_{t} =  x_{t-2} + \xi_{t-1}+ \xi_t .  
\end{equation}
Continuing and substituting for $x_{t-2}$, followed by $x_{t-3}$ and so on (a process known as ``back substitution'') gives
\begin{equation}
x_{t} =  \xi_{1} +\xi_{1}+ \cdots  + \xi_{t-1}+ \xi_t ,  
\end{equation}
\noindent where $x_{t}$ is written as the sum of the current noise term $\xi_t $
and the past noise terms.

This result can be generalised by writing $x_{t}$ as the  linear combination of the current white noise term and the $q$ most recent past  noise terms
\begin{equation}
x_{t} =  \xi_{t} +  \beta_1 \xi_{t-1}+ \beta_2 \xi_{t-2} +  \cdots  + \beta_q \xi_{t-q}.  
\label{eq:MAq}
\end{equation} 
This defines a moving average or \textbf{MA model} of order $q$, shortly MA($q$).

Putting all together, we can write the general expression:
\begin{equation}
x_{t} = \alpha_1 x_{t-1} + \alpha_2 x_{t-2} + \cdots +  \alpha_p x_{t-p} +  \xi_{t} +  \beta_1 \xi_{t-1}+ \beta_2 \xi_{t-2} +  \cdots  + \beta_q \xi_{t-q}.  
\label{eq:ARMApq}
\end{equation}
The time series is said to follow an autoregressive moving average or \textbf{ARMA model} of order $(p, q)$, denoted ARMA($p, q$).

By expanding  on the former, a great deal of models can be conceived.
Also, a variety of methods, algorithms, and related software, is at hand for estimating the model parameters from time series data. Cowpertwait and Metcalfe \cite{metcalfe2009introductory} provide a thorough introduction with R language examples, for R fans. The book edited by Barber, Cemgil and Chiappa \cite{barber2011bayesian} offers a comprehensive picture of modern time series techniques, specifically those based on Bayesian probabilistic modelling.  Time series modelling  is a fast-growing  trend in neuroscience data analysis, which is addressed in-depth by Ozaka \cite{ozaki2012time}.
\end{svgraybox} 
\end{table}

Going back to stochastic processes, if a process has no memory beyond the last transition then it is called a \emph{Markov process} and the transition probability enjoys the property:
\begin{equation}
P(\mathbf{x}_n, t_n | \mathbf{x}_{n-1}, t_{n-1}; \cdots ; \mathbf{x}_{1}, t_{1}) = P(\mathbf{x}_{n}, t_{n} | \mathbf{x}_{n-1}, t_{n-1})
\label{eq:markovfull}
\end{equation}
\noindent with $t_{1} < t_{2}< \cdots < t_{n}$.

A Markov process is fully determined by the two densities $P(\mathbf{x}_{1}, t_{1})$ and $P(\mathbf{x}_{2}, t_{2} \mid \mathbf{x}_{1}, t_{1})$; the whole hierarchy can be reconstructed from them. For example, 
from Eq. (\ref{eq:update}) 
using the Markov property $P(3 \mid 1,2) = P(3 \mid 2)$:
\begin{equation}
P(1,2,3)=  P( 1) P( 2 \mid  1) P(3 \mid 2).
\end{equation}
The  factorisation of the joint pdf can thus be explicitly written in full notation as
\begin{equation}
P(\mathbf{x}_{n}, t_{n}; \mathbf{x}_{n-1}, t_{n-1} ; \cdots ;   \mathbf{x}_{1}, t_{1})= P(\mathbf{x}_{1}, t_{1} )\prod_{i=2}^{n} P( \mathbf{x}_{i}, t_{i} \mid  \mathbf{x}_{i-1}, t_{i-1}),
\end{equation}
\noindent with the propagator $P( \mathbf{x}_{i+1}, t_{i+1} \mid  \mathbf{x}_{i}, t_{i})$ carrying the system forward in time, beginning with the initial distribution $P(\mathbf{x}_{1}, t_{1} )$.

\begin{figure}[t]
\sidecaption[t]
\includegraphics[scale=0.20,keepaspectratio=true]{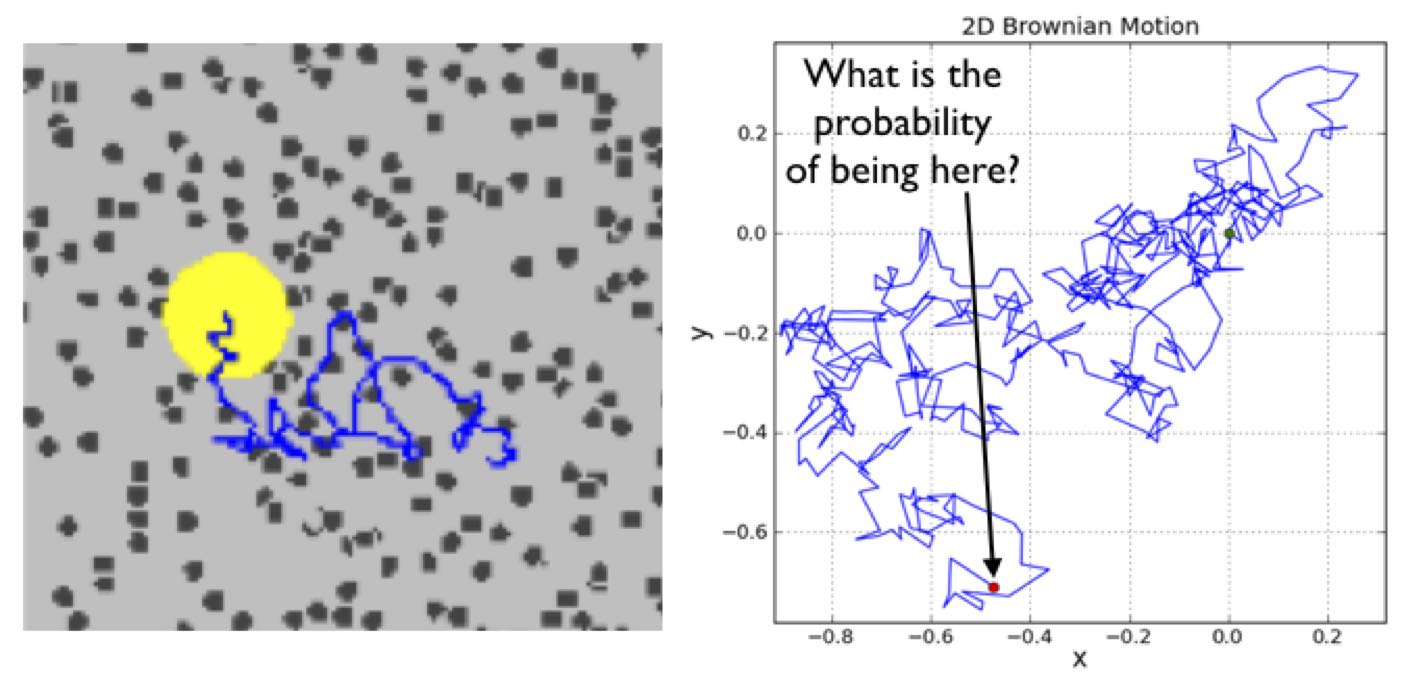}
\caption{Motion of a Brownian particle. Left: The physical mechanism of the displacement (in blue):  the bigger particle performs a \textbf{Brownian motion} (Bm) as a result of the collisions with small particles (figure is not to scale). Right: Sample path of the Bm performed by the bigger particle. Here the fundamental question is: What is the probability of the particle being at location $\mathbf{x} = (x,y)$ at time $t$?  (cfr. Box \ref{tab:hall})}
\label{Fig:brown2D}
\end{figure}

A well known example of  Markov process is the \textbf{Wiener-L\'evy process} describing the position of a Brownian particle (Fig.\ref{Fig:brown2D}). 

The fact that a Markov process is fully determined by  $P( 1)$ and $P( 2 \mid  1)$ does not mean that such two functions can be chosen arbitrarily, for they must also obey two important identities.

The first one is Eq.~(\ref{eq:prop}) that in explicit form reads: 
\begin{equation}
P( \mathbf{x}_{2}, t_{2} ) = \int_{\mathbf{x}_{1}} P( \mathbf{x}_{2}, t_{2} \mid  \mathbf{x}_{1}, t_{1}) P(  \mathbf{x}_{1}, t_{1}) d\mathbf{x}_{1}.
\end{equation}
\noindent This equation simply constructs the one time probabilities in the future $t_{2}$ of $t_{1}$, given the conditional probability $P( \mathbf{x}_{2}, t_{2} \mid  \mathbf{x}_{1}, t_{1}) $.

The second  property can be obtained by marginalising the joint pdf $P(\mathbf{x}_{3}, t_{3} , \mathbf{x}_{2}, t_{2} \mid  \mathbf{x}_{1}, t_{1})$ with respect to $\mathbf{x}_{2}$ and by using   the definition of conditional density under the Markov property: 
\begin{equation}
P(\mathbf{x}_{3}, t_{3} \mid  \mathbf{x}_{1}, t_{1})= \int_{\mathbf{x}_{2}} P(\mathbf{x}_{3}, t_{3} \mid \mathbf{x}_{2}, t_{2})P( \mathbf{x}_{2}, t_{2} \mid  \mathbf{x}_{1}, t_{1}) d\mathbf{x}_{2},
\label{eq:CK}
\end{equation}
\noindent Equation (\ref{eq:CK}) is known as the \textbf{Chapman-Kolmogorov Equation} (C-K equation, from now on). It is ``just'' a statement saying that to move  from position $ \mathbf{x}_{1}$ to $\mathbf{x}_{3}$ you just need to average out all possible intermediate positions $\mathbf{x}_{2}$ or, more precisely, by marginalisation over the nuisance variable $\mathbf{x}_{2}$.

\begin{figure}[t]
\sidecaption[t]
\includegraphics[scale=0.12,keepaspectratio=true]{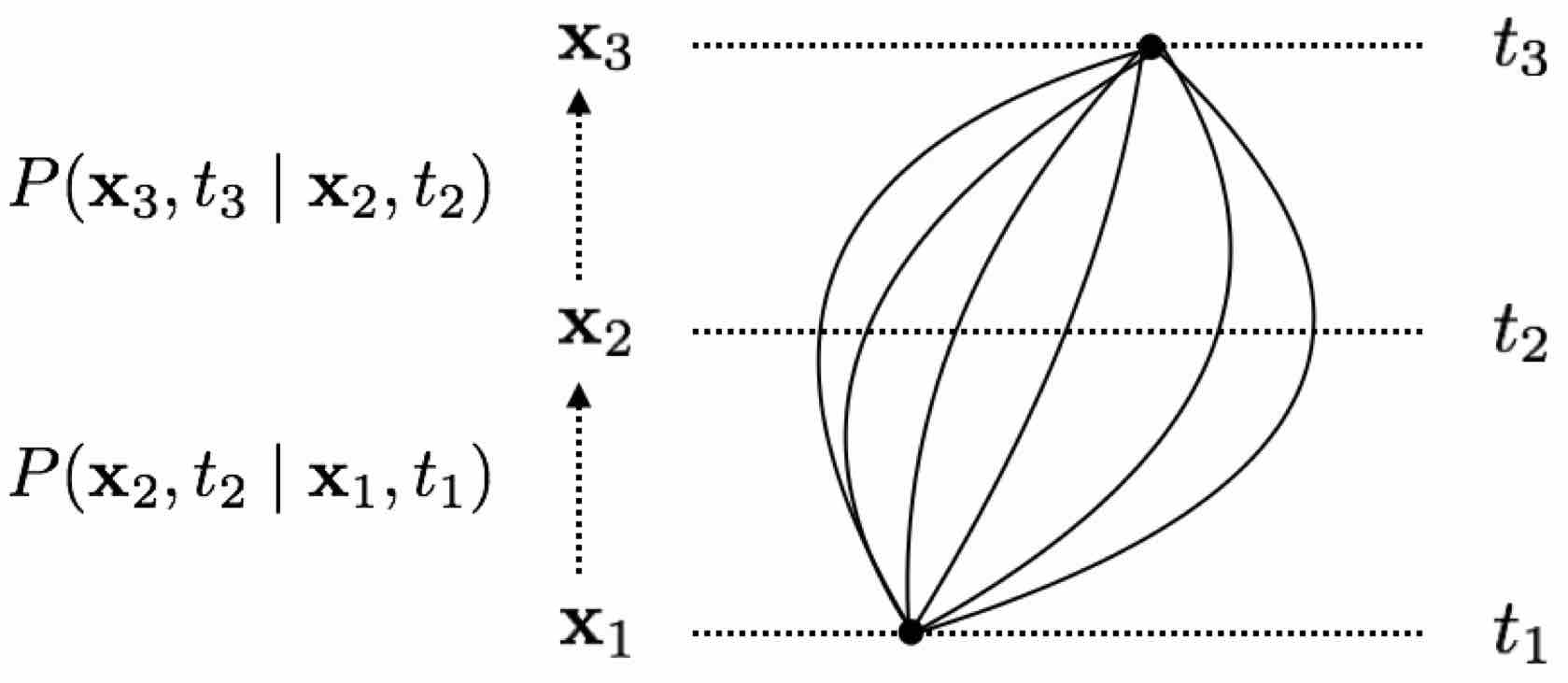}
\caption{The Chapman-Kolmogorov equation at work: the probability of transition from the event $(\mathbf{x}_{1}, t_{1})$  to $(\mathbf{x}_{3}, t_{3})$  is broken into a subprocess from $(\mathbf{x}_{1}, t_{1})$  to an intermediate, nuisance event $(\mathbf{x}_{2}, t_{2})$ (which is not observed in practice)  and then from $(\mathbf{x}_{2}, t_{2})$ to $(\mathbf{x}_{3}, t_{3})$, by considering all the paths from  $\mathbf{x}_{1}$ to $\mathbf{x}_{3}$.}
\label{Fig:chap}
\end{figure}

Such equation is a consistency equation for the conditional probabilities of a Markov process and the starting point for deriving the equations of motion for Markov processes. Aside from providing a consistency check, the real importance of the C-K equation is that it enables us to build up the conditional probability densities over the ``long'' time interval $[t_{1},t_{3}]$ from those over the ``short'' intervals $[t_{1},t_{2}]$ and $[t_{2},t_{3}]$.

%

The C-K equation is a rather complex nonlinear functional equation  relating all conditional probabilities $P(\mathbf{x}_{i}, t_{i} \mid \mathbf{x}_{j}, t_{j})$ to each other. Its solution would give us a complete description of any Markov process, but unfortunately, no general solution to this equation is known: in other terms, it  expresses the Markov character of the process, but containing no information about any particular Markov process.

The idea of  forgetting the past so to use  the present state  for determining  the next one might seem an oversimplified assumption when dealing, for instance, with eye movements performed by an observer engaged in some overt attention task.
However, this conclusion may not in fact be an oversimplification. This was discussed by Horowitz  and Wolfe \cite{horowitz1998visual}, and will be detailed in the following subsection.

\subsection{Case study: the Horowitz  and Wolfe  hypothesis of amnesic visual search}
\label{sec:memory}
Serial and parallel theories of visual search  have in common the memory-driven assumption that efficient search is based on accumulating information about the contents of the scene over the course of the trial.

Horowitz  and Wolfe in their seminal Nature paper  \cite{horowitz1998visual}  tested the hypothesis whether visual search relies on memory-driven mechanisms. They designed their stimuli so that, during a trial, the scene would be constantly changing, yet the meaning of the scene (as defined by the required response) would remain constant. They asked human observers to search for a letter ``T'' among letters ``L''. This search demands visual attention and normally proceeds at a rate of $20-30$ milliseconds per item. In the critical condition, they randomly relocated all letters every $111$ milliseconds. This made it impossible for the subjects to keep track of the progress of the search. Nevertheless, the efficiency of the search was unchanged. 

On the basis of achieved results they proposed  that visual search processes are ``amnesic'': they act on neural representations that are continually rewritten and have no permanent existence beyond the time span of visual persistence. 

In other terms, the visual system does not accumulate
information about object identity over time during a search
episode. Instead, the visual system seems to exist in a sort of eternal
present. Observers are remarkably oblivious
to dramatic scene changes when the moment of change is obscured
by a brief flicker or an intervening object.  

Interestingly enough, they claim that an amnesic visual system may be a handicap only
in the laboratory. The structure of the world makes it unnecessary to
build fully elaborated visual representations in the head. 
 Amnesia can be an efficient strategy for a visual system operating in the real world.

\begin{table}
\begin{svgraybox}
\caption{\textbf{The hall of fame of Markov processes}}
\label{tab:hall}
The most famous Markov process is the Wiener-L\'evy process describing the position of a Brownian particle. Brownian particles can be conceived as a bodies of microscopically-visible size suspended in a liquid,  performing movements of such magnitude that they can be easily observed in a microscope, on account of the molecular motions of heat \cite{EinsteinBrown}. Figure~\ref{Fig:brown2D} shows an example of the 2-dimensional motion of one such particle. 

A probabilistic description of the random walk of the Brownian particle must answer the question: What is the probability $P(\mathbf{x},t)$ of the particle being at location $\mathbf{x} = (x,y)$ at time $t$?

In the $1-$dimensional case, the probability $P(x,t)$ and its evolution law  are defined for  $-\infty < x < \infty$, $t >0$ by the densities
\begin{equation}
P(x,t)= \frac{1}{\sqrt{4 \pi D t}} \exp\left( -\frac{x^2}{4 D t} \right),
\label{eq:gaussdiff}
\end{equation}
\begin{equation}
P( x_{2}, t_{2} \mid  x_{1}, t_{1})= \frac{1}{\sqrt{4 \pi D (t_{2} -t_{1} )}} \exp\left( -\frac{(x_{2}-x_{1})^2}{4 D (t_{2} -t_{1} )} \right).
\end{equation}
that satisfy  the Chapman-Kolmogorov equation. In both equations,  $D$ denotes a \textbf{diffusion coefficient}.
The diffusion concept  has deep roots in statistical physics: indeed,    Einstein was the first to show  in his  seminal work on Brownian motion \cite{EinsteinBrown}  that the coefficient $D$ captured the average or mean squared displacement in time of a moving Brownian particle  (``[...] a process of diffusion, which is to be looked upon as a result of the irregular movement of the particles produced by the thermal molecular movement'', \cite{EinsteinBrown}). We will further discuss this important concept  in Section~\ref{sec:levels}.

%

\end{svgraybox} 
\end{table}


\subsection{Stationary Markov processes and Markov chains}

Recall that for stationary Markov processes the transition probability $P( \mathbf{x}_{2}, t_{2} \mid \mathbf{x}_{1}, t_{1})$ only depends  on the time interval. 
For this case one can introduce the special notation
\begin{equation}
P( \mathbf{x}_{2}, t_{2} \mid \mathbf{x}_{1}, t_{1})= T_{\tau}( \mathbf{x}_{2} \mid  \mathbf{x}_{1})  \text{   with   $\tau = t_{2} -  t_{1}$}.  
\end{equation}

The Chapman-Kolmogorov equation then becomes 
\begin{equation}
T_{\tau+ \tau^\prime}( \mathbf{x}_{3} \mid  \mathbf{x}_{1}) = \int_{\mathbf{x}_{2}}   T_{\tau^\prime}( \mathbf{x}_{3} \mid  \mathbf{x}_{2}) T_{\tau}( \mathbf{x}_{2} \mid  \mathbf{x}_{1}) d\mathbf{x}_{2}.
\label{eq:discSCK}
\end{equation}

If one reads the integral as  the product  of two matrices or integral kernels, then
\begin{equation}
T_{\tau+ \tau^\prime} =    T_{\tau^\prime} T_{\tau} \;\;\; (\tau, \tau^\prime > 0)
\label{eq:discSCK2}
\end{equation}

A simple but important class of stationary Markov processes are the \textbf{Markov chains} defined by the following properties:
\begin{enumerate}
\item the state space of $\mathbf{x}$ is a \emph{discrete set of states};
\item the \emph{time variable is discrete};
\end{enumerate}

\begin{figure}[t]
\sidecaption[t]
\includegraphics[scale=0.15,keepaspectratio=true]{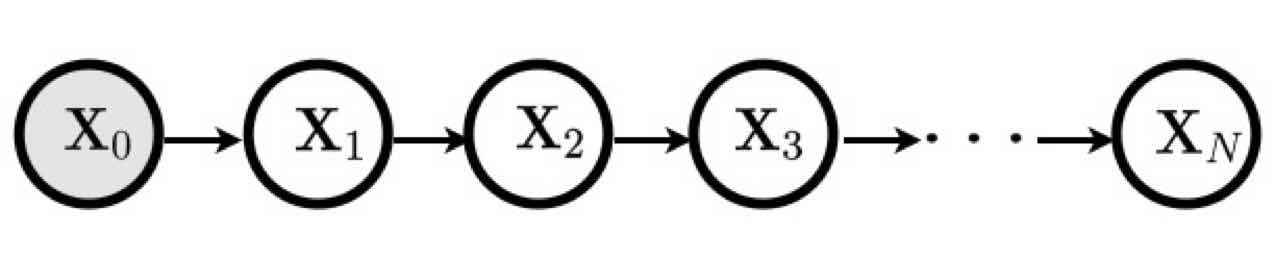}
\caption{The PGM of a Markov chain: given the prior or initial condition $P(\mathbf{x}_{1})$, the behaviour of the system  is determined by the conditional  probability $P(\mathbf{x}_{t} \mid \mathbf{x}_{t-1})$}
\label{Fig:Markov}
\end{figure}

In this case the dynamics can be represented  as the PGM in Fig. \ref{Fig:Markov}
The PGM shows that the joint distribution for a sequence of observations $P(\mathbf{x}_{0}, \mathbf{x}_{1}, \mathbf{x}_{2},\cdots , \mathbf{x}_{N})$ can be written as the product:
\begin{equation}
P(\mathbf{x}_{1})P( \mathbf{x}_{2} \mid \mathbf{x}_{1})\cdots  P(\mathbf{x}_{N} \mid \mathbf{x}_{N-1}) = P(\mathbf{x}_{1}) \prod_{t=2}^{N} P(\mathbf{x}_{t} \mid \mathbf{x}_{t-1})
\label{eq:markovjoint}
\end{equation}
This is also known as an \textbf{observable Markov process}.

A \textbf{finite Markov chain} is one whose range consists of  a finite number of $N$ states. In this case the first probability distribution is an $N$ component vector. The transition probability $T_{\tau}(\mathbf{x}_{2} \mid\mathbf{x}_{1})$ is an $N \times N$ matrix.

Thus, the C-K equation, by using the form in Eq. \ref{eq:discSCK2}, leads to the matrix equation
\begin{equation}
T_{\tau} =    (T_{1})^{\tau}
\label{eq:discSCK3}
\end{equation}

Hence the study of finite Markov chains amounts to investigating the powers and the properties of the $N \times N$ transition matrix: this is a \emph{stochastic matrix} whose elements are nonnegative and each row adds up to unity (i.e., they represent transition probabilities). One seminal application of  Markov chains to scan paths has been provided by Ellis and Stark \cite{ellistark}.

\subsubsection{Case study: modelling gaze shifts as observable finite Markov chains}
\label{sec:markovshift}

\begin{figure}[t]
\sidecaption[t]
\includegraphics[scale=0.23,keepaspectratio=true]{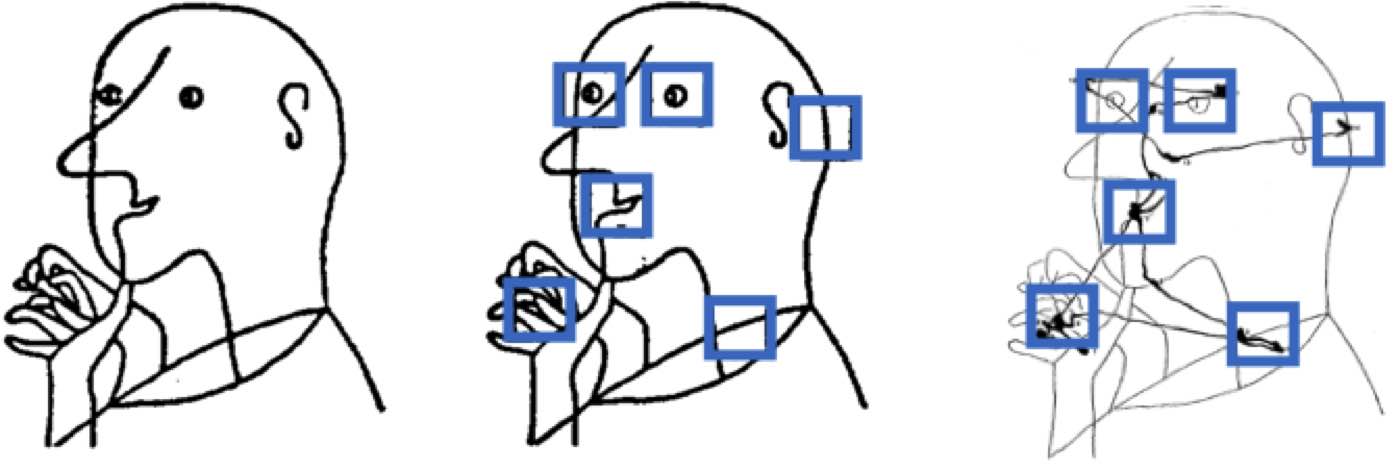}
\caption{Markov analysis of eye movements made by a subject viewing for the first time a drawing
adapted from the Swiss artist Klee (left). Centre: ROIs superimposed on the drawing, defining the states of the Markov chain: $S= \{ s_1=\text{``left eye''}, s_2=\text{``right eye''}, s_3=\text{``nose''}, s_4=\text{``mouth''} , s_5=\text{``hand''}, s_6=\text{``neck''} \}$. Right: saccades represented as state transitions in the state-space.  Modified after \cite{hacisalihzade1992visual,ellistark}  }
\label{Fig:stark}
\end{figure}

Ellis and Stark pioneered the use of Markov analysis for characterising scan paths \cite{ellistark} in an attempt to go beyond visual inspection of the
eye movement traces and application of a subjective test for
similarity of such traces. In particular, they challenged the assumption of what they defined ``apparent randomness'', that many studies at the time were supporting in terms of either simple random or stratified random sampling \cite{ellistark}. To this end (see Fig. \ref{Fig:stark}), they defined regions  of interest (ROI) defined on the viewed picture, each ROI denoting a
 \emph{state} into which the fixations can be located.  By
postulating that the transitions from one state to another have certain probabilities, they effectively described the generating
process for these sequences of fixations as Markov processes.
This way, they were able to estimate the marginal probabilities of viewing a point of interest $i$, i.e., $P(X = s_i)$, and the  conditional probability of viewing a point of interest $j$ given a previous viewing of a point of interest $i$, i.e., $T(X = s_j \mid X = s_i)$, where $s_i, s_j$ are states in the state-space $S$ (see Fig. \ref{Fig:stark}).

By comparing expected frequency of transitions according to random sampling models with  observed transition frequencies, they were able to assess the statistically significant differences that occurred (subject-by-subject basis with a chi-square goodness-of-fit test on the entire distribution of observed and expected transitions).  Thus, they concluded that ``there is evidence that something other than stratified random sampling is taking place during the scanning'' \cite{ellistark}. In a further study \cite{hacisalihzade1992visual},  examples have been provided for exploiting the observable Markov chain  as a generative machine apt to sample simulated scan paths, once the transition matrix has been estimated / learned from data.

\subsection{Levels of representation of the dynamics of a stochastic process}
\label{sec:levels}

Up to this point, we have  laid down  the basis for handling eye movements in the framework of stochastic processes (cfr., Section~\ref{sec:stoch} and previous subsections of the present one). Now we are ready for embracing a broader perspective.

Let us go back to Fig.~\ref{Fig:ensemble} showing an ensemble of  scan paths recorded from different observers while viewing  the same image (in a similar vein, we could take into account an ensemble of  scan paths recorded from the same observer in different trials).  Our assumption is that such ensemble  represents the outcome of a stochastic process and  modelling/analysis should  confront with the fundamental question: What is the probability $P(\mathbf{x}, t)$ of gazing at location $\mathbf{x}$ at time $t$?

There are three different levels to represent and to deal with such question;  for grasping the idea it is useful to take a physicist perspective.
Consider each trajectory (scan path) as the trajectory of a particle (or a random walker). Then, Figure~\ref{Fig:ensemble}  provides a snapshot of the evolution of a many-particle system. In this view, the probability $P(\mathbf{x}, t)$ can be interpreted as the density $\rho(\mathbf{x}, t)$ (number of particles per unit length, area or volume) at point $\mathbf{x}$ at time $t$. In fact the density $\rho(\mathbf{x}, t)$ can be recovered by multiplying the probability density   $P(\mathbf{x}, t)$   by the number of particles.

The finest grain of representation  of a many-particle system is the individual particle, where each stochastic trajectory becomes the basic unit of the probabilistic description of the system. This is the \textbf{microscopic level}. In its modern form, it was first proposed   by the french physicist Paul Langevin, giving rise to the notion of random walks where the single walker dynamics is governed by both regular and stochastic forces  (which resulted in a new mathematical field of  \textbf{stochastic differential equation}, briefly SDE, cfr. Box \ref{tab:Ito}). At this level, $P(\mathbf{x}, t)$ can be be obtained by considering the collective statistical behaviour as given by the individual simulation of many individual particles (technically a Monte Carlo simulation, see Box \ref{tab:Sample} and refer to Figure~\ref{Fig:bw20} for an actual example)

In the opposite way, we could straightforwardly consider, in the large scale limit,  the equations governing the evolution of the space-time probability density $P(\mathbf{x}, t)$ of the particles. Albert Einstein basically followed this path when he derived the diffusion equation for Brownian particles \cite{einstein1905motion}. This coarse-grained representation is the \textbf{macroscopic level} description.

A useful analogy for visualising both levels is provided by structure formation on roads such as jam formation in freeway traffic.
At the microscopic scale, one can study the motion of an individual vehicle, taking into account many peculiarities, such as motivated driver behaviour and physical constraints. On the macroscopic scale,  one can directly address phase formation phenomena  collectively displayed by the car ensemble.

How do we relate the macroscopic and the microscopic levels? The crucial link is provided by the intermediate \textbf{mesoscopic level}  of description.  Pushing on with the traffic analogy, instead of following the motion of each vehicle (microscopic level), a stochastic cluster (mesoscopic) of congested cars is considered by starting to average or integrating microscopic fundamental laws. Then, further coarse-graining (and related approximations) allow to reach the macroscopic description where single vehicle behaviour has no place.

In comparison with the microscopic approach based on SDEs, the mesoscopic description does not allow to get individual realisations of the process but yet keeps the whole amount of statistical information of the underlying microscopic process. Technically it consists in finding integral or integro-differential equations for the probability that governs the evolution of the system.  In the picture we have so far outlined the mesoscopic level is represented  by the Chapman-Kolmogorov equation. At the microscopic level we actually  consider the dynamics of the particle as governed by a regular force  plus a  fluctuating, random force due to the incessant molecular impacts. The C-K equation coarse-grains the landscape stating that the probability of a particle being at a point $\mathbf{x} + \Delta \mathbf{x}$ at time $t + \Delta t$ is given by the probabilities of all possible ``pushes'' $\Delta \mathbf{x}$, multiplied by the probability of being at  $\mathbf{x} $ at time $t$. This assumption is based on the independence of the ``push'' of any previous history of the motion; it is only necessary to know the initial position of the particle at time  $t$ not at any previous time. This is the Markov postulate, and the C-K equation is the central dynamical equation for all Markov processes.

Summing up, the passage from a microscopic description  to a macroscopic one, can be envisioned as a coarse-graining operation.
At the mesoscopic level an appropriately chosen coarse-graining of the observation time-scale, permits the physical process to be described as Markovian, and such coarse-graining allows to switch from the individual particle to a density of particles  $\rho(\mathbf{x}, t)$. Subsequently, under the same coarse-graining, the C-K equation can be reduced to a master equation or a diffusion equation describing the evolution of the system at the macroscopic level. More details are given in the following subsections.




\subsubsection{The microscopic level}
The microscopic description of a system amounts to writing down the evolution equations or \textbf{differential equations} (see Box \ref{tab:deq})   that  describe the fine-grained dynamics of  individual trajectories: e.g., the path of  a Brownian particle   or  the scan path of an eye-tracked observer.
A simple form of such equations is the following:
\begin{equation}
 \overbrace{\frac{dx}{dt}}^{\text{state-space rate of change}}=\overbrace{a(x,t)}^{\text{deterministic comp.}} + \overbrace{b(x,t) \xi(t)}^{\text{stochastic comp.}} ,
\label{eq:Langevin}
\end{equation} 
\noindent which we call the \textbf{Langevin equation}, in analogy with the well known equation that in statistical physics describes  the time evolution of the velocity of a Brownian particle. In Eq. (\ref{eq:Langevin}) the drift term $a(x,t)$ represents the  deterministic component  of the process; the diffusive component $b(x,t) \xi(t)$   is the stochastic component, $\xi(t)$ being the ``noise''  sampled from some probability density, i.e. $\xi(t) \sim P(\xi)$, usually a zero-mean Normal distribution.

Equation (\ref{eq:Langevin}) is an SDE,  which  in a more formal way can be written in the It\^o form of Eq. (\ref{eq:LangevinIto}) as detailed in Box \ref{tab:Ito}.

\begin{table}
\begin{svgraybox}
\caption{ \textbf{Dynamical systems and differential equations}}
\label{tab:deq}
A system that changes with time is called a \textbf{dynamical system}. A dynamical system consists of a \textbf{space of states} and entails a law of motion between states, or a \textbf{dynamical law}.
The deterministic component of Langevin equation (\ref{eq:Langevin})
\begin{equation}
 \frac{dx(t)}{dt}= a(x(t),t)
\label{eq:Langdet}
\end{equation} 
\noindent is one such law, the variable $x(t)$ being the variable that, moment to moment, takes values in the state space of positions. Equation (\ref{eq:Langdet}) is a \textbf{differential equation} describing the rate of change of state-space variable $x$. 

In simple terms, a dynamical law is a rule that tells us the next state given the current state. This can be more readily appreciated if we recall the definition of derivative given in Box \ref{tab:int}, but avoiding  the shrinking operation ($\lim_{\Delta t \rightarrow 0}$), i.e. we approximate the derivative as a discrete difference 
$$\frac{dx(t)}{dt} \approx \frac{x(t + \Delta t)- x(t) }{\Delta t}.$$  By assuming for simplicity  a unit time  step, i.e., $\Delta t  = 1$ and substituting in Eq.  (\ref{eq:Langdet}):
\begin{equation}
\overbrace{ x(t + 1)}^{\text{next state}} = \overbrace{x(t)}^{\text{current state}} + a(x(t),t)
\label{eq:Langdetdisc}
\end{equation} 
Equation (\ref{eq:Langdetdisc}) is  the discrete-time version of the differential equation (\ref{eq:Langdet}), namely a finite-difference equation. The model in discrete time emphasises the predictive properties of the law: indeed,  with the scientific method we seek to make predictions about phenomena that are subject to change. Caveat: we should always be cautious about how predictable the world is, even in classical physics. Certainly, predicting the future requires a perfect knowledge of the dynamical laws governing the world but at the same time  entails  the ability to know the initial conditions with almost perfect precision. However, perfect predictability is not achievable, simply because we are limited in our resolving power. There are  cases in which the tiniest differences in the initial conditions (the starting state), leads to large eventual differences in outcomes. This phenomenon is called \emph{chaos}. 

The law formalised in Eqs. (\ref{eq:Langdet}) or (\ref{eq:Langdetdisc}) are deterministic. In stark contrast,  the Langevin equation (\ref{eq:Langevin}) ``corrupts'' the deterministic law of motion with the ``noise'' introduced by the RV $\xi(t)$. Thus, the eventual outcome is not deterministic but stochastic (though it may be predictable in probability).  Langevin equation is but one example of \textbf{stochastic differential equation} (SDE).

\end{svgraybox} 
\end{table}

\begin{figure}[t]
\sidecaption[t]
\includegraphics[scale=0.15,keepaspectratio=true]{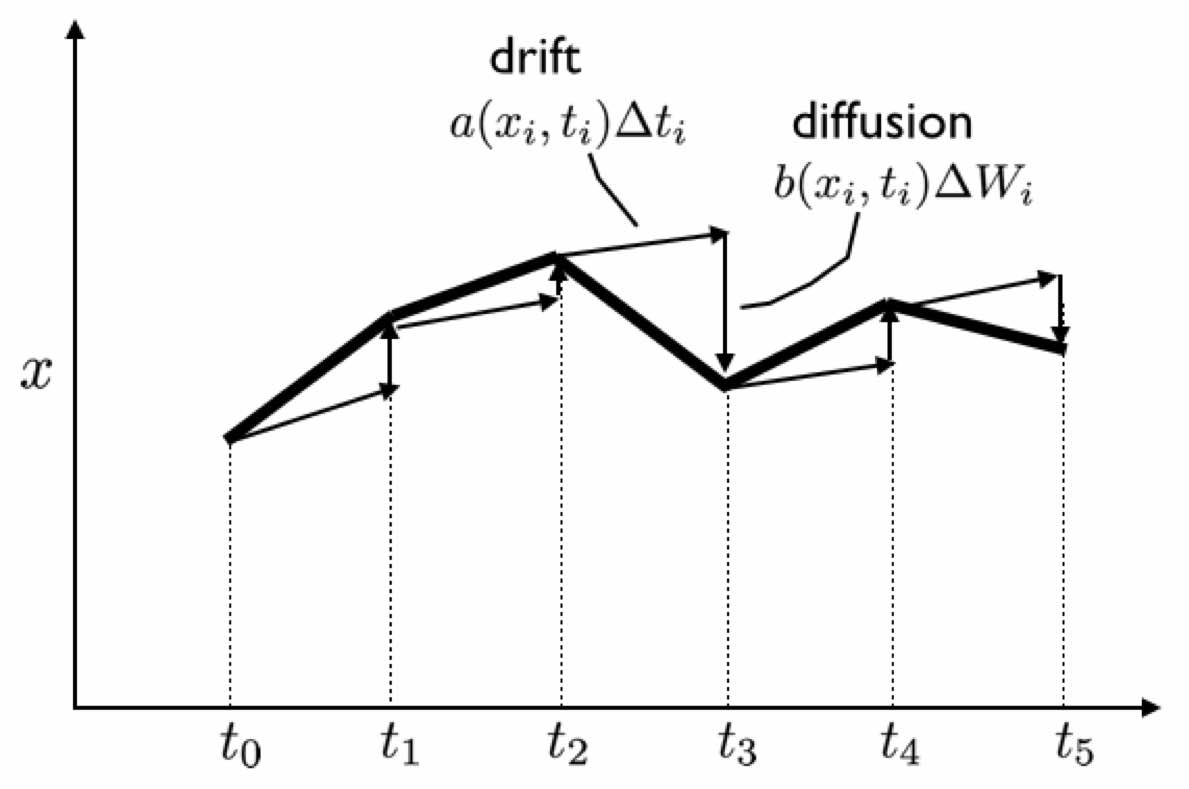}
\caption{ The Cauchy-Euler procedure for constructing an approximate solution of
the Langevin SDE in the It\^o form (cfr. Box \ref{tab:Ito})}
\label{Fig:cauchy}
\end{figure}

Concretely, the construction of a trajectory (a solution) can be performed by refining the intuitive  discretisation approach presented in Box \ref{tab:deq}.   Eq. (\ref{eq:Langevin}) is discretised as in  Eq. (\ref{eq:LangeItodiscrete}) by executing a sequence of drift and diffusion steps as illustrated in Fig. \ref{Fig:cauchy}.

In continuous time a $2$-dimensional random motion of a particle, with stochastic position/state  $\mathbf{x}(t)$, under the influence of an external force field \cite{siegert2001},  can be   described by the Langevin stochastic equation 
\begin{equation}
d\mathbf{x}(t) =  \mathbf{A}( \mathbf{x},t)dt+ {\mathbf  B}({\mathbf x}, t) \boldsymbol \xi dt.
\label{eq:LangND}
\end{equation}
\noindent  As in the one-dimensional case, the trajectory of  ${\mathbf x}$ is determined
by a deterministic part $ \mathbf{A}$, the drift,  and a stochastic part 
 ${\mathbf  B}({\mathbf x}, t) \boldsymbol \xi dt$, where  $\boldsymbol \xi$ is a random vector and  
$\mathbf{B}$ is a diffusion factor.

By simulating from  Langevin-like equations, for suitable choice of parameters   $\mathbf{A}$, $\mathbf{B}$ and noise distribution $P(\boldsymbol \xi)$ (which is likely to be non Gaussian as we will see) it is possible to obtain trajectories that are similar to the individual trajectories shown in the left panel of Fig. \ref{Fig:coarse}.

\subsubsection{The mesoscopic level}
\begin{figure}[t]
\sidecaption[t]
\includegraphics[scale=0.20,keepaspectratio=true]{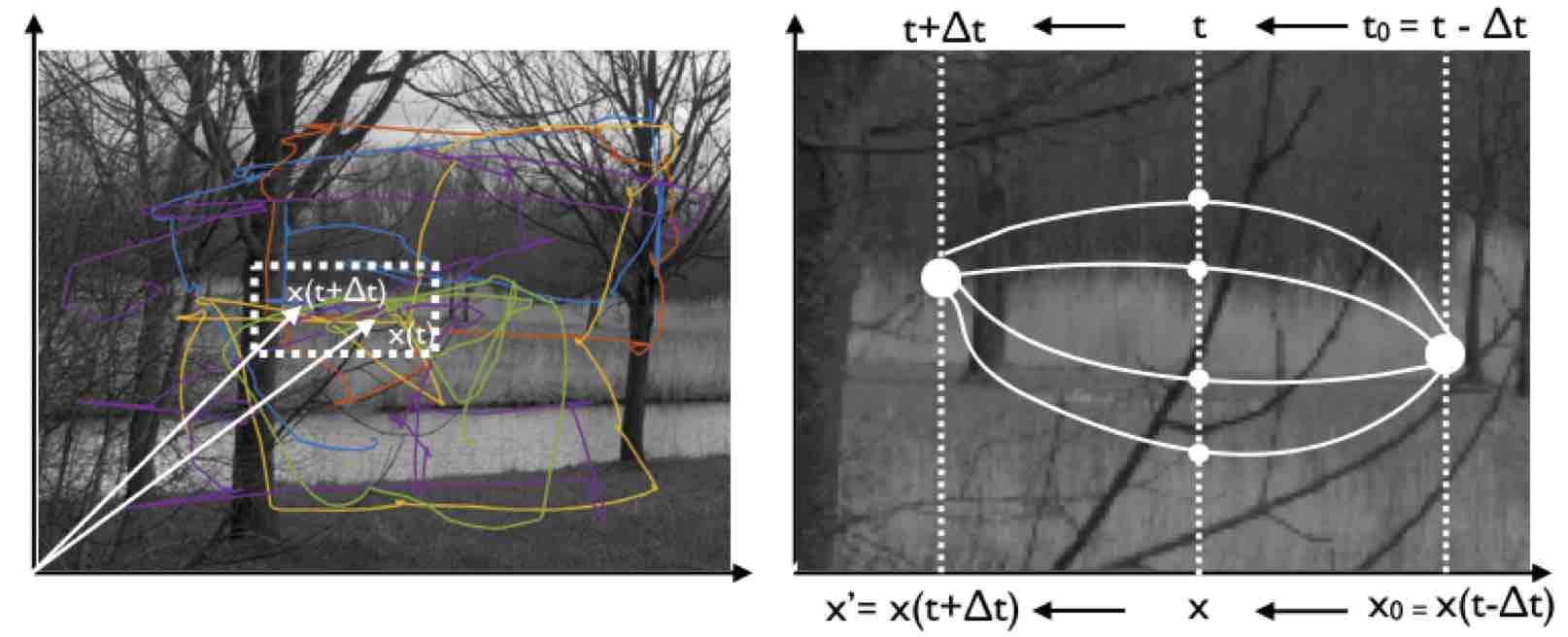}
\caption{Left panel: the  ensemble of  scan paths that was presented in Fig.~\ref{Fig:ensemble}. At the microscopic level, each trajectory can be seen as the output of the simulation of Langevin equation.   Right panel: a close-up of the  window selected from the image in the left panel (dotted line).  The mesoscopic representation used by the C-K equation for computing the conditional probability of shifting the  gaze from  $(\mathbf{x}_0, t_0)$  to  $(\mathbf{x}^{\prime}, t + \Delta t)$. Here the individual trajectories are replaced by the particle densities around such points. $\Delta t$ is a small time interval} 
\label{Fig:coarse}
\end{figure}

If we select an appropriate spatio-temporal scale we can coarse-grain our description by summarising single particle dynamic behaviours in terms of  variations on particle densities at different points, say  $\mathbf{x}_{0}$ and $\mathbf{x}^{\prime}$,   as illustrated in the right panel of Figure~\ref{Fig:coarse}. This density based description to an actual image  the abstract C-K construction we previously presented in Fig. \ref{Fig:chap}. In this case, by considering the state $\mathbf{x} = (n,m)$ as indexing a discrete grid of image pixels and, admittedly,  to avoid burdening mathematical details,  we can rewrite C-K equation (\ref{eq:CK}) in discrete form:
\begin{equation}
P(\mathbf{x}^{\prime}, t + \Delta t \mid  \mathbf{x}_{0}, t_{0})= \sum_{\mathbf{x}} P(\mathbf{x}^{\prime}, t + \Delta t  \mid \mathbf{x}, t) P( \mathbf{x}, t \mid  \mathbf{x}_{0}, t_{0}).
\label{eq:CKdisc}
\end{equation}

Now, the microscopic individual trajectories are replaced  on a coarse-grain scale by the particle densities around such points and  single particle dynamics  is summarised by the Markov-based transition probability $P(\mathbf{x}^{\prime}, t + \Delta t \mid  \mathbf{x}_{0}, t_{0})$.  Note that even in the discrete case where the C-K equation has an intuitive form, it is equally a complicated equation that a function of four parameters namely has to fulfil.

However,  from this level it is  possible  to compute the evolution of particle densities in a larger scale limit, where we consider  the coarse-grained dynamics of  the overall pdf of the many-particle system, as detailed below. 

\subsubsection{The macroscopic level}

There are two possible macroscopic limits from mesoscopic equations: the macroscopic limit in time or in space.  When we consider the macroscopic limit in time of the C-K equation, we obtain the \textbf{Master equation};   when we consider the macroscopic limit both in time and in the state space of the C-K equation, we obtain the famous \textbf{Fokker-Planck equation} (F-P). 

To illustrate how this works we present a simple derivation of the Master equation starting from the discrete C-K equation (\ref{eq:CKdisc}).
First of all, we assume that $\Delta t$ is a small time interval. Then we characterise the short time properties of the conditional probability $P(\mathbf{x}^{\prime}, t + \Delta t  \mid \mathbf{x}, t)$, where $\mathbf{x}$ denote a number of intermediate states between the initial state $\mathbf{x}_0$  and the ending state $\mathbf{x}^{\prime}$ as usually considered in the C-K construction (cfr. Fig.~\ref{Fig:coarse}, right).

We also assume  that a transition from  $\mathbf{x}$  to state $\mathbf{x}^{\prime}$ is proportional to time when time gets small and that such transition occurs at  a transition rate which we denote $w(\mathbf{x}^{\prime} \mid \mathbf{x})$ (density variation per unit time). Thus:
\begin{equation}
P(\mathbf{x}^{\prime}, t + \Delta t  \mid \mathbf{x}, t) \approx \Delta t \times w(\mathbf{x}^{\prime} \mid \mathbf{x}), 
\label{eq:mast1}
\end{equation}
\noindent which obviously holds only for $\mathbf{x}^{\prime} \neq \mathbf{x}$ since $w(\mathbf{x} \mid \mathbf{x}) = 0$.

However,  to be complete we must also consider  the probability that no state transition occurs in the small time interval: 

\begin{equation}
Q(\mathbf{x}) = 1 - \Delta t \sum_{\mathbf{x}^{\prime} \neq \mathbf{x}} w(\mathbf{x}^{\prime} \mid \mathbf{x}). 
\label{eq:notrans}
\end{equation}

Eventually,
\begin{equation}
P(\mathbf{x}^{\prime}, t + \Delta t  \mid \mathbf{x}, t) \approx \Delta t \times w(\mathbf{x}^{\prime} \mid \mathbf{x}) + Q(\mathbf{x}) \delta_{\mathbf{x}^{\prime} , \mathbf{x}}, 
\label{eq:mast2} 
\end{equation}
\noindent where $\delta_{\mathbf{x}^{\prime} , \mathbf{x}}$ is the Kroenecker symbol:  $\delta_{\mathbf{x}^{\prime} , \mathbf{x}} =0$ when $\mathbf{x}^{\prime} \neq \mathbf{x}$, and $\delta_{\mathbf{x}^{\prime} , \mathbf{x}}=1$ when $\mathbf{x}^{\prime} =\mathbf{x}$.

If we plug the right hand side of Eq.~(\ref{eq:mast2}) in C-K equation~(\ref{eq:CKdisc}), after some (tedious and unrelevant) algebra we obtain the following result:
\begin{eqnarray}
P(\mathbf{x}^{\prime}, t + \Delta t \mid  \mathbf{x}_{0}, t_{0}) =  P(\mathbf{x}^{\prime}, t  \mid  \mathbf{x}_{0}, t_{0}) + \\ \nonumber
\Delta t  \sum_{\mathbf{x}} \left[ w(\mathbf{x}^{\prime} \mid \mathbf{x}) P( \mathbf{x}, t \mid  \mathbf{x}_{0}, t_{0}) -
w(  \mathbf{x} \mid \mathbf{x}^{\prime}) P( \mathbf{x}^{\prime}, t \mid  \mathbf{x}_{0}, t_{0})  \right].
\label{eq:ME}
\end{eqnarray}

Note that in the limit $\Delta t \rightarrow 0$, we can write
$$ \lim_{\Delta t \to 0} \frac{P(\mathbf{x}^{\prime}, t + \Delta t \mid  \mathbf{x}_{0}, t_{0}) - P(\mathbf{x}^{\prime}, t  \mid  \mathbf{x}_{0}, t_{0}) }{\Delta t } = \frac{\partial P( \mathbf{x}^{\prime}, t \mid  \mathbf{x}_{0}, t_{0})}{\partial t}$$
This is the definition of a \emph{partial derivative} of first order with respect to time, denoted by the symbol $\frac{\partial}{\partial t}$. \footnote{If we have a function of more than one variable, e.g., $f(x,y, z,\cdots)$, we can calculate the derivative with respect to one of those variables, with the others kept fixed. Thus, if we want to compute $\frac{\partial f(x,y, z,\cdots)}{\partial x}$, we define the increment $\Delta f= f(\left[x + \Delta x\right], y,z,\cdots) - f(x,y, z,\cdots)$ and we construct the partial derivative as in the simple derivative case as $\frac{\partial f(x,y, z,\cdots)}{\partial x}=\lim_{\Delta x \to 0} \frac{\Delta f}{\Delta x}$. By the same method we can obtain the partial
derivative with respect to any of the other variables.}

Using such definition in Eq. (\ref{eq:ME}), in the limit $\Delta t \rightarrow 0$:
\begin{equation}
\frac{\partial P( \mathbf{x}^{\prime}, t \mid  \mathbf{x}_{0}, t_{0})}{\partial t} =    \sum_{\mathbf{x}} \left[ w(\mathbf{x}^{\prime} \mid \mathbf{x}) P( \mathbf{x}, t \mid  \mathbf{x}_{0}, t_{0}) -
w(  \mathbf{x} \mid \mathbf{x}^{\prime}) P( \mathbf{x}^{\prime}, t \mid  \mathbf{x}_{0}, t_{0})  \right].
\label{eq:ME2}
\end{equation}

This equation can be further simplified by using  the marginalisation rule 
$$ \sum_{\mathbf{x}_{0}}P( \mathbf{x}^{\prime}, t \mid  \mathbf{x}_{0}, t_{0}) P( \mathbf{x}_{0}, t_{0}) = P( \mathbf{x}_{0}, t_{0})$$

Multiplying equation (\ref{eq:ME2}) by $P( \mathbf{x}_{0}, t_{0})$ and summing over $\mathbf{x}_{0}$, we eventually obtain
\begin{equation}
\frac{\partial P( \mathbf{x}^{\prime}, t)}{\partial t} =    \sum_{\mathbf{x}} \left[ w(\mathbf{x}^{\prime} \mid \mathbf{x}) P( \mathbf{x}, t ) -
w(\mathbf{x} \mid \mathbf{x}^{\prime}) P( \mathbf{x}^{\prime}, t )  \right].
\label{eq:MEfinal}
\end{equation} 

This is our final Master equation and has a very simple interpretation. The density $P( \mathbf{x}^{\prime}, t)$ is the probability of the ensemble of observers (the many particle system) of being in (gazing at)  state $\mathbf{x}^{\prime}$ at time $t$.  How does this density change over time? The system can be in any state $ \mathbf{x} \neq \mathbf{x}^{\prime} $ and move into the state $\mathbf{x}^{\prime} $. The system is in state $\mathbf{x}$  with probability $P( \mathbf{x}, t )$ and from any of these states $\mathbf{x}$ it moves to the state $\mathbf{x}^{\prime} $ with probability $w(\mathbf{x}^{\prime} \mid \mathbf{x}) P( \mathbf{x}, t )$. But simmetrically, it could also already be in state $\mathbf{x}^{\prime} $ and move to one of the other states $\mathbf{x}$ with probability $w(\mathbf{x} \mid \mathbf{x}^{\prime}) P( \mathbf{x}^{\prime}, t )$.

In other terms the Master equation is a probability flux balance equation, best understood if $P(\mathbf{x}^{\prime},t)$ is interpreted as a particle density: the rate of change in that density (derivative) is the difference of what comes in and what goes out.

Technically speaking the Master equation (\ref{eq:MEfinal}) is a stochastic \textbf{partial differential equation} (PDE) defining the ``law of motion'' of the density $P(\mathbf{x}^{\prime},t)$ in probability space and it has been obtained by taking the limit in time of the C-K equation.

When we consider the macroscopic limit both in time and in the state space of the C-K equation, we obtain another PDF,  namely the \textbf{Fokker-Planck }(F-P) equation.We omit the formal derivation because beyond the scope of this Chapter. 

%
%
In the simple 1-dimensional case, the F-P equation for diffusive  processes is the following:
\begin{equation}
\frac{\partial P(x,t)}{\partial t}= - \frac{\partial }{\partial x} [a(x,t)P(x,t)] + \frac{1}{2}\frac{\partial^{2}}{\partial x^{2}} [b(x,t)^2 P(x,t)]
\label{eq:FP}
\end{equation}

The symbols $\frac{\partial}{\partial x}$  and $\frac{\partial^{2}}{\partial x^{2}} = \frac{\partial}{\partial x} ( \frac{\partial}{\partial x}) $ denote partial derivatives with respect to space of first and second-order, respectively.

What is important to note is that there is a formal link between the microscopic description provided by the Langevin equation (cfr., the 1-dimensional case of Eq. \ref{eq:Langevin})   and the macroscopic description addressed by the F-P equation (\ref{eq:FP}), which is established via 
$a(x,t)$ and  $b(x,t)$.
The term $a(x,t)$ represents a \textbf{drift} which is related to the average deviation of the process
\begin{equation}
a(x,t) = \lim_{\Delta t \to 0} \frac{\langle \Delta x \rangle}{\Delta t}.
\label{eq:drift}
\end{equation}
\noindent over a small time interval $\Delta t$; the bracket operator $\langle  \rangle$ denotes the average or  expectation  value of any function $f(X)$.\footnote{This is physicists' preferred notation, which you are likely to most frequently  run into when dealing with these problems. In other more mathematically inclined papers and books you will find the expectation notation $E\left[f(X)\right]$  or $E f(X)$.  }. $\Delta x$ is a deviation or displacement in state-space  

The term $b^2(x,t)$ represents a \textbf{diffusion} term, which is related to the mean square deviation of the process:
\begin{equation}
b^2(x,t)= \lim_{\Delta t \to 0} \frac{ \langle(\Delta x)^2 \rangle}{\Delta t}.
\label{eq:diff}
\end{equation}

Eventually, at the macroscopic level, by knowing the evolution of $P(x,t)$ in time,   one can obtain  statistical ``observables'' as the moments, correlations, etc. These obviously lack  microscopic details from the underlying stochastic process, which for some specific purposes may be important.

We will turn now to the fundamental example of the Wiener process to make clear the connections between the macroscopic and microscopic levels of description. 

\begin{table}
\begin{svgraybox}
\caption{\textbf{Stochastic Differential Equations}}
\label{tab:Ito}
The Langevin equation written in the form (\ref{eq:Langevin}) poses some formal problems. Since $\xi(t)$  is  noise it consists of a set of points that in some cases can be even uncorrelated. As a  consequence $\xi(t)$ is often non-differentiable. Thus, $x(t)$ should be non-differentiable too, so that the left hand side of (\ref{eq:Langevin}) is incoherent from this point of view. To overcome this problem, the $1-$dimensional Langevin equation is usually presented in the mathematically sound form: 
\begin{equation}
dx(t) = a(x(t),t)dt + b(x(t),t) \xi(t)dt = a(x(t),t)dt + b(x(t),t) dW(t)
\label{eq:LangevinIto}
\end{equation}
with $W(t) = \int_0^t \xi(t^\prime) dt^\prime$,  so that the integration of the stochastic component $\int b(x,t) dW(t)$ can be performed according to the rules of stochastic calculus (in the It\^{o} or Stratonovich approach \cite{higham2001algorithmic}). Throughout this chapter we shall use with a certain liberality both forms (\ref{eq:Langevin}) and  (\ref{eq:LangevinIto}) at our convenience.
Thus,  a stochastic quantity $x(t)$ obeys an It\^o SDE written as in (\ref{eq:LangevinIto}), 
if for all $t$ and $t_0$,
\begin{equation}
 x(t) = x(t_0) + \int_{t_0}^{t} a(x(t^\prime),t^\prime)dt^\prime + \int_{t_0}^{t} b(x(t^\prime),t^\prime) dW(t^\prime) 
\label{eq:LangeItoint}
\end{equation}


A  discretised version of the SDE can be obtained by taking a mesh of points $t_i$(Fig.\ref{Fig:cauchy})
$$t_0 < t_1 < t_2<\cdots <t_{n-1}<t_{n}=t$$
and writing the equation as
\begin{equation}
 x_{i+1} = x_{i} + a(x_{i},t_{i}) \Delta t_{i} + b(x_{i},t_{i}) \Delta W_{i} 
\label{eq:LangeItodiscrete}
\end{equation}
Here, $x_{i}= x(t_{i})$ and 
\begin{equation}
\Delta t_{i}= t_{i+1}-t_{i} ,
\end{equation}  
\begin{equation}
\Delta W_{i}= W(t_{i+1}) - W(t_{i}) \propto \sqrt{\Delta t_{i} }\xi_i.
\label{eq:deltaW}
\end{equation}  

The approximate procedure for solving the equation is to calculate $x_{i+1}$ from the knowledge of $x_{i}$ by adding a deterministic term $a(x_{i},t_{i}) \Delta t_{i}$ and a stochastic term $b(x_{i},t_{i}) \Delta W_{i}$, which contains the element $\Delta W_{i}$, namely the increment of the Wiener process. The solution is then formally constructed by letting the mesh size go to zero. The method of constructing a solution outlined above is called the Cauchy-Euler method, and can be used to generate simulations. By construction the time development of $x(t)$ for $t > t_0$ is independent of $x(t_0)$ for $t > t_0$ provided $x(t_0)$ is known. Hence, $x(t)$ is a Markov process.

For an intuitive, Matlab based, thorough introduction to SDEs see Higham \cite{higham2001algorithmic}. 

\end{svgraybox} 
\end{table}

\subsubsection{Example: the Wiener process}
Recall again  the most famous Markov process: the Wiener process describing Brownian motion (Bm), cfr., Box \ref{tab:hall}. The SDE defining the  motion of a particle undergoing 1-dimensional Brownian motion can be obtained by setting to zero the drift component $a(x,t)$ and letting $b(x,t) =\sqrt{2D}$, where $D$ is the diffusion coefficient. Thus:
\begin{equation}
 dx =  \sqrt{2D}  dW(t)
\label{eq:LangevinItoWien}
\end{equation}

\begin{figure}[t]
\sidecaption[t]
\includegraphics[scale=0.13,keepaspectratio=true]{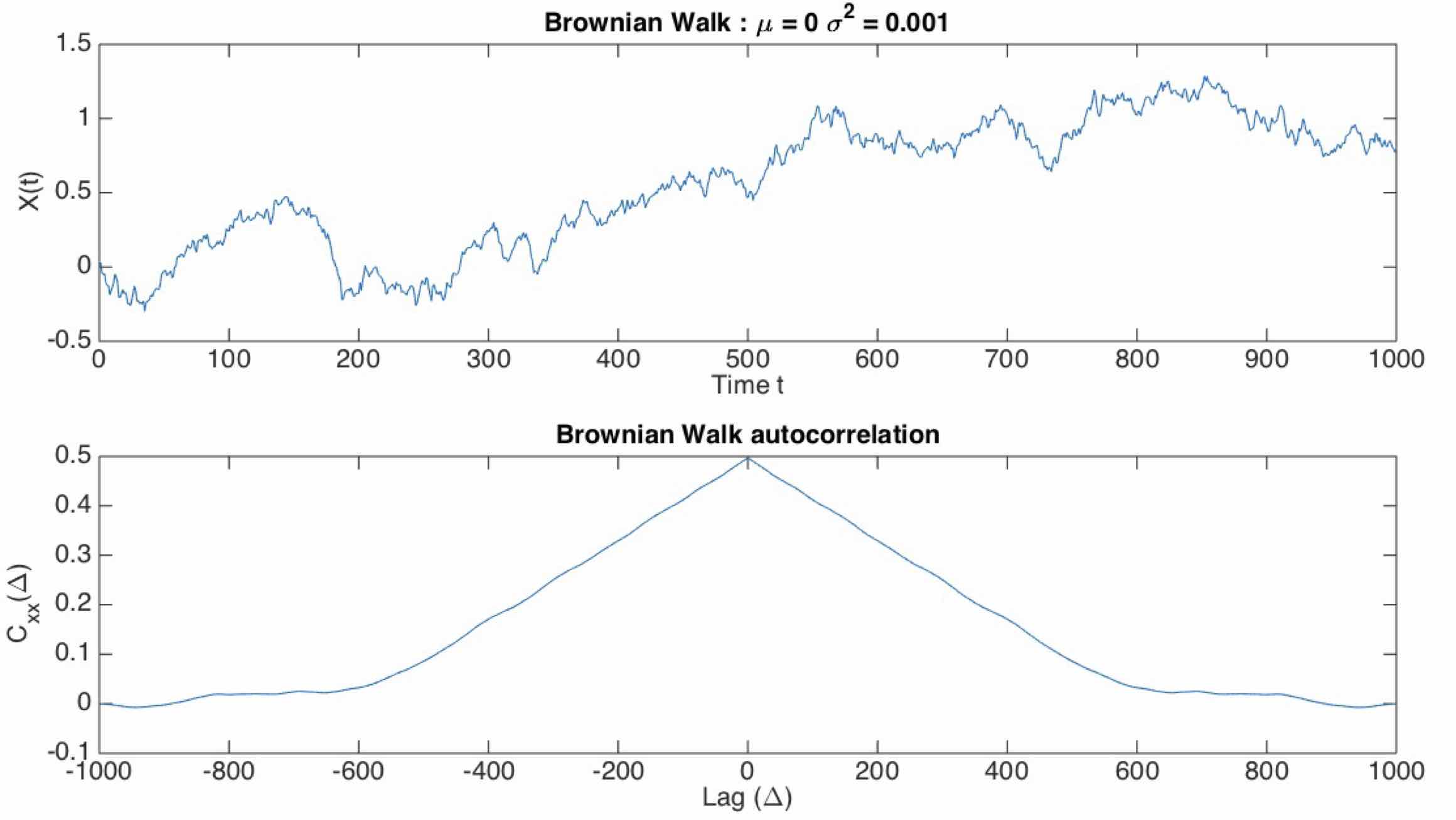}
\caption{One dimensional Brownian motion. Top: the random walk process; bottom: the autocorrelation of the process. (cfr. Box \ref{tab:quant})}
\label{Fig:bw}
\end{figure}

By using Eqs. (\ref{eq:LangeItodiscrete}) and  (\ref{eq:deltaW}) introduced in  Box \ref{tab:Ito},  the discretised version of the Wiener process (\ref{eq:LangevinItoWien})   over a small but finite time interval $\Delta t = T/N$, $N$ being the discrete number of integration steps, can be written as
\begin{equation}
 x_{i+1} = x_{i} +   \sqrt{2D} \Delta W_{i} = x_{i} + \sqrt{2D \Delta t_{i} }\xi_i
 \label{eq:windisc}
\end{equation}
with $\xi$ sampled from a zero-mean Gaussian distribution of unit variance $\mathcal{N}(0,1)$

Equation \ref{eq:windisc}  shows that
  the  system describes a refined version of the simple additive random walk.
Once again, it is worth noting that   since the $\xi(t)$ are sampled i.i.d,   then  the differences in sequential observations  are i.i.d, namely, $ x_{i+1} - x_{i}  = \Delta x_{i}$, rather than the observations themselves.
In fact, if we compute the auto-correlation function of the $\{x(t)\}$ time series, it exhibits  a slower decay ---differently from the white noise process---, which shows how this simple random walk exhibits memory (Fig. \ref{Fig:bw}). 

%

\begin{figure}[t]
\sidecaption[t]
\includegraphics[scale=0.20,keepaspectratio=true]{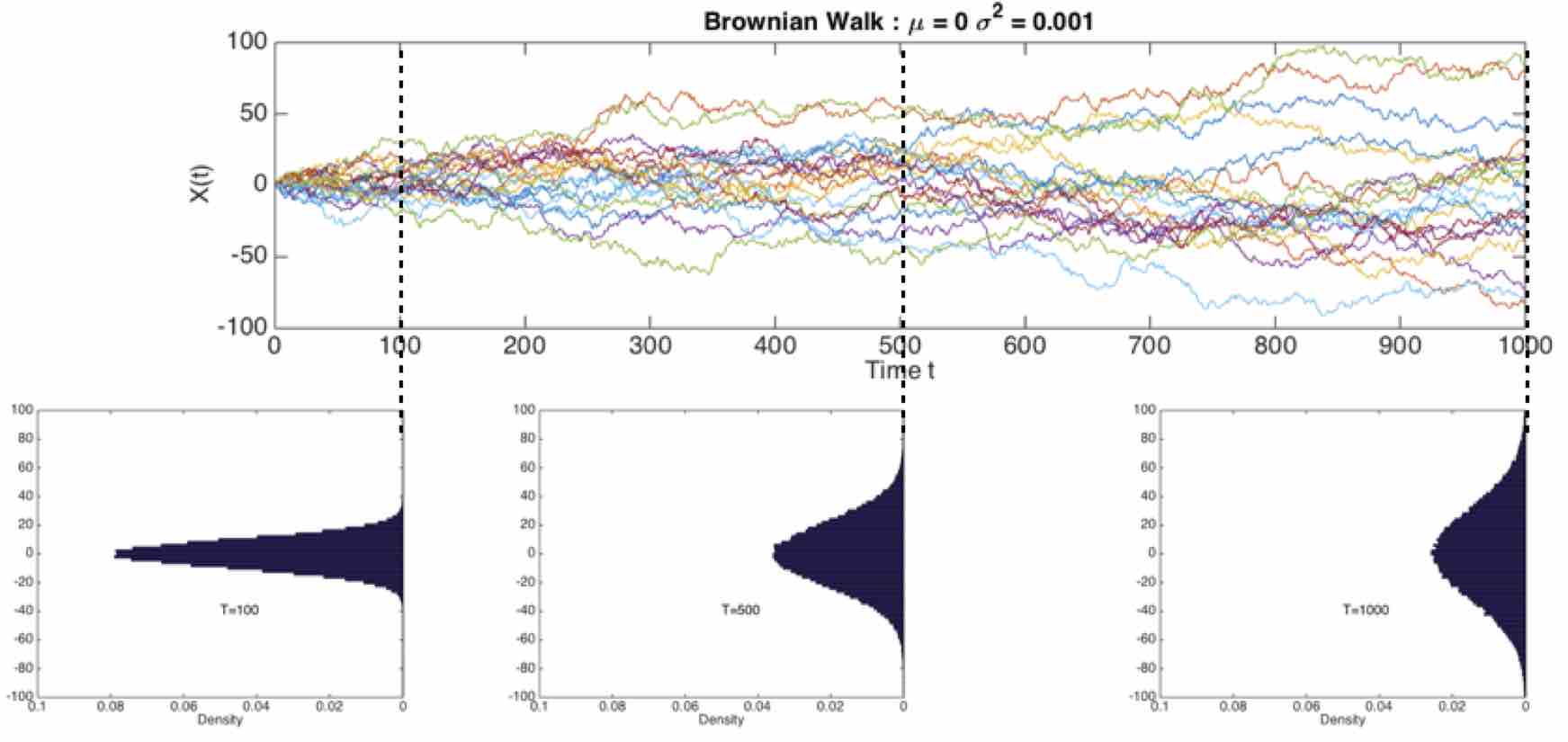}
\caption{A Monte Carlo simulation (cfr. Box~\ref{tab:Sample}) of the macroscopic dynamics  of $P(x,t)$ in the case of Brownian motion. Top: The simulation of individual trajectories of $10^5$  random walkers: only $20$ are shown for visualisation purposes. Bottom: The distributions (histograms) of the  walkers, after $T = 100$,  $T = 500$ and $T = 1000$ time steps. The  distribution initially concentrated at a point takes later the Gaussian form, whose width grows in time as $t^{1/2}$. This kind of diffusion is called the \emph{normal diffusion}.}
\label{Fig:bw20}
\end{figure}

Let us   simulate the Brownian motion  of a large number of particles, say $10^5$ particles. Resorting to a Monte Carlo approach, we can obtain this result by running in parallel $10^5$ random walks, each walk being obtained by iterating Eq.(\ref{eq:windisc}). Fig. \ref{Fig:bw20} (top) shows an example of $20$ such trajectories. In probabilistic terms each trajectory is a realisation, a sample of the stochastic process $\{X(t)\}$. 

We may be interested in gaining some statistical insight of the collective behaviour of all such random walkers. This can be obtained by considering the the dynamics of the  pdf $P(x,t)$ describing the probability of finding a particle at position $x$ at time $t$. Empirically, we can estimate $P(x,t)$ at any time $t$ by computing the density of particles occurring within a certain bin $(x-\delta, x+\delta)$ centred on $x$, that is by computing the histogram $h(x,t)$ and normalising it with respect to the total number of particles. This procedure is shown in the bottom of Fig. \ref{Fig:bw20}: the empirical pdf has a nice bell shape, i.e. it is a Normal distribution, which spreads as time increases.

This insight can be given a formal justification by resorting to  the macroscopic level of description of process dynamics as provided by F-P equation (\ref{eq:FP}). By setting again $a(x,t)=0$ and $b(x,t) =\sqrt{2D} $:
\begin{equation}
\frac{\partial P(x,t)}{\partial t}= D\frac{\partial^{2}P(x,t)}{\partial x^{2}}
\label{eq:Diff}
\end{equation}
This is the well-known  \textbf{heat} or \textbf{diffusion  equation}. Thus, the pdf $P(x,t)$ of finding a particle at position $x$ at time $t$ evolves in time according to the diffusion equation when the underlying microscopic dynamics is such that the particle position corresponds to a Wiener process. 

The solution to the heat equation (\ref{eq:Diff})  is the time-dependent Gaussian pdf, as anticipated in Box \ref{tab:hall}:
\begin{equation}
P(x,t)= \frac{1}{\sqrt{4 \pi D t}} \exp\left( -\frac{x^2}{4D t} \right),
\label{eq:gaussdiff2}
\end{equation}
By comparing the  Gaussian pdf variance $\sigma^2 = 4D t$  to the definition of $b^2(x,t)$ given in Eq. (\ref{eq:diff}),  we can set the following correspondences:
\begin{equation}
\sigma^2 = 2 D t = b^2 t \approx \langle x^2\rangle
\label{eq:D}
\end{equation}
In other terms for Bm, the average square deviation of the walk, and thus the spread of the Gaussian,  grows linearly with time, as it can be intuitively appreciated from Fig. \ref{Fig:bw20}.

More precisely, define  the \textbf{Mean Square Displacement} (MSD) of a walk that starts at position $x_0$ at time $t_0$:
\begin{equation}
MSD = \langle | x - x_0 |^2\rangle,
\label{eq:MSD}
\end{equation}
\noindent which is, quoting Einstein,  the square of the displacement in the direction of the  $x$-axis ``that a particle experiences on the average'' \cite{EinsteinBrown}. Here $x_0$ denotes the initial position.  In the case of Brownian motion, Einstein \cite{EinsteinBrown} was the first to show that:
\begin{equation}
MSD = 2 D t
\label{eq:MSD2}
\end{equation}
Note that $\langle | x - x_0 |^2\rangle = \langle x^2\rangle + x_0 ^2 - 2 x_0 \langle x\rangle $. Hence, when the initial position is at $x_0=0$, $MSD = \langle x^2\rangle \propto t $.

Equation  (\ref{eq:MSD2}) is sometimes more generally written  in terms of the \textbf{Hurst exponent} $H$
\begin{equation}
MSD = k t^{2H}
\label{eq:MSD_H}
\end{equation}
\noindent with $H=\frac{1}{2}$ for Bm. This is useful for characterising different kinds of diffusions, like hyperdiffusion or subdiffusion, as discussed in Box \ref{tab:Hurst}.


\begin{table}
\begin{svgraybox}
\caption{\textbf{The Hurst exponent: the Swiss army knife of diffusion processes (without SDE pain) }}
\label{tab:Hurst}
%
%
The Hurst exponent, $H$, is related to the signal correlation behaviour and it allows the detection of the long-range dependences. 
In general, properties of Gaussian diffusion may be expressed in terms of the MSD of $x$ and its relation to time:
\begin{equation}
MSD = \langle | x(t) -x(0) |^2 \rangle =kt^{2H}
\label{eq:MSD_H2}
\end{equation}
When $H=0.5$, MSD is linear in time:
\begin{equation}
MSD = kt,
\label{eq:linMSD}
\end{equation}
\noindent which exemplifies the ordinary condition of Bm, the derivative of Bm being additive white Gaussian noise.
When $H > 0.5$, increments are positively correlated, i.e. the random walk shows the tendency to continue to move in the current direction. This behaviour is called \textbf{persistence}.  In this case, MSD increases nonlinearly with respect to time, indicative of \textbf{hyperdiffusion}. In particular, for $H=1$,
\begin{equation}
MSD =  kt^2.
\label{eq:sqMSD}
\end{equation} 
\noindent In this case diffusion follows correlated \textbf{fractional Brownian motion} (fBm), whose derivative is fractional Gaussian noise.

In the case $H < 0.5$, the random walk generates negatively correlated increments and is \textbf{anti-persistent}.

It is important to note that for $H \neq 0.5$, the increments are not independent, thus  the fBm is a Gaussian process but it is not a Markov process.


Interesting work has been reported in the recent literature on the use of the $H$ exponent to analyse eye movements, e.g. by Engbert and colleagues on random walk analysis of fixational  eye movements \cite{engbert2006microsaccades}.
For such purposes the scaling exponent $H$ can be estimated as follows \cite{engbert2006microsaccades}. Consider a time series of gaze positions of length $N$, $\mathbf{x}_1, \cdots, \mathbf{x}_N$. Define the displacement estimator
\begin{equation}
\overline{\delta^2(\Delta_m)} = \frac{1}{N-m} \sum_{i=1}^{N-m}  \| \mathbf{x}_{i+m} - \mathbf{x}_{i} \|^2 ,
\label{eq:D2}
\end{equation}
\noindent namely the \textbf{time averaged MSD}.
By recalling that $MSD  \approx t^{2H}$, the Hurst exponent $H$ can be obtained by calculating the slope in the plot of $\log \overline{\delta^2(\Delta_m)} $ as a function of   $\log \Delta_m$, where  $\Delta_m  = mT_0$ is the \emph{time lag},  $T_0$  (ms) being the sampling time interval and $m=1,2,3,\cdots$.

More recently, Engbert has proposed more sophisticated estimation framework based on the Bayesian approach \cite{engbert2012bayesianHurst}.

A last, but important remark: the equivalence
\begin{equation}
\overline{\delta^2(\Delta)} \approx \langle x(\Delta) ^2 \rangle 
\label{eq:D3}
\end{equation}
holds when the process is \textbf{ergodic}, where
ensemble averages and long-time averages are equivalent in the limit of long measurement times.

\end{svgraybox} 
\end{table}

\subsubsection{Case study: from random walks to saccade latency}

\begin{figure}[t]
\sidecaption[t]
\centering
     \subfigure[Latencies]{\includegraphics[width=0.52\textwidth]{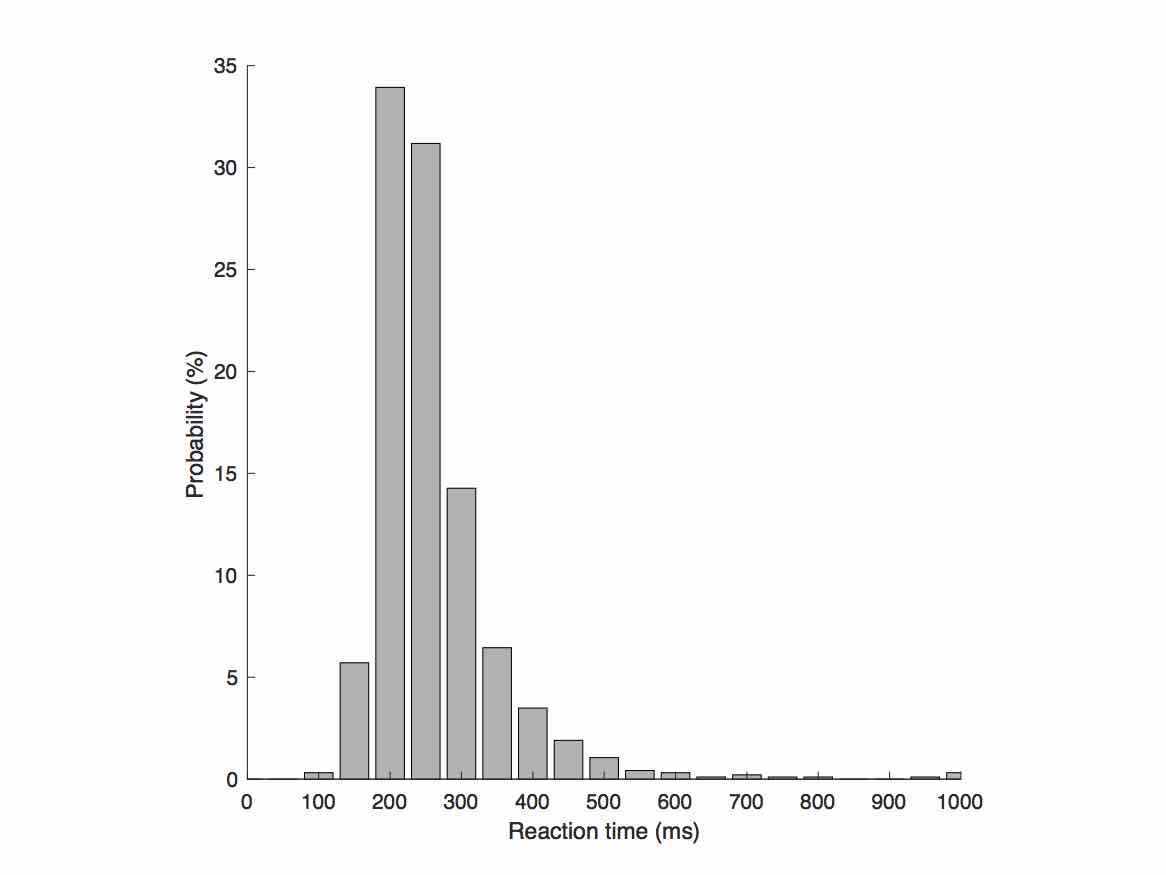}\label{Fig:RThist}}
   \subfigure[Reciprocals]{\includegraphics[width=0.52\textwidth]{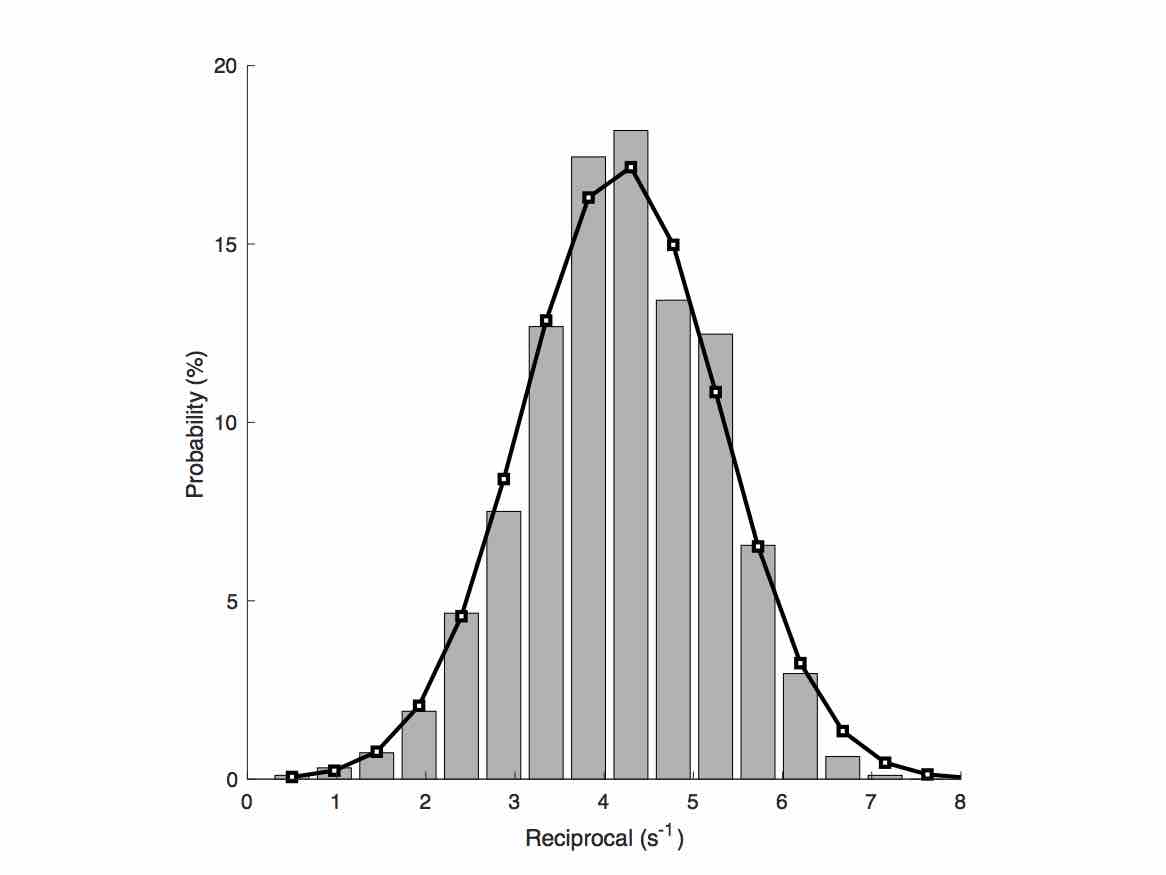}\label{Fig:RTRechist}}
\caption{Empirical distributions (histograms) of latencies (left panel) and reciprocals (right panel). The latencies/reaction times themselves are typically skewed, with a tail towards long reaction times.  When plotted as a function of reciprocal reaction time (inverse of the reaction times or promptness), the distribution looses its skewness, and it turns out to obey the most fundamental of all stochastic laws, the normal or Gaussian distribution.  The reciprocal histogram can thus be approximated by a normal distribution  (the black line superimposed to the histogram, with the same mean and standard deviation of actual reciprocals)\cite{carpenter1995neural,Noorani2016} }
\label{Fig:RT}
\end{figure}

A saccade represents the output of a decision, a choice of where to look, and reaction time, or latency, can be regarded as an experimental ``window'' into decision processes.   Typical latencies are around $200$ms (see Fig. \ref{Fig:RThist}) whereas the sensorimotor components of a saccade are only around $60$ms: this suggests that
reaction time is composed of more than the simple sum of times of for sensory input and motor output, and  such extra time (also called neural procrastination)  reflects the time taken for the brain to choose a response. Put simply, reaction time is  a useful indicator of decision time.
In experimental paradigms, reaction time varies stochastically between one trial and the next, despite standardized experimental conditions. Furthermore, distribution of reaction times is typically skewed, with a tail towards long reaction times (Fig. \ref{Fig:RThist}). However, if we take the reciprocal of the latencies and plot these in a similar fashion, the resulting distribution appears Gaussian (Fig. \ref{Fig:RTRechist}). 

\begin{table}
\begin{svgraybox}
\caption{\textbf{The probit function}}
\label{tab:Probit}
When confronting with reaction times and in particular with the well known Carpenter's LATER model~\cite{carpenter1995neural,Noorani2016}, one tipically has to deal with reciprobit plots, a tool that sometimes students find difficult to conceptualise. Indeed, the subtle probit function lies  behind such graphs. 

Recall from your elementary probability and statistics class that a quantile function returns the value $x$ such that $F_X(x)= P(X \leq x)= p$ where $0 < p < 1$ is a probability value and  $F_X(\cdot)$ is the \textbf{cumulative distribution function} (CDF) of random variable $X$. In simple words, it returns a threshold value $x$ below which random draws from the given CDF would fall $p$ percent of the time.

The probit function $probit(\cdot)$ is  exactly the quantile function associated with the standardised normal distribution $\mathcal{N}(0,1)$ and the standard CDF $\Phi(\cdot)$, formally
\begin{equation}
 probit(p) = \Phi^{-1}(p).
 \label{eq:probit}
\end{equation}
\noindent where $\Phi^{-1}(\cdot)$ is the inverse CDF of the standard normal.
Clearly, by the above definition, the following properties hold:
$$  \Phi(probit(p)) = \Phi(\Phi^{-1}(p))=p $$
and
$$probit(\Phi(z)) = \Phi^{-1}(\Phi(z)) = z$$
\noindent where $z \sim \mathcal{N}(0,1)$ is a standardised random variable (the famous ``z-score'') sampled from the standard normal.

Unfortunately, the standard  CDF $\Phi$  and its inverse $\Phi^{-1}$ are not available in closed form, and computation requires careful use of numerical procedures (unless we do not exploit good old tables). To such end, $\Phi(z)= \frac{1}{\sqrt{2\pi}}\int_{-\infty}^{z} e^{\frac{-t^2}{2}}dt$ is related to the error function $erf(x)=\frac{1}{\sqrt{\pi}}\int_{-x}^{x} e^{-t^2}dt$ as  $\Phi(z)= \frac{1}{2}\left[ 1 + erf \left( \frac{z}{\sqrt{2}}\right)\right]$. However, we need not worry  too much about these mathematical details since numerical calculations of the error and the inverse error  functions are widely available in software for statistics and probability modelling.

Eventually, given a probability value $\Phi(z)= p$  from the normal CDF,  the probit of such value can be computed in terms of the inverse error function $erf^{-1}(\cdot)$ as
\begin{equation}
 probit(p) = \sqrt{2}erf^{-1}(2p -1),
 \label{eq:probit2}
\end{equation}
\noindent which, for instance, in Matlab code boils down to  the simple  statement
\begin{center}
\textsf{prob_p    =  sqrt(2) * erfinv(2 * p -1)}
\end{center}

This is all we need for producing beautiful reciprobit plots (see Fig.~\ref{Fig:Reciprob}).
\end{svgraybox} 
\end{table}

\begin{figure}[t]
\sidecaption[t]
\centering
     \subfigure[Reciprocal ECDF]{\includegraphics[width=0.53\textwidth]{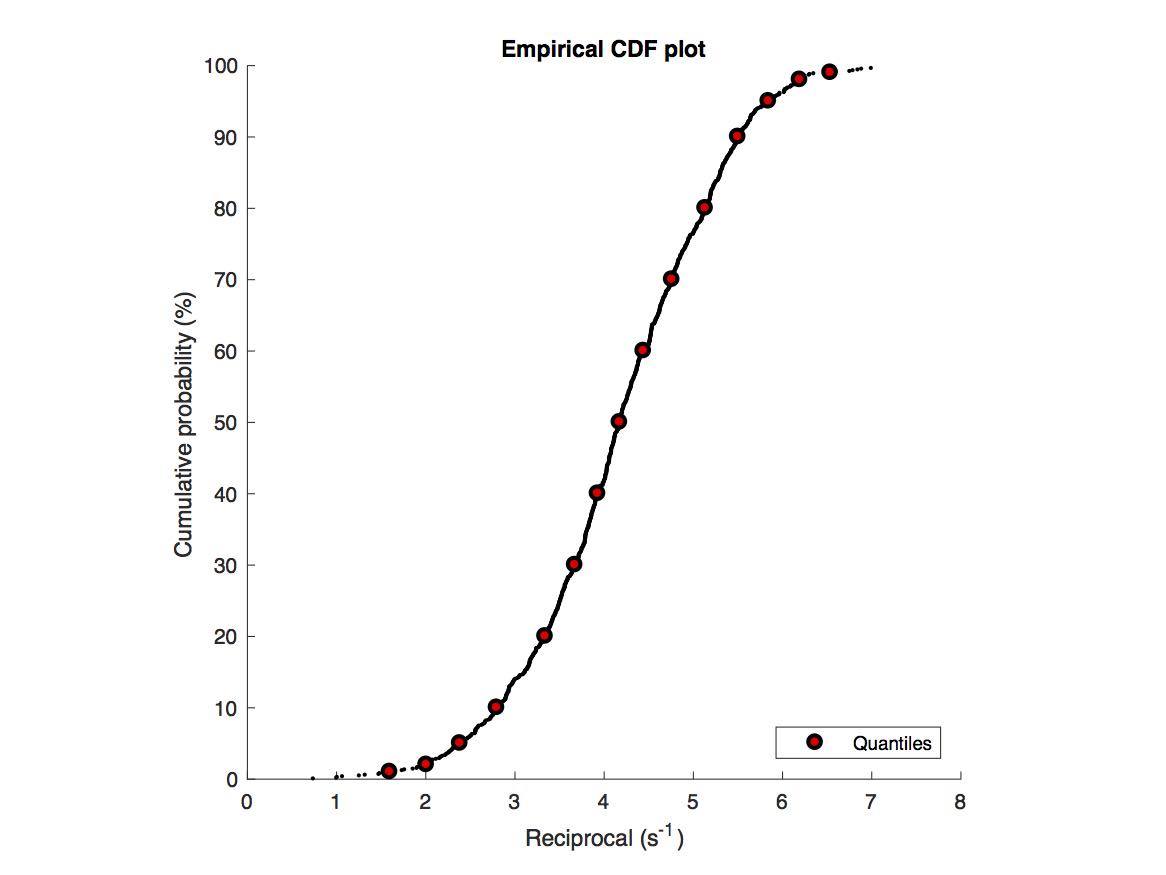}\label{Fig:ReciCDF}}
   \subfigure[Reciprobit plot]{\includegraphics[width=0.53\textwidth]{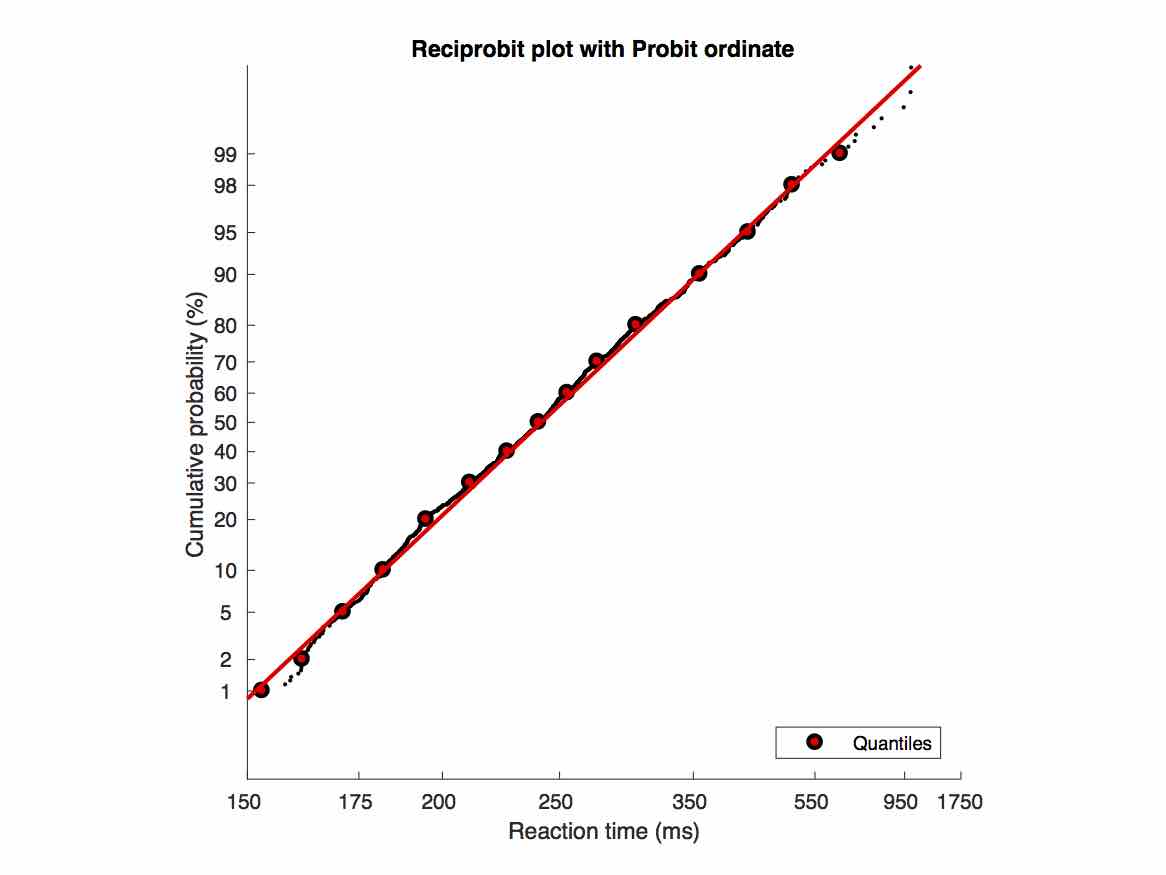}\label{Fig:Reciprob}}
\caption{Empirical CDF  of reciprocals (left panel) and reciprobit plot (right panel) of the same data presented in Fig. \ref{Fig:RT}.  In the cumulative probability of inverse reaction time, every dot is an actual individual data point, while red markers indicate quantiles at probabilities $p=0.01, 0.02, 0.05, 0.1, 0.2, 0.3, \cdots, 0.7, 0.8, 0.9, 0.95, 0.98, 0.99$. In the right panel, the same data as in the left panel plotted as a cumulative histogram but using a probit scale. Note that the latency uses a reciprocal scale resulting in a reciprobit graph}
\label{Fig:Reci}
\end{figure}

Histograms, however, are in some sense  problematic since their shape depends on the bin size and they have an effectively arbitrary vertical scale. As an alternative,  cumulative histograms or empirical Cumulative Distribution Function (ECDF) are normalized (running from $0$ to $1$ or $100\%$), can represent all data without binning (continuous), thus facilitating comparisons between data sets. The ECDF $\widetilde{F_N}(x)$, is an unbiased estimator of the CDF, the Cumulative Distribution Function $F_X(x)$,  of a RV $X$ (cfr. Box \ref{tab:Probit} for refreshing these basic concepts). In practice, given $N$ data points $x_i, i=1, \cdots, N$ the computation of $\widetilde{F_N}(x)$ boils down to two steps: (1) count the number of data less than or equal to $x$; divide the number found in (1) by the total number of data in the sample. The ECDF of inverse reaction times previously presented in Fig. \ref{Fig:RT} is illustrated in Fig. \ref{Fig:ReciCDF}.

An even  better result can be achieved by using a non-linear probit scale (an inverse error function transformation,  see Box \ref{tab:Probit}, Eq. \ref{eq:probit2}) for the vertical frequency axis, rather than  a linear scale    \cite{carpenter1995neural,Noorani2016}.  Such transformation stretches the ends of the ordinate axis in such a way as to generate a straight line if the data is indeed Gaussian, as  it can be seen in Fig. \ref{Fig:Reciprob}.  Since the latency uses a reciprocal scale, this is known as a  \textbf{reciprobit plot} (because it combines a \emph{reci}procal and a \emph{probit} scale).  Such plot provides at a glance the visual impression of the two parameters describing the normal distribution of reciprocals, the mean $\mu_r$ and the variance $\sigma^2_r$: $\mu_r$ corresponds to the median of the plot (where the line intersects $p = 50\%$), and the standard deviation $\sigma_r$ to its slope (a steep line has a small variance, a shallow one a greater one).

\begin{figure}[t]
\sidecaption[t]
\includegraphics[scale=0.19,keepaspectratio=true]{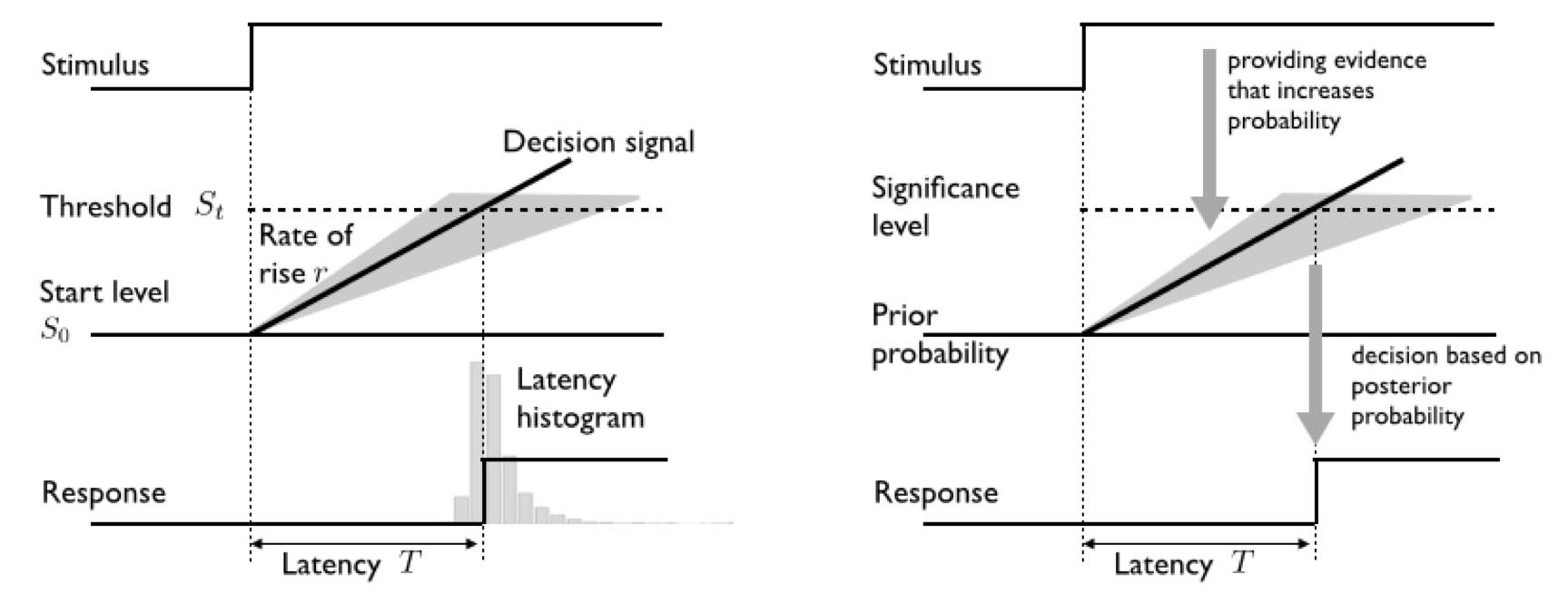}
\caption{The LATER model \cite{carpenter1995neural}. Left: the essential model. Right: LATER as an ideal Bayesian decision-maker (see Box \label{tab:Later}, for a formal discussion)}
\label{Fig:Later}
\end{figure}

The basic idea that  reciprocal latencies follow a Gaussian or normal distribution and that these reciprocals have equal variability around a mean value $\mu_r$,  as captured by the representations we have introduced above, lies at the heart of the  LATER model.
The LATER model (``Linear Approach to Threshold with Ergodic Rate'' (\cite{carpenter1995neural}, but see \cite{Noorani2016} for a recent review) is one of the simplest, and yet one of the most elegant and powerful models of reaction time distributions in decision tasks: it is assumed that  some decision signal is accumulated over time at a constant rate of rise $r$ until a threshold is reached, at which point a response is triggered (Fig. \ref{Fig:Later}, left panel). Crucially,  the rate $r$ at which such decision signal accumulates is normally distributed across trials. In mathematical terms, the model is easily specified. If 
\begin{enumerate}
\item the response is triggered when the evidence - starting from a resting level $S_0$ - reaches a threshold level $S_T$, and
\item evidence accumulates at a constant rate $r$ which, across trials, follows a normal distribution, $\mathcal{N}(\mu_r, \sigma_r^2)$,
\end{enumerate}
then the response latency $T$ is determined by:
\begin{equation}
\label{eq:later}
T=\frac{S_T - S_0}{r}.
\end{equation}
If one further assumes that both $S_0$ and  $S_T$ are relatively constant across trials, then the distribution of the times is the reciprocal of a normal distribution:
\begin{equation}
\label{eq:recinormal}
\frac{1}{T}=\mathcal{N}\left(\frac{\mu_r}{S_T - S_0},\left( \frac{\sigma_r}{S_T - S_0}\right)^2\right),
\end{equation}
\noindent which Carpenter terms the \textbf{Recinormal distribution}~ \cite{carpenter1995neural,Noorani2016}.

The stochasticity behind the LATER model can be described  at the \emph{macroscopic level}, in terms of full probability distribution functions,   by interpreting LATER  as an optimal model of Bayesian decision making, see Box \ref{tab:Later} and Fig. \ref{Fig:Later} (right).

\begin{figure}[t]
\sidecaption[t]
\includegraphics[scale=0.20,keepaspectratio=true]{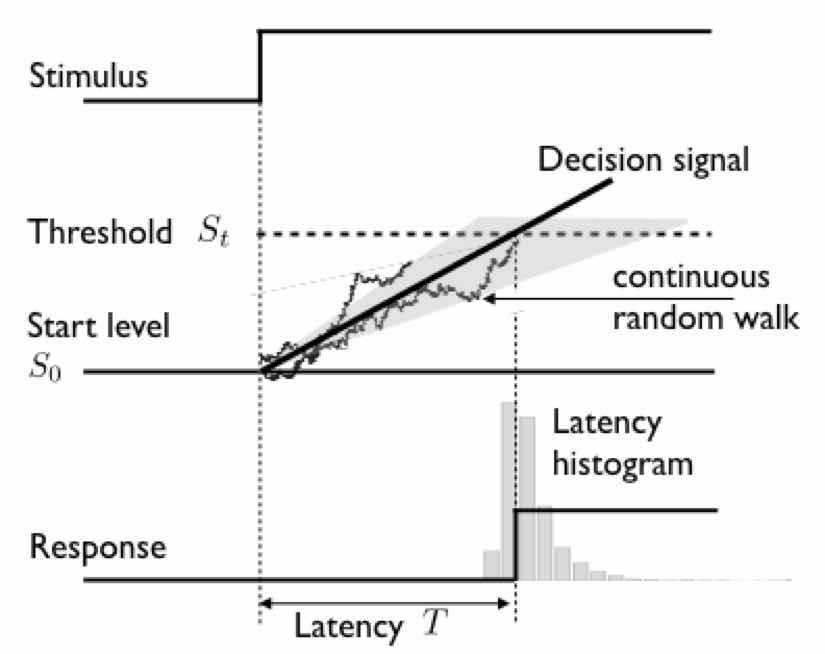}
\caption{The LATER model  as a diffusion model of decision making based on continuous random walks generated through the  Langevin-type SDE of Eq.~\ref{eq:laterSDE}. Two trajectories only are shown for graphical clarity.}
\label{Fig:LaterRW}
\end{figure}

The corresponding description at the microscopic level explains the LATER model in terms of a random walk model.
In this perspective, we  consider the function $S(t)$, as the evidence  accumulated  at time $t$ starting from prior level $S(0)=S_0$ in the process of reaching the threshold $S_T$. The random ``trajectory'' in time of $S(t)$ is that of a drifting gaussian random walker as shown in Fig. \ref{Fig:LaterRW}.  Indeed, it can be formally shown that the average accumulation of evidence $<S(t)>$  with mean rate $r$ is  described by a normal distribution centred at a mean $S_0 + r t$ having   variance at time $t$ equal to $\sigma_r^2 t^2$ \cite{moscoso2008theory}. Then the microscopic behaviour of the random walker $S(t)$ is described by a Langevin-type SDE of drift  $r$ and diffusion coefficient $\sigma_r \sqrt{2t}$:
\begin{equation}
\label{eq:laterSDE}
d S(t) = r dt + \sigma_r \sqrt{2t} dW(t),
\end{equation}
\noindent where $W(t)$ is the standard Wiener process with linear drift. Thus LATER can be considered a non-linear version of the Drift Diffusion Model \cite{ratcliff2008} of decision making. 
Note how the random walk model  provides the microscopic dynamics of how the  long-tail skewed  distribution of reaction times is originated (Fig. \ref{Fig:LaterRW}).

One  general implication of all this is that the large random variation observed in  latencies is not the result of noise at the input, as has commonly been supposed, but represents a gratuitous, ``deliberate'' randomising device (summarised by the random term $dW(t)$ in Eq. \ref{eq:laterSDE}). It has been surmised that its purpose is  to prevent the generation of undesirably stereotyped behaviour - a roulette-wheel within the brain, that one may or may not care to think of as the basis for the sense of free will and creativity \cite{Noorani2016}.

Eventually, it is worth pointing out that the predictions of latencies from random walk models and from the original LATER model are very similar, and in fact can only really be distinguished when there are a vast number of data points for latencies.

\begin{table}
\begin{svgraybox}
\caption{\textbf{LATER as a Bayesian decision-maker}}
\label{tab:Later}
Classic, \textbf{frequentistic hypothesis testing} considers the \textbf{likelihood ratio} $\frac{P(\mathcal{D} \mid \mathcal{H})}{1- P(\mathcal{D} \mid \mathcal{H})}$, where $P(\mathcal{D} \mid \mathcal{H})$ is the likelihood of the hypothesis $\mathcal{H}$  being tested, e.g., $\mathcal{H} = \textsf{``the stimulus is  present''}$,  and $\mathcal{D}$ is the evidence provided by the stimulus. In \textbf{Bayesian  hypothesis testing},  posterior and prior probabilities need to be taken into account. Denote $P(\mathcal{H})$ the prior probability of the hypothesis $\mathcal{H}$  and  $P(\mathcal{H} \mid \mathcal{D})$ the posterior probability of the hypothesis given the evidence $\mathcal{D}$.

At the macroscopic description level, LATER can be directly interpreted  as an optimal Bayesian model of hypothesis testing as sketched in the right panel of Fig. \ref{Fig:Later}.

To such end, simply rewrite  Eq. (\ref{eq:later}) as
\begin{equation}
\label{eq:later2}
S(T) = S(0) + rT
\end{equation}
\noindent Then, by using Bayes' rule, we can rewrite the LATER parameters in Eq. (\ref{eq:later2}) in terms of \textbf{log-odds} (namely, the log-ratio of the probability that an event  will happen to the probability that the event will not happen):
\begin{equation}
\label{eq:laterodds}
S(T) =  \log \frac{P(\mathcal{H} \mid \mathcal{D})}{1- P(\mathcal{H} \mid \mathcal{D})} = \overbrace{  \log \frac{P(\mathcal{H})}{1- P(\mathcal{H})} }^{S(0)} 
+ \int_0^T     \overbrace{ \log \frac{P(\mathcal{D} \mid \mathcal{H})}{1- P(\mathcal{D} \mid \mathcal{H})}}^{r} dt = S(0) + rT,
\end{equation}
\noindent where the starting level $S(0)$ denotes  the \textbf{log-prior odds} $ \log \frac{P(\mathcal{H})}{1- P(\mathcal{H})} $  and the rate of information intake $r$ is the \textbf{log Bayes factor} of the stimulus.

Thus, the accumulated evidence $S(T)$ is an optimal estimate of the \emph{log-posterior odds} $\log \frac{P(\mathcal{H} \mid \mathcal{D})}{1- P(\mathcal{H} \mid \mathcal{D})}$ (a.k.a., the logit of the posterior probability) of the hypothesis being tested
\end{svgraybox} 
\end{table}

\subsection{Walking on the safe side: the Central Limit Theorem}

Once more, consider   Eq. (\ref{eq:windisc}), namely the discretised version of the Wiener process. We rewrite here - with step index  $n = i+1$ -  for the reader's convenience:
\begin{equation}
 x_{n} = x_{n-1} +   \sqrt{2D} \Delta W_{n-1}. 
  \label{eq:windisc2}
\end{equation}

It is easy to see that by repeated substitution and by assuming the initial condition $x_0=0$, after $n$ integration steps: 
\begin{equation}
\begin{split}
x_{n} &= x_{n-1} +   \sqrt{2D} \Delta W_{n-1}  \\
&= x_{n-2} + \sqrt{2D}\Delta W_{n-2}+ \sqrt{2D}\Delta W_{n-1}  \\
&= x_{n-3} + \sqrt{2D}\Delta W_{n-3} + \sqrt{2D}\Delta W_{n-2}+ \sqrt{2D}\Delta W_{n-1}  \\
&=\cdots \\
&=  \sum_{i=0}^{n-1} \sqrt{2D} \Delta W_{i} 
\end{split}
\label{eq:windisc3}
\end{equation}
Note that by definition of Brownian motion, the Wiener increment  $\Delta W_{i}$ is zero-mean Gaussian distributed with variance $\sigma^2 = \Delta t_{i}$, that is $\Delta W_{i} \sim \mathcal{N}(0, \Delta t_{i})$. By using elementary properties of Gaussian RVs, the increment $\Delta W_{i}$  can be computed by i) sampling  $\xi_{i} \sim \mathcal{N}(0, 1) $ and   ii) multiplying  $\xi_{i}$  by the standard deviation $\sigma = \sqrt{\Delta t_{i}}$, namely:
$$\Delta W_{i} =  \sqrt{\Delta t_{i}}\xi_{i}.$$
Then, Eq. (\ref{eq:windisc3}) can be simply written as
\begin{equation}
x_{n} =  \sum_{i=0}^{n-1} \sqrt{2D\Delta t_{i}} \xi_{i}. 
\label{eq:windisc4}
\end{equation}
This result is nothing but the answer to  our fundamental question, ``What is the probability $P(x_n)$ of being at point $x_n$ after $n$ steps?'' As before, the final position $x_{n}$ is a zero-mean Gaussian random variable with  
variance $\sigma^2 =  2D t_{n}$ (assuming initial time $t_{0}=0$). Thus, $P(x_n) =  \mathcal{N}(0, 2D t_{n})$.

In short, by exploiting the simple property that the sum of independent Gaussian RVs (here the particle displacements) is a Gaussian RV, we have derived (in discrete form) the result discussed in Box \ref{tab:hall}: the probability $P(\mathbf{x},t)$ of a Brownian  particle being at location $\mathbf{x}$ at time $t$ is a Gaussian distribution $\mathcal{N}(0, 2D t)$. Indeed, the Monte Carlo simulation presented in Figure~\ref{Fig:bw20}, which illustrates both the microscopic trajectories  of  a large number of Brownian particles and the macroscopic evolution of  the solution of the diffusion equation, is a procedure based on Eq.~(\ref{eq:windisc4}): 1) find the cumulative sums up to $x_{n}$ for a large number of walkers, and 2) compute their histogram to approximate  $P(\mathbf{x},t)$.

Though intuitive, this view is an overly simplified picture of the whole story. To see why, consider a very simple kind of discrete random walker, an instantiation  of the basic model presented at the very beginning in Box~\ref{tab:simple}, Eq.~(\ref{eq:simpleRW}).  

At fixed time intervals $\Delta t$ the walker either jumps to the right with probability $p$ or to the left with probability $q=1-p$. The intuition is that for $p=\frac{1}{2}$, at each step the walker is tossing a fair coin to make the left/right decision.

The microscopic behaviour of the walker is governed by the following discrete equation:
\begin{equation}
 x_{n} = x_{n-1} + (2 \xi_{n}-1) , \textsf{      } n= 1,2, \cdots,
\label{eq:simpleBinRW}
\end{equation}
\noindent where $x_{0}$ is the initial position, e.g. $x_{0}=0$.
The discrete RV $\xi_{n} \in \{0, 1\}$   stands for the coin toss at step $n$, and it is sampled from the Bernoulli distribution, namely $\xi_{n} \sim Bern(p) = p^{\xi_{n}} q^{1-\xi_{n}}$. In other terms $\{ \xi_{n} \}_{n\geq 1}$ is a Bernoulli process.  

\begin{figure}[t]
\sidecaption[t]
\centering
     \subfigure[Microscopic behaviour]{\includegraphics[width=0.51\textwidth]{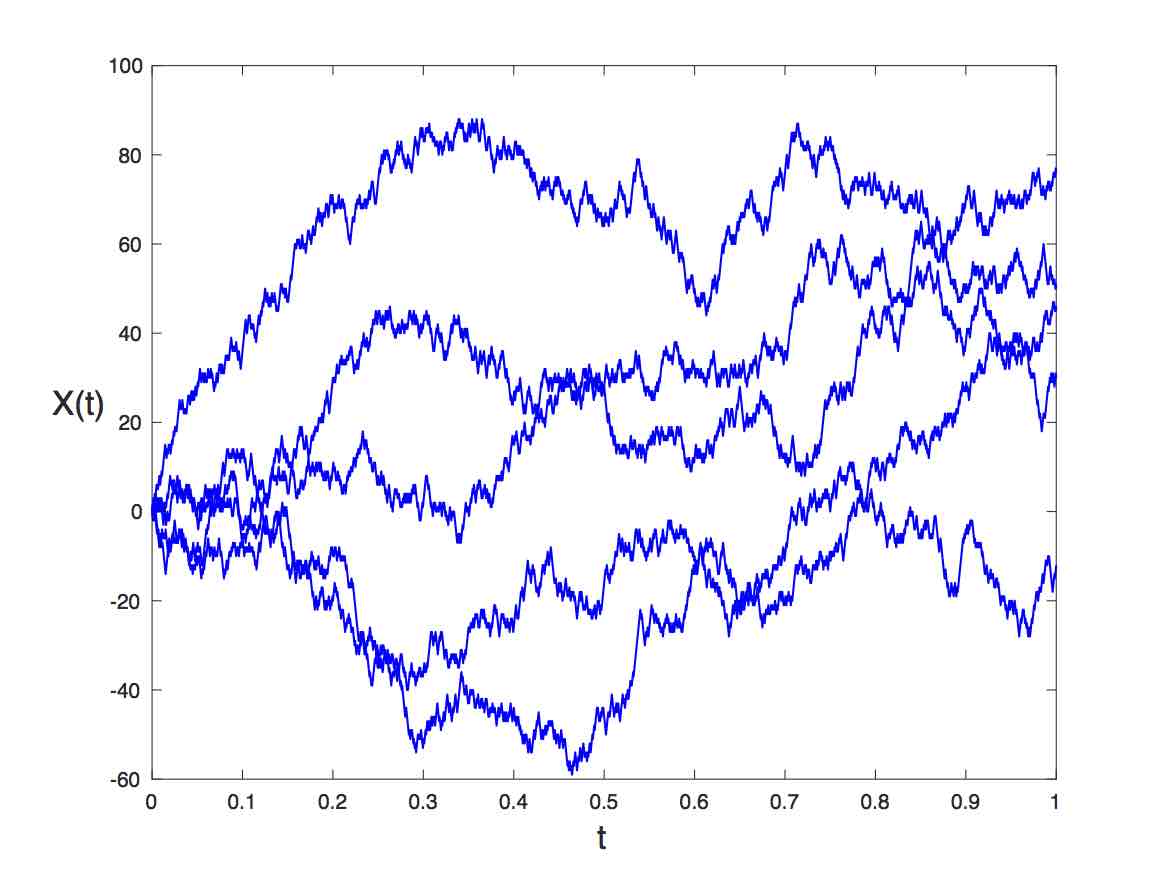}\label{Fig:microRW}}
   \subfigure[Macroscopic behaviour]{\includegraphics[width=0.51\textwidth]{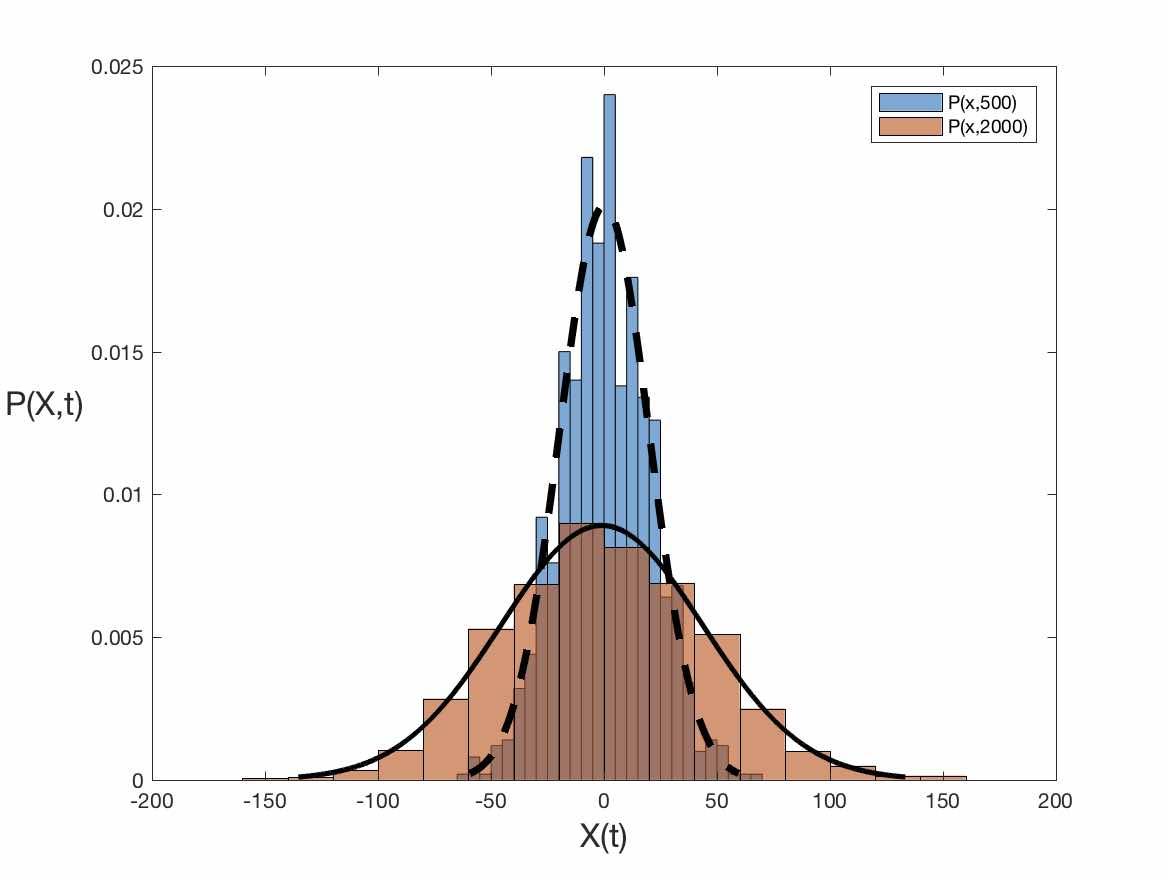}\label{Fig:macroRW}}
\caption{Microscopic and macroscopic behaviour of the simple random walker. Panel~\ref{Fig:microRW} shows five  among $1000$ trajectories simulated  via  Eq.~(\ref{eq:simpleBinRW}), for $2000$ iteration steps. Panel~\ref{Fig:macroRW}: the macroscopic behaviour of the many-walker system captured  by computing the empirical distributions $P(x_{n})$ at steps $n=500$ and $n=2000$. Both distributions are overlaid with  the Gaussian pdfs fitted at the same time steps (dotted  and continuous line, respectively)}. 
\label{Fig:SimpleBinRW}
\end{figure}

The dynamics behind Eq. (\ref{eq:simpleBinRW}) is simple:  at each step, if $\xi_{n}=0$, then Eq.(\ref{eq:simpleBinRW}) gives $x_{n} = x_{n-1} + 1$ (right step); otherwise, $x_{n} = x_{n-1} - 1$ (left step).    Figure~\ref{Fig:microRW} shows five trajectories of the simple walker  drawn by iterating Eq. (\ref{eq:simpleBinRW}) $2000$ steps. For such a long sequence,  trajectories look similar to those produced by a Gaussian random walker (Brownian particle). This becomes evident by inspecting the macroscopic behaviour as illustrated in Figure~\ref{Fig:macroRW}.  The collective behaviour of $1000$ walkers at a step $n$ is obtained by computing for each walker the sum of a number of i.i.d displacements/increments up to step $n$, as in Eq. (\ref{eq:windisc4}), then the (empirical) distribution $P(x_{n})$ is computed.  The figure shows two ``snapshots'' of the evolution of $P(x_{n})$ at  $n=500$ and $n=2000$: it is easy to see that its behaviour is that of a Gaussian distribution spreading in time, much like  the Brownian diffusion behaviour shown in Figure~\ref{Fig:bw20}. To better visualise such trend, we also plot the Gaussian distributions fitted at the same time steps.

This result is not surprising: the Gaussian or Normal  distribution correctly describes an amazing variety of phenomena. Most important for our purposes, the bell-shaped curve appears in nature ubiquitously due to the wide applicability of the \textbf{Central Limit Theorem} (CLT). 

The key idea behind the classical CLT (but see Box~\ref{tab:CLT} for more details) is the following. If we sum a large number $n$ of  RVs $X_i$ that are
\begin{enumerate}
\item \emph{statistically independent} and 
\item \emph{identically distributed} (i.i.d),  and that  
\item have a \emph{finite variance}, 
\end{enumerate}
the distribution $P(S_n)$ for the sum $S_n = \sum_{i=1}^{n} X_i $ converges to a Gaussian distribution with mean $n\mu$ and variance $n \sigma^2$, namely, $P(S_n) \approx \mathcal{N}(n\mu,n \sigma^2)$. 

%
%

The CLT theorem generalises the thorough intuition we initially gained by exploiting the Gaussian nature of a Brownian random walk.
If a trajectory consists of a set of independent displacements with finite variance,  then the total distance covered (the sum of  these displacements) follows a Gaussian law whose variance  is proportional to the number of these displacements (and, in consequence, proportional to time).

If the   three conditions required by the CLT are fulfilled, then the MSD will behave at large times like $\langle x(t) ^2 \rangle \propto t $ (no matter how complicated the motion pattern is) in the limit $t \to  \infty$.
Such scaling law  $\langle x(t) ^2 \rangle \propto t $, as previously stated,  is characteristic of the diffusion equation, but  it also arises asymptotically in many other cases. This is not a coincidence but a direct consequence of the baseline CLT.  

A tenet of this Chapter, following Paul and Baschnagel \cite{paul2013stochastic}, is exploiting the  classical CLT as the pivotal concept to  distinguish stochastic behaviours that occur within its limits and behaviours that violate its limits. Examples of the latter are considered in the following Section.


\begin{table}
\begin{svgraybox}
\caption{\textbf{The Gaussian bell tolls for thee: The Central Limit Theorem}}
\label{tab:CLT}
In a nutshell, the simplest form of the CLT   states that  if one sums together many \textbf{independent  and identically distributed} (i.i.d) random variables with \textbf{finite variance}, then the probability density of the sum will be close to a Gaussian. 

Formally, consider $n$ i.i.d RVs  $X_1, X_2, \cdots, X_n$, with mean
$$ \langle X_i \rangle = \mu$$
and finite variance
$$ Var(X_i)  = \sigma^2 < \infty$$
\noindent for all $i=1,\cdots,n$.
Consider then the RV $S_n$, which is the sum 
\begin{equation}
S_n =  X_1 + X_2+ \cdots +X_n.
\end{equation}
Then, as $n \to \infty$, the distribution of the normalised sum $\frac{S_n - n \mu}{\sigma \sqrt{n}}$ converges  to the standard Gaussian distribution. More precisely, for $-\infty < a < \infty$
\begin{equation}
\lim_{n \to \infty} P\left(\frac{S_n - n \mu}{\sigma \sqrt{n}} \leq a \right) =  \frac{1}{\sqrt{2\pi}} \int_{-\infty} ^{a} e^{-\frac{x^2}{2}} dx,
\label{eq:CLT}
\end{equation}
\noindent where $\frac{1}{\sqrt{2\pi}} e^{-\frac{x^2}{2}} = \mathcal{N}(0,1)$ is the standard Normal PDF with zero mean and unit variance. 

The two most common manifestations of the CLT are the following. 
First, as $n \to \infty$, the sum $S_n$ ``tends'' to a Gaussian variable with distribution $\mathcal{N}(n \mu,n \sigma^2)$. Second,  the same holds for $\frac{S_n}{n}$, where 
$$\frac{S_n}{n}= \frac{X_1 + X_2+ \cdots +X_n}{n} = \frac{1}{n} \sum_{i=1}^{n} X_i,$$
which is nothing but the empirical mean or sample average. Namely, $\frac{S_n}{n}$ ``tends'' to the distribution $\mathcal{N}\left(\mu, \frac{\sigma^2}{n}\right)$. 

If these issues might seem too abstract or of limited interest for you, recall that you have been trained in your daily lab work to collect results from several repeated  measurements, to take mean values and to eventually estimate confidence intervals by using tables from the Gaussian distribution. Your data  do not necessarily  follow a Gaussian distribution, yet you are not worried about. Any problem where the final output results from the average over a set of identical and independent variables with finite variance leads to a Gaussian PDF. More often than not, the CLT is your unconscious safety belt.

This is the ``baseline'' or classical  CLT (a.k.a. the Lindeberg-L\'evy CLT).  However, it is worth mentioning that the CLT comes in various stronger and weaker forms \cite{schuster2016stochasticity}.
Under certain conditions,  the RVs $X_i$ are not required  to be identically distributed. In such case we have different variances, 
$Var(X_i)  = \sigma_i^2$ but the CLT still holds if  the contribution of any individual random variable  to the overall variance $\sigma_{n}^{2} = \sum_{i=1}^{n} \sigma_{i}^{2}$ is arbitrarily small for $n \to \infty$ (Lindeberg's condition). 

What happens if the $X_i$ are originated from a distribution whose variance is not finite? It is indeed the case that there are experimental time series (e.g., saccades in free viewing, go back to Fig. (\ref{Fig:ampl})),  for which the distribution of independent increments exhibit much fatter tails than the normal, and sometimes considerable skewness as well.
In Figure~\ref{Fig:FigSample} we have seen one such example: the Cauchy distribution with a power-law tail (see Box \ref{tab:pow}).  The Cauchy distribution does not have a finite variance.   
The Generalized CLT - due to Gnedenko, Kolmogorov and L\'evy - states that the sum of a number of RV with a power-law tail  will tend to a non Gaussian $\alpha$-stable distribution (stable L\'evy noise) as $n \to \infty$ \cite{paul1954theorie,kolmogorov1954limit}. We will see in Section~\ref{sec:levy}, that such kind of  heavy-tailed distributions play an important role in the modelling of saccadic eye movements.

%
%

\end{svgraybox} 
\end{table}


\section{Walking on the wild side: eye movements beyond the CLT}
\label{sec:wild}

In spite of the nice behaviour of RWs patrolled by the CLT, when dealing with eye movements most interesting cases happen when the CLT is violated:
\begin{description}
\item[(i) Violation of independency:] Long-range correlations are present, so once the random walker decides moving in one direction it keeps on doing the same for a long time (this will lead to superdiffusion) or, alternatively, once it stops it remains resting for an arbitrarily long time (then subdiffusion will emerge)
\item[(ii) Violation of identity:] motion consists of non-identical displacements that become gradually shorter (subdiffusion) or longer (superdiffusion) probably because of external constraints, and non stationarity arises.
\item[ (iii) Violation of moment finiteness:] The displacements forming the trajectory can be fitted to a PDF with non-finite mean or variance, so as a result arbitrarily large displacements are likely with a certain frequency (long tail distributions)
\end{description}

When one of the three conditions is violated then the process is said to exhibit \textbf{anomalous diffusion}. A simple way to define anomalous diffusion or an anomalous random walk is when $ \langle  x(t) ^2 \rangle $ does not increase linearly with time.
In such case,  the corresponding MSD shows a power-law behaviour
\begin{equation}
MSD =  kt^{\gamma},
\label{eq:linMSD}
\end{equation}
\noindent with $\gamma \neq 1$


\subsection{A first violation: i.i.d denied}


 One way to anomalous diffusion is by introducing ``memory'' effects in the process. This gives rise to \textbf{ long-range} power--law autocorrelations in the underlying noise that drives the random walk. Long-range memory effects violate the condition of independent random variables.
 
\begin{figure}[t]
\sidecaption[t]
\includegraphics[scale=0.19,keepaspectratio=true]{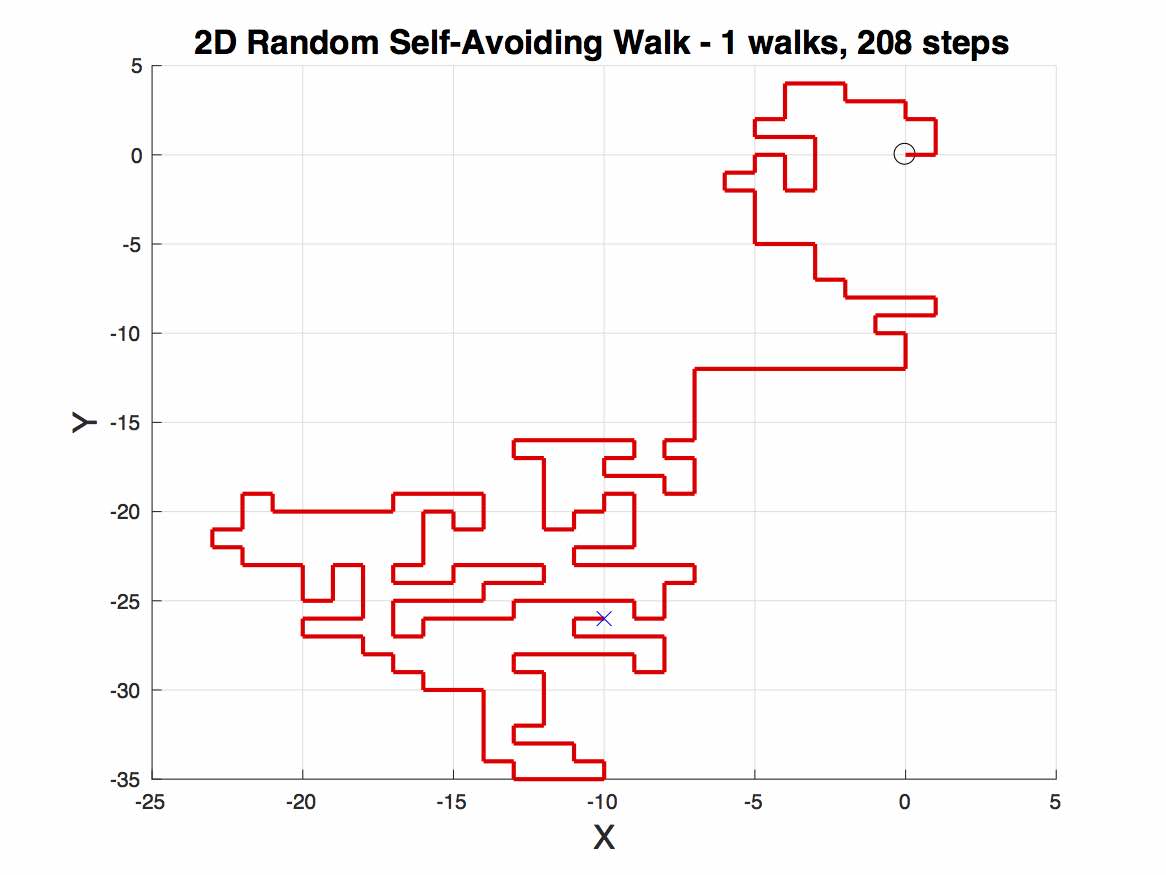}
\caption{A single trajectory of a 2-dimensional Self-Avoiding Walk (SAW). Whereas a random walk can intersect its path arbitrarily often,  a SAW may occupy each site only once. This leads to an increase in the distance between the start and the end,  compared to the classic random walk. Self-avoidance has been proposed as the key mechanism driving the drift observed in fixation tasks \cite{engbert2011integrated}.}.
\label{Fig:SAW}
\end{figure}

 One intuitive example is the  \textbf{self-avoiding walk} (SAW). In this process the random walker has to keep track of the whole history of his path while he moves along, since he is not allowed to visit a site twice. Sites already visited therefore act like a repulsive potential for the continuation of the walk; Fig. \ref{Fig:SAW} illustrates a $2$-dimensional SAW. Intuitively, this ``long--range repulsive interaction'' along the path should make the overall displacement grow stronger with increasing $t$ than in the case of the Bm \cite{paul2013stochastic}.
\textbf{Fractional Brownian motion} (fBm) is another example. It was introduced  by Mandelbrot and van Ness \cite{mandelbrot1968fractional}  to account for processes obeying a scaling law of the functional form
$MSD =  kt^{2H}$, with $0<H<1,  \; \; H \neq \frac{1}{2}$, where $H=\frac{1}{2}$ is the special case of Bm (cfr. Box \ref{tab:Hurst}).

It is described by the time-varying pdf
\begin{equation}
P(x,t)  = \frac{1}{\sqrt{4 \pi D t^{2H}}} \exp \left[ - \frac{x^2}{4  D t^{2H}}    \right].
\end{equation}
  

fBm has been used  as a mathematical reference for random-walk analysis  of  fixational eye movements (FEMs) and for studying their correlations  across time (e.g., \cite{engbert2011integrated}). Also, properties of persistence / antipersistence have been exploited, for instance, in analysing optokinetic nystagmus (OKN,  \cite{trillenberg2001random}).

\subsubsection{Case study: random walk analysis of microsaccades}
\label{sec:microsacc}


In a number of studies, Engbert and colleagues, e.g., \cite{engbert2006microsaccades,engbert2011integrated,engbert2012bayesianHurst} have shown that a  typical trajectory generated by the eyes during FEMs  exhibits clear features of a random walk.

For instance, on a short time scale ($2$ to $20$ ms), the RW is persistent, whereas on a long time scale ($100$ to $800$ ms) it exhibits anti-persistent behaviour. Thus, they observed a time-scale separation with two qualitatively different types of motion.
On the short time scale,  
drift produces persistence and this tendency is increased by the presence of microsaccades. On the long time scale, the anti--persistent behaviour is specifically created by microsaccades. 
Since the persistent behaviour on the short time scale helps to prevent perceptual fading and the anti-persistent behaviour on the long time scale is error-correcting and prevents loss of fixation, they  concluded that microsaccade are optimal motor acts to contribute to visual perception

A more recent model of FEMs has also incorporated self-avoidance as the key mechanism driving drifts observed in fixation tasks \cite{engbert2011integrated}. The Self-Avoiding walk model encodes history by treating space as a lattice and recording the number of visits to each site: the SAW proceeds by choosing the least-visited neighbour at each step. A SAW simulation is provided in Figure~\ref{Fig:SAW}. The model proposed in \cite{engbert2011integrated} also includes a confining potential to keep the random walk near the origin, which is needed for the long-time subdiffusive nature of fixation tasks, as well as a mechanism for triggering microsaccades when occupying highly-visited sites.

\subsubsection{Case study: optokinetic nystagmus}
OKN is a reflexive eye movement with target-following slow phases (SP) alternating with oppositely directed fast phases (FP). For a quick grasp of this kind of eye movement, a video is worth a thousand words, e.g., visit \url{https://www.youtube.com/watch?v=KSJksSA6Q-A}.
Trillenberg \emph{et al} \cite{trillenberg2001random} by measuring FP beginning and ending positions, amplitudes, and intervals and SP amplitudes and velocities,  tried to  predict future values of each parameter on the basis of past values, using state-space representation of the sequence (time-delay embedding) and local second-order approximation of trajectories. Since predictability is an indication of determinism, this approach allows  to investigate the relative contributions of random and deterministic dynamics in OKN. FP beginning and ending positions showed good predictability, but SP velocity was less predictable. FP and SP amplitudes and FP intervals had little or no predictability. FP beginnings and endings were as predictable as randomised versions that retain linear auto-correlation; this is typical of random walks. Predictability of FP intervals did not change under random rearrangement, which also is a characteristic of a random process. They concluded that there is undoubtedly a gross level of deterministic behaviour in OKN. Yet within this range, there is apparently significant random behaviour, with a small amount of predictability. The random behaviour has overlaid on it a form of long-term correlation in the form of anti-persistence. This mixture of dynamics is intriguing and provides a challenge for mathematical modelling efforts, though the physiological meaning of these dynamics is open to conjecture.

\begin{table}
\begin{svgraybox}
\caption{\textbf{Power--law distribution}}
\label{tab:pow}
Many empirical quantities cluster around a typical value: speeds of cars on a highway, the temperature in Freiburg at noon in February, etc. Distributions of these quantities place a negligible amount of probability far from the typical value, making the typical value representative of most observations. In short, the underlying processes that generate these distributions fall into the general class well-described by the CLT.

Not all distributions fit this pattern, however, and in some cases the deviation is not a defect or problem, but rather an indication of interesting underlying complexity in the generating process. Complex social, biological and technological systems give rise to countless example of ``non-normal'' and \textbf{heavy-tailed} distributions.
A power-law distribution is one such  kind of probability distribution (see Newman \cite{newman2005power} for a nice review). When the probability of measuring a particular value of some quantity varies inversely as a power of that value, the quantity is said to follow a power law, also known  as the \textbf{Pareto distribution}. There are several ways to define them mathematically, one way, for a continuous random variable is the following~\cite{newman2005power}: 
\begin{equation}
P(x) = C x ^{-\mu},  x \geq x_{min},
\label{eq:powlawdef}
\end{equation}
\noindent where  $C= (\mu -1) x_{min}^{\mu-1}$. Note that this
expression only makes sense for $\mu > 1$, which is indeed a requirement for a power--law form to normalize.

Power--law distributions have many interesting mathematical properties. Many of these come from the extreme right-skewness of the distributions and the fact that only the first $(\mu-1)$ moments of a power-law distribution exist; all the rest are infinite. In general, the $k$-th moment is defined as
\begin{equation}
\langle x^k \rangle = \int_{x_{min}}^{\infty}  x^{k}P(x) dx 
= C \int_{x_{min}}^{\infty} x^{-\mu+ k} dx = x^{k}_{min} \left(\frac{\mu -1 }{\mu -1 -k}\right), 
\label{eq:powlawmoment}
\end{equation}
\noindent with  $\mu > k+1$. Thus, when $1 < \mu < 2$, the first moment (the mean or average) is infinite, along with all the higher moments. When $2 < \mu < 3$, the first moment is finite, but the second (the variance) and higher moments are infinite! In contrast, all the moments of the vast majority of other pdfs are finite.

Another interesting property of power-law distributions is \textbf{scale invariance}. If we compare the densities at $P(x)$ and at some $P(c x)$, where $c $ is some constant, they are always proportional, i.e. $P(cx) \propto P(x)$. This behaviour shows that the relative likelihood between small and large events is the same, no matter what choice of ``small'' we make. That is, the density ``scales.'' It is easy to see that if  we take the logarithm of both sides of Eq. (\ref{eq:powlawdef}):
\begin{equation}
\ln P(x) = \ln C -\mu \ln x, 
\label{eq:logpowlawdef}
\end{equation}
\noindent the rescaling $x \rightarrow cx$ simply shifts the power-law up or down on a logarithmic scale. The result shows another  well-known property of a power-law distribution: it appears as a straight line on a $\log-\log$ plot, as opposed to the strongly curved behaviour of  an exponential distribution (see Fig. \ref{Fig:powlaw}).


Inspiring analyses of eye movements and visual search in terms of power--law behaviour and power spectra have been conducted by Deborah Aks \emph{et al.} \cite{aks2002memory} suggesting  that our oculomotor system may produce a complex and self-organising search pattern providing maximum coverage with minimal effort.

\end{svgraybox} 
\end{table}

\begin{figure}[t]
\sidecaption[t]
\includegraphics[scale=0.17,keepaspectratio=true]{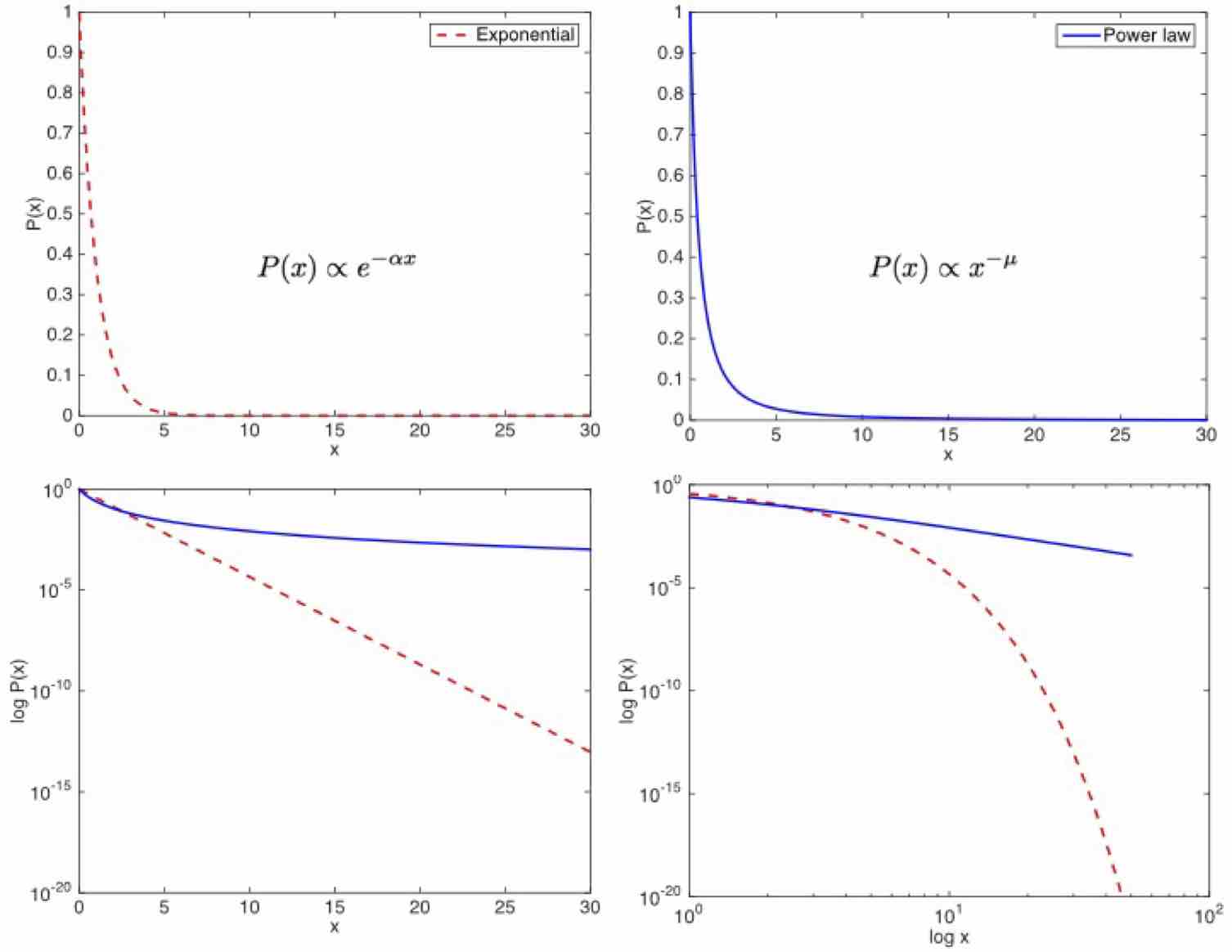}
\caption{Exponential distribution  $P(x) \propto e ^{-\alpha x}$ (top-left) vs. power--law distribution $P(x) \propto x ^{-\mu}$ (top-right): pdf shapes look apparently similar. However,  the exponential pdf is represented as a straight line on a semilog graph of $\log P(x)$ versus $x$ (bottom-left), whilst  the power-law shapes as a straight line on a log-log graph (bottom-right), a signature of the heavy-tail behaviour}.
\label{Fig:powlaw}
\end{figure}

\subsection{A second violation: loosing your moments}
\label{sec:levy}

Even in the absence of correlations, a mechanism for disrupting convergence to Brownian motion in the long time limit  is using power--law tailed distributions in the random walk steps (i.e., power--law distributed noise rather than Wiener or similar noise). \textbf{L\'evy flights} (LFs) are one such mechanism. LFs are stochastic processes characterised by the occurrence of extremely long jumps, so that their trajectories are not continuous anymore. The length of these jumps is distributed according to a L\'evy stable statistics with a power--law tail and divergence of the second moment. This peculiar property strongly contradicts the ordinary Bm, for which all the moments of the particle coordinate are finite.

\begin{figure}[t]
\sidecaption[t]
\includegraphics[scale=0.16,keepaspectratio=true]{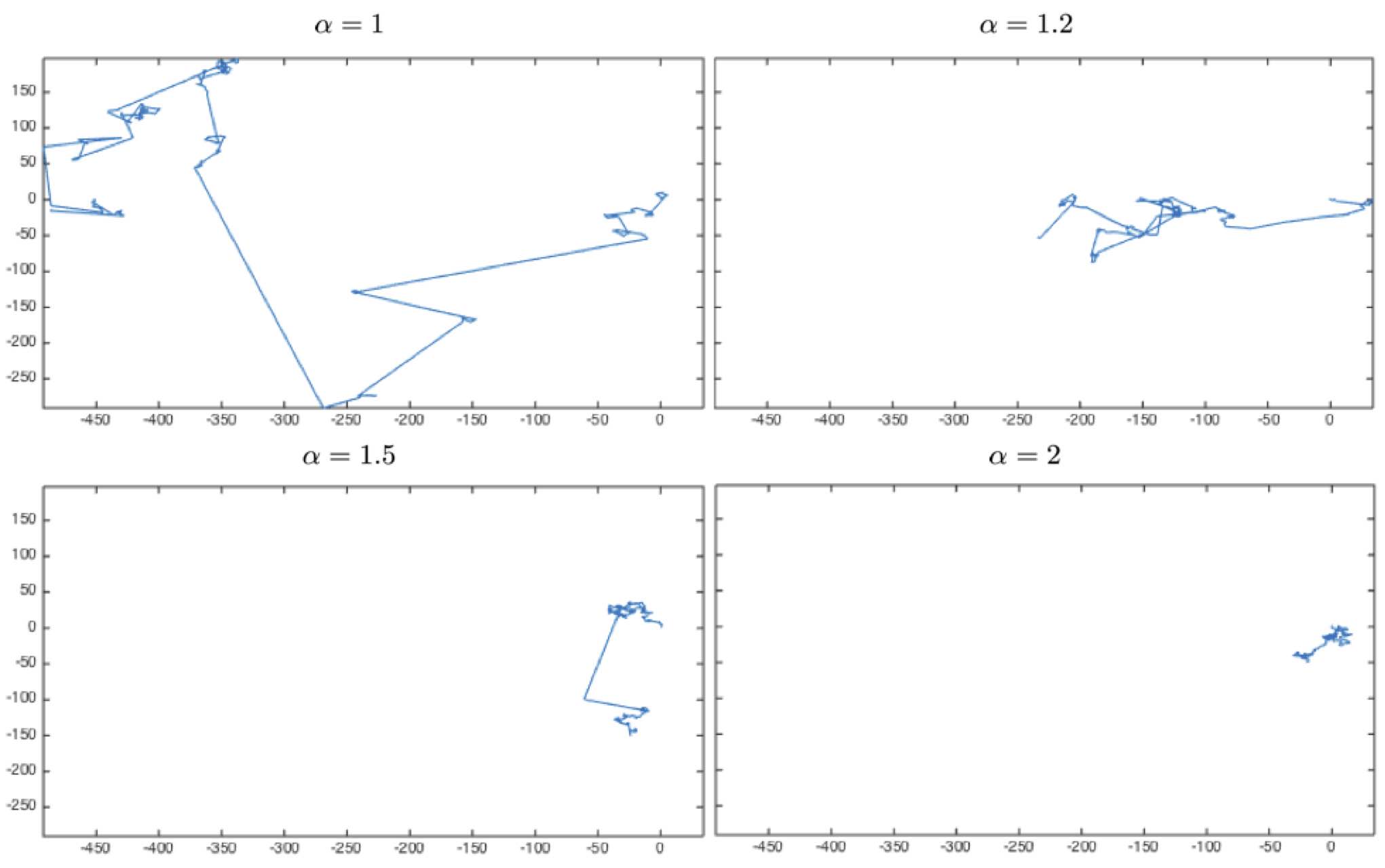}
\caption{Different 2-dimensional $\alpha$-stable  motions obtained by sampling the ``noise'' component $\boldsymbol \xi \sim f( \boldsymbol \xi; \alpha,  \beta, \gamma, \delta)$ in Eq. (\ref{eq:LevyND}) for different values of the characteristic index parameter $\alpha$. The plots shown in the four panels -- left to right, top to bottom --, have been generated via $\alpha=1, \alpha=1.2, \alpha=1.5, \alpha=2$, respectively. The same number of discrete steps ($\#500$) has been fixed for all the examples, but note how the ``scale'' of the exploration restricts as $\alpha \rightarrow 2$, eventually reaching  the limit case $\alpha=2$ where classic Bm is generated (bottom right plot).}
\label{Fig:fly}
\end{figure}

For a random walker who takes steps of size  $l$ according to a probability density function  
\begin{equation}
P(l) \approx l ^{-\mu},
\label{eq:powlaw}
\end{equation}
the resulting type of diffusion depends on the value of $\mu$. In particular:
\begin{description}
\item[a) $\mu > 3$:] the CLT guarantees convergence to normal diffusion and Brownian regime holds;
\item[b) $\mu \to 1$:] the ballistic motion limit is reached;
\item[c) $1<\mu < 3$:] superdiffusive behaviour occurs.
\end{description}

LFs  arise in the super diffusive regime, when the jump size distribution has a power--law tail with $\mu < 3$.  As discussed in Box \ref{tab:pow}, for such values of the power--law exponent, the RVs can have diverging variance. The necessary and sufficient conditions of the classical CLT do not hold in this case. L\'evy flight patterns comprise sequences of randomly orientated straight-line movements. Frequently occurring but relatively short straight-line movement randomly alternate with more occasionally occurring longer movements, which in turn are punctuated by even rarer, even longer movements, and so on with this pattern repeated at all scales. Some examples of LF patterns are provided in Figure~\ref{Fig:fly}. As a consequence, the straight-line movements have no characteristic scale, and LFs are said to be \textbf{scale-free}.

At the microscopic level, the simulation of individual LF trajectories   do not require complex calculations to execute.   They  are a  Markovian process and can be easily obtained    from  Eq. (\ref{eq:LangND}), by setting $\mathbf{A}=0$
\begin{equation}
d\mathbf{x}(t) =   {\mathbf  B}({\mathbf x}, t) \boldsymbol \xi(t) dt = {\mathbf  B}({\mathbf x}, t) d {\mathbf L_{\alpha}}(t)
\label{eq:LevyND}
\end{equation}

The  form of this equation is that of the Wiener process,  however in this case the stochastic increment $d {\mathbf L_{\alpha}}(t)= \boldsymbol \xi(t) dt$ is sampled from an $\alpha$-\textbf{stable distribution} $f( \boldsymbol \xi; \alpha,  \beta, \gamma, \delta)$ (cfr. Box \ref{tab:alpha}):
\begin{equation}
 \boldsymbol \xi(t) \sim f( \boldsymbol \xi; \alpha,  \beta, \gamma, \delta).
\label{eq:sampleLevyND}
\end{equation}
\noindent In other terms, $d {\mathbf L_{\alpha}}(t)$ in the context of Eq.~(\ref{eq:LevyND}) represents an infinitesimal L\'evy motion.

The macroscopic description of the pdfs for particles undergoing a L\'evy flight can be modeled using a generalised version of the Fokker-Planck equation. The equation requires the use of fractional derivatives and we will not discuss it  here since really beyond the scope of an introductory chapter.

By discretising and iterating Eq. (\ref{eq:LevyND}), over a large number of trials a L\'evy  flight will be distributed much farther from its starting position than a Brownian random walk of the same length (see again  Figure~\ref{Fig:fly}). Indeed, the $MSD$ of a Brownian walker has a linear dependence on time whereas that of a L\'evy  flier grows faster and depends on time raised to some power $>1$.  This result gives a precise meaning to their characterisation as  ``super-diffusive.''  The probability density function for the position of the walker converges to a L\'evy  $\alpha$-stable distribution with L\'evy index $\alpha = \mu -1$, with $0<\alpha \leq 2$ (with the special case $\alpha = 2$ corresponding to normal diffusion). The L\'evy index is an important feature. For instance, it has been shown that it is suitable to characterise the variation or activity of fMRI signals of different networks in the brain  under resting state condition. Visual and salience networks seem to present a definite L\'evy motion-like behaviour of activity, whereas areas from the cerebellum exhibit Brownian motion~\cite{costa2016foraging}.

The Hurst exponent $H$, the characteristic index $\alpha$, and the power--law exponent  $\mu$  are related as follows:
\begin{equation}
H = \frac{1}{\alpha} = \frac{1}{\mu -1 }.
\label{eq:exponents}
\end{equation}

Thus, rephrasing the conditions that have been discussed for the $\mu$ exponent,  for $\alpha < 2$  one cannot define the MSD  because it diverges. Instead, one can study moments of order lower than $\alpha $ because they do not diverge.  Nevertheless, one can define some ``empirical'' width, such as half widths at half maximum, and show that a \emph{pseudo}-MSD grows as $ \approx t^{\frac{1}{\alpha}}$ for L\'evy flights.



\begin{table}
\begin{svgraybox}
\caption{\textbf{Stable distributions}}
\label{tab:alpha}
The family of $\alpha$-stable distributions \cite{gnedenko1954limit} form a four-parameter family of continuous probability densities, say $f( \boldsymbol \xi; \alpha,  \beta, \gamma, \delta)$.  The parameters are  the skewness $ \beta $ (measure of asymmetry), the scale $\gamma$ (width of the distribution) and the location   $\delta$ and, most important, the \textbf{characteristic exponent} $\alpha$, or index of the distribution that specifies the asymptotic behaviour of the distribution.
The relevance of $\alpha$ derives from the fact that  the pdf of  jump length scales, asymptotically, 
  as $l^{-1-\alpha}$.  Thus, relatively long jumps are more likely when $\alpha$ is small. By sampling  $\boldsymbol x \sim f( \boldsymbol x; \alpha,  \beta, \gamma, \delta)$,
 for $\alpha \geq 2$ the usual Bm occurs; if $\alpha < 2$ ,  the distribution of  lengths is ``broad''
and the so called Le\'vy flights take place.

One example of  $\alpha$-stable motions generated for varying the  $\alpha$ index is illustrated in Fig. \ref{Fig:fly}.

A  random variable $X$  is said to have a stable distribution if:  (1) the  parameters of its probability density function   $f(x; \alpha,\beta,\gamma,\delta)$ are  in the following ranges $\alpha \in \left(0; 2\right]$,  $ \beta \in \left[-1; 1\right]$, $\gamma > 0$,  $\delta \in \mathbb{R}$ and (2) if its characteristic function $ E\left[\exp(i t x)\right]=\int_{\mathbb{R}} \exp(i t x) dF(x)$, $F$ being the CDF, can be written as
\begin{equation}
E\left[\exp(i t x)\right]=
\begin{cases} 
\exp \left[-|\gamma t|^{\alpha} \left(1-i\beta \frac{t}{|t|})\tan(\frac{\pi\alpha}{2}\right)+i\delta t \right] \\ 
\exp \left[-|\gamma t| \left(1+i \beta  \frac{2}{\pi} \frac{t}{|t|}  \ln |t| \right)+i \delta t\right] 
\end{cases} \nonumber
\end{equation}
\noindent The first expression holds if $\alpha \neq 1$, the second if $\alpha =1$.

Special cases of stable distributions  whose pdf can be written analytically, are given for $\alpha=2$,  the \textbf{Normal distribution}  with 
\begin{equation}
f( x; 2,  0, \frac{\sigma}{\sqrt{2}}, \mu)= \mathcal{N}(x; \mu, \sigma^2), 
\label{eq:normd}
\end{equation}
for $\alpha=1$, the \textbf{Cauchy} or \textbf{Lorentz distribution} 
\begin{equation}
f( x; 1,  0, \gamma, \delta)= \frac{1}{\pi \gamma} \left[ \frac{\gamma ^2}{(x- \delta)^2 + \gamma^2} \right], 
\label{eq:cauchyd}
\end{equation}
\noindent and for $\alpha=0.5$, the \textbf{L\'evy distribution}
\begin{equation} 
f( x; 0.5,1,\gamma,\delta)= \sqrt{\frac{\gamma}{2 \pi}}\frac{\exp{-(\frac{\gamma}{2(x-\delta)})}}{(x-\delta)^{3/2}}.
\label{eq:levyd} 
\end{equation}
For all other cases, only the characteristic function is available in closed form, and numerical approximation techniques must  be adopted for both sampling and parameter estimation \cite{chamb,nolan1997,koutrouvelis1980}. A very nice  and simple to use Matlab package for parameter inference and computation of $\alpha$-stable distributions  is freely downloadable at Mark Veilette's homepage \url{http://math.bu.edu/people/mveillet/html/alphastablepub.html}.

Some examples of $\alpha$-stable pdfs and related \textbf{complementary cumulative distribution function}  (CCDF) are given in Fig.\ref{Fig:levyplots}. The use of the CCDF, or upper tail,  of jump lengths is the standard convention in the literature, 
for  the sake of a more precise description of the tail behaviour, i.e. the laws governing
the probability of large shifts.   The CCDF is defined   as $\overline{F}_{X}(x)= P(X>x)=1 - F_{X}(x)$, where $F_{X}(\cdot)$ is the CDF (cfr., Box~\ref{tab:Probit}).  

\end{svgraybox} 
\end{table}

\begin{figure}[t]
\sidecaption[t]
\includegraphics[scale=0.15,keepaspectratio=true]{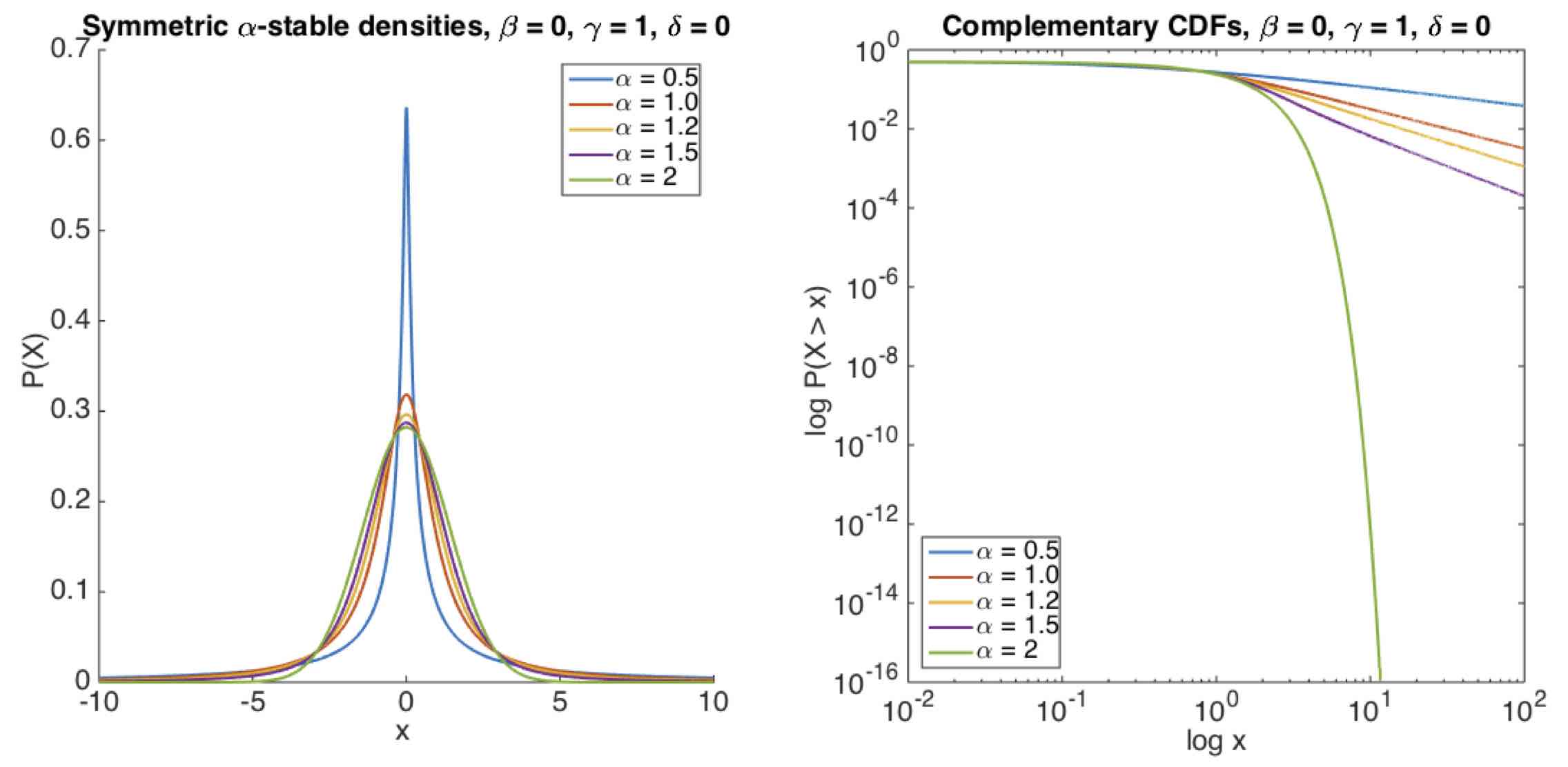}
\caption{Plots of symmetric $\alpha$-stable  distributions (left) and their complementary CDF (CCDF) on $\log-\log$ axes (right) for different values of the characteristic index parameter $\alpha=0.5, 1, 1.2, 1.5, 2$. The CCDF shows the rapid fall-off of the tail of the Gaussian case ($\alpha=2$) as opposed  to the power--law tail behaviour of actual L\'evy flights $\alpha < 2$}
\label{Fig:levyplots}
\end{figure}

\subsubsection{Case study: the L\'evy flight of saccades}


Brockmann and Geisel \cite{brockgeis} have  assumed a power--law dependence in the tail of the saccade amplitude distribution, for which they found empirical support in free viewing of natural scenes. Minimisation of the time needed to scan the entire visual space then led them to predict that eye movement trajectories behave as L\'evy flights, as opposed to more common diffusive random walks, which would result from a Gaussian  amplitude distribution. But in order to obtain simulated eye trajectories that look like their observed scan paths, an empirical determination of a salience field for the correspondingly viewed scene is still needed. Brockmann and Geisel derived that salience field from the spatial distribution of fixations made by observers throughout the scene (a picture of a party). As to the amplitude distribution they considered the Cauchy distribution (Eq. \ref{eq:cauchyd})

The stochastic assumptions of saccade generation made by Brockmann and Geisel \cite{brockgeis} involve a Markovian process, consistent with an interpretation of visual search originally proposed by Horowitz and Wolfe \cite{horowitz1998visual}. However, the predictions and results of the Brockmann and Geisel model do not change substantially if those assumptions are relaxed so as to allow a sufficiently rapidly decaying correlation in the saccade sequences.

Further evidence and characterisation of Lévy-like diffusion in eye movements associated with spoken-language comprehension have recently been provided by Stephen, Mirman, Magnuson, and Dixon \cite{stephen2009}. 

\subsubsection{Case study: the microsaccade conundrum}
Martinez-Conde and colleagues \cite{martinez2013impact,otero2013oculomotor} have put forward the proposal that microsaccades and saccades are the same type of eye movement (the ``continuum hypothesis'').
The microsaccade--saccade continuum is sustained by evidence that saccades of all sizes share a common generator

In this respect, a straightforward hypothesis on the function  microsaccades is that they help to scan fine details of an object during fixation. This hypothesis would imply that fixational eye movements represent a search process. According to this analogy, the statistics of microsaccades can be compared to other types of random searches, namely inspection saccades during free picture viewing \cite{brockgeis}.

Given these assumptions, Engbert \emph{et al.} \cite{engbert2006microsaccades} checked whether the amplitude distribution of microsaccades and saccades follows a similar law.

To investigate the distribution of microsaccade amplitude in a data set of $20,000$ microsaccades, they analysed the tail of the distribution  on a double logarithmic scale. They obtained a power--law decay of the tail   with  exponent $\mu = 4.41$, which would  reject the hypothesis of a L\'evy flight for microsaccades (requiring $\mu < 3$), if compared with numerical results obtained by Brockmann and Geisel \cite{brockgeis}. In turn, this apparently would also lead to reject the ``continuum hypothesis''. However there are many subtleties that should be taken into account in order to fairly compare two such different analyses (e.g., sampling rate and discretisation of the eye tracking raw data) before reaching a conclusion. But at least this nice piece of work is useful to show how advanced statistical methods for eye movement analysis can address big questions in the field.  

\begin{figure}[t]
\sidecaption[t]
\includegraphics[scale=0.13,keepaspectratio=true]{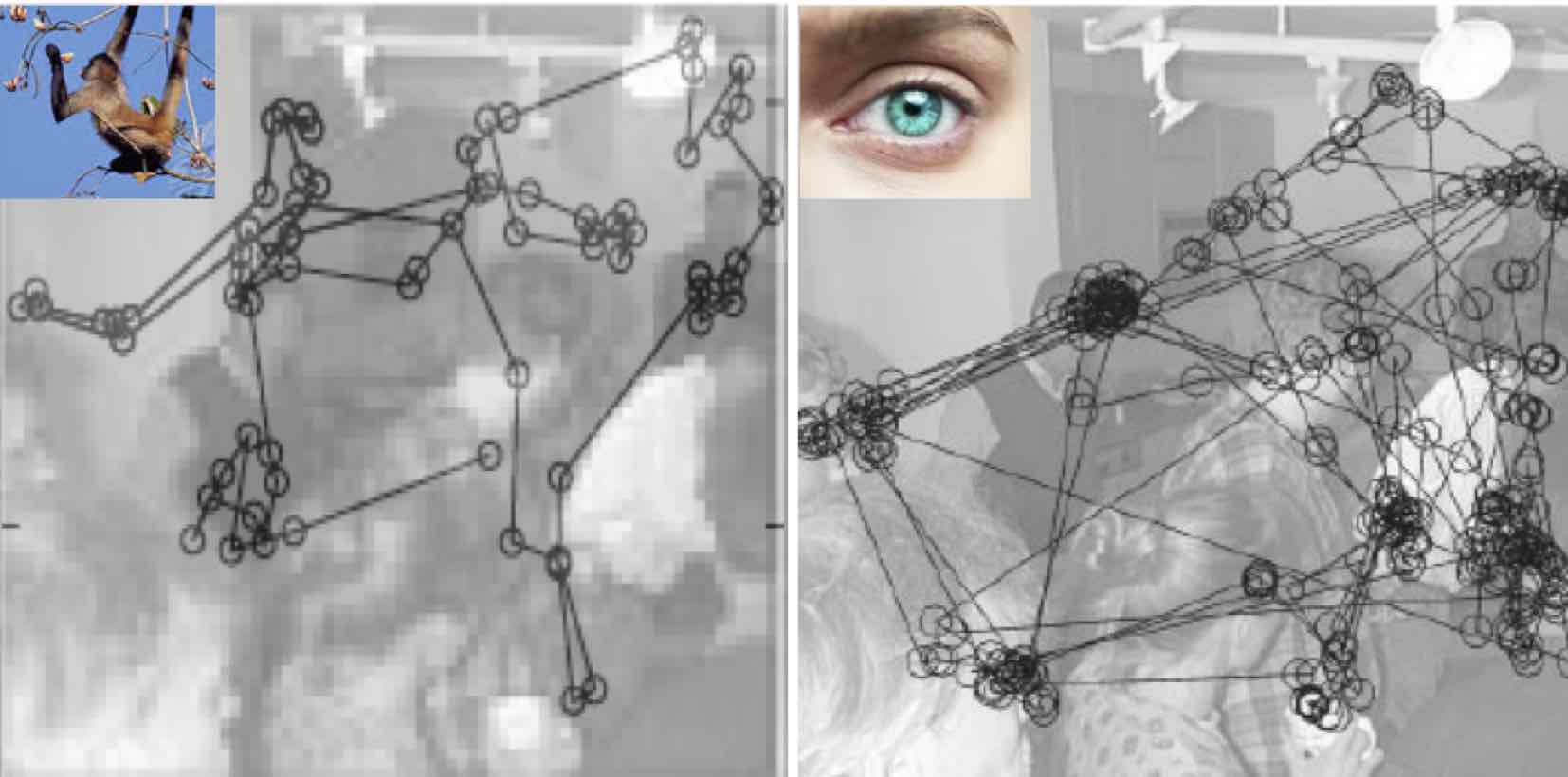}
\caption{Monkey or human: can you tell the difference? The left image has been obtained by superimposing a typical trajectory of spider monkeys foraging in the forest of the Mexican Yucatan, as derived from \cite{ramos2004levy}, on the ``party picture'' used in \cite{brockgeis}. The right image is an actual human scan path (modified after \cite{brockgeis}) }
\label{Fig:spider}
\end{figure}

\subsection{The foraging perspective}

Consider Fig. \ref{Fig:spider}:  \emph{prima facie}, is seems to illustrate  a bizarre jest.
However, from eye-movements  studies \cite{canosa2009real,tatler2006long,tatler2008systematic,tatler2009prominence,dorr2010variability,over2007coarse,TatlerBallard2011eye},  there is evidence that eye movement trajectories and their statistics  are strikingly similar, with respect to the resulting movement patterns and their statistics to those exhibited by foraging animals,  \cite{viswa08,codling2008,plank2008,reynolds2008optimal}.
In other terms, eye
movements and animal foraging address in some way a similar problem \cite{brockgeis}.  Under the foraging metaphor, the eye (and the brain modules controlling the eye behaviour) is the forager, the input visual representation $\mathcal{D}$ is the foraging landscape. Points attracting fixations are foraging sites (in the case of static images) or moving preys (time-varying scenes); gaze shifts occur due to local exploration moves, prey pursuit and long relocation from one site to another.

An intriguing issue is whether the foraging theory underpinning the proposed analyses just provides a useful computational theory metaphor,  or  constitutes a more substantial ground. 
Interestingly enough, Hills \cite{hills2006animal} has argued that what was once foraging in a physical space for tangible resources became, over evolutionary time, foraging in cognitive space for information related to those resources. Adaptations that were selected for during ancestral times are, still adaptive now for foraging on the internet or in a supermarket, or for goal-directed deployment of visual attention \cite{wolfe2013time}.  In these terms, the foraging approach may set a broader perspective for discussing fundamental themes in eye movement behaviour,  e.g., the ``continuum hypothesis'' of Martinez-Conde and colleagues \cite{martinez2013impact,otero2013oculomotor}.

Building on  this rationale,   
gaze shift models have been proposed coping with different levels of visual representation complexity \cite{bfpha04,BocFerAnnals2012,BocFerSPIC2012,BocFerSMCB2013,BocCOGN2014,napboc_TIP2015} and eye movement data analyses have been performed in terms of foraging efficiency \cite{wolfe2013time,cain2012bayesian}.

\begin{figure}[t]
\sidecaption[t]
\includegraphics[scale=0.20,keepaspectratio=true]{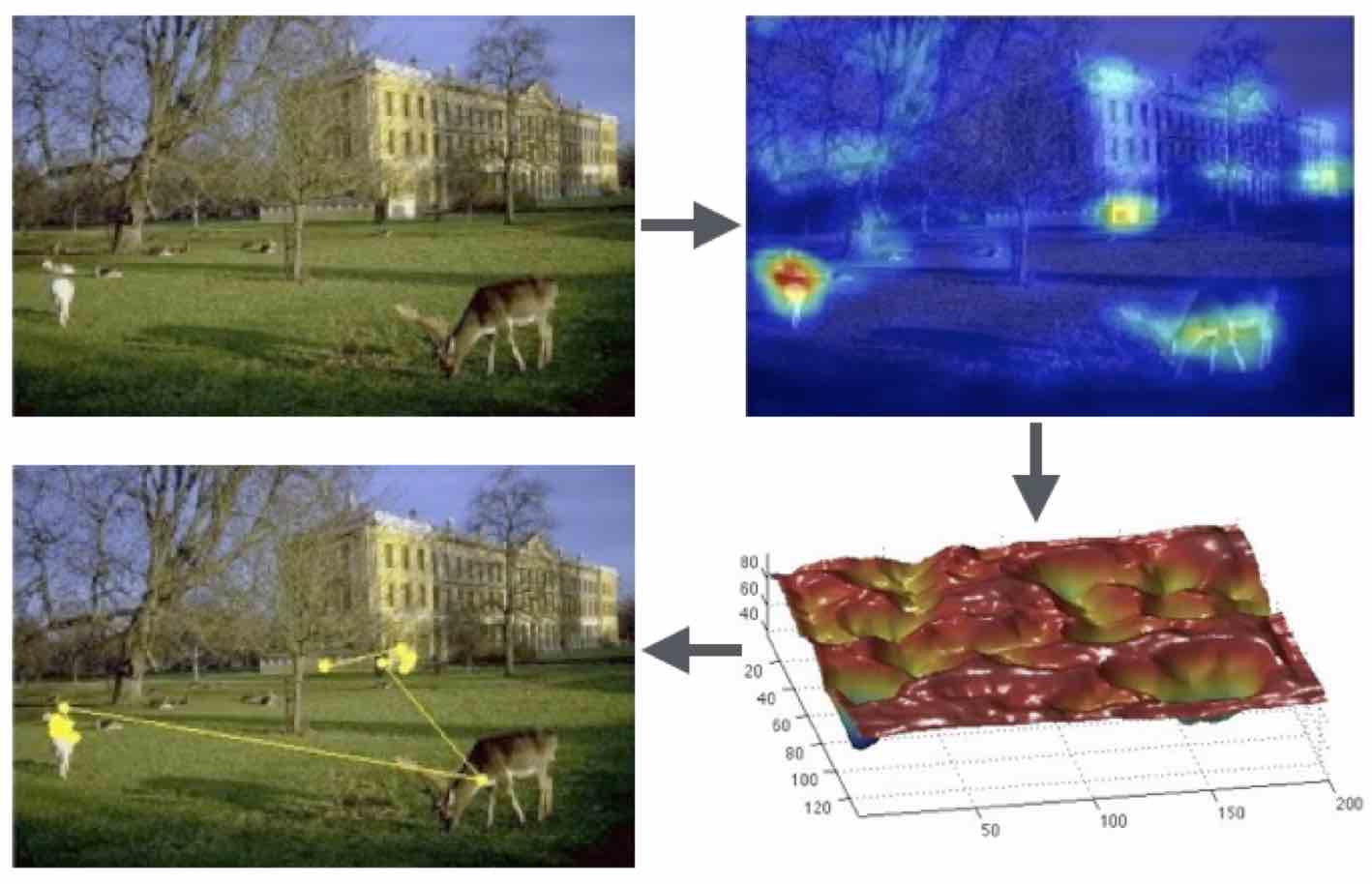}
\caption{The L\'evy model used as a generative model. Top left: the original image. Top right: the salience map. Bottom right: the potential $V( \mathbf{x_i},t_i)$ computed from saliency: \emph{potential wells}  represent informative regions that can attract gaze. Bottom, left: the final scan path superimposed on the original image.}
\label{Fig:simul}
\end{figure}

More formally, rewrite the 2-dimensional Langevin equation (\ref{eq:LangND}), interpreting the deterministic component  $\mathbf{A}( \mathbf{x},t)$ as an external force field due to a potential  $V( \mathbf{x},t)$ \cite{bfpha04} (see Fig. \ref{Fig:simul}), that is  
 $ \mathbf{A}( \mathbf{x},t)= - \nabla V( \mathbf{x},t)$, where the ``del'' (or ``nabla'') symbol $\nabla$ denotes the gradient operator\footnote{A salience map, and thus the potential field $V$ derived from salience, varies in space (as shown in Fig. \ref{Fig:simul}). The  map of such variation, namely the rate of change of $V$ in any spatial direction,  is captured by the vector field $\nabla V$. To keep things simple, think of   $\nabla$  as a ``vector'' of components $(\frac{\partial}{\partial x}, \frac{\partial}{\partial y})$. When $\nabla$ is applied to the field $V$, i.e. $\nabla V=(\frac{\partial V}{\partial x}, \frac{\partial V}{\partial y})$, the gradient of $V$ is obtained}.

 Then, 
\begin{equation}
d\mathbf{x}(t) =  - \nabla V( \mathbf{x},t)dt +  {\mathbf  B}({\mathbf x}, t) d {\mathbf L_{\alpha}}(t).
\label{eq:LangLevyND}
\end{equation}

Equation (\ref{eq:LangLevyND}) now provides a microscopic description (trajectory) of a RW biased by an external force field.

We can thus generalise to two dimensions the discretisation method used to obtain the $1-$dimensional Eq (\ref{eq:LangeItoint}) so to gain an operative definition of the SDE (\ref{eq:LangLevyND})
\begin{equation}
\overbrace{\mathbf{x}_{i+1}}^{\text{new gaze location}}  =  \overbrace{\mathbf{x}_{i}}^{\text{current gaze location}}  - \overbrace{\nabla V( \mathbf{x_i},t_i) \Delta t_i}^{\text{external force}}+\overbrace{ {\mathbf  B}({\mathbf x}_i, t_i)  (\Delta t_i)^{\frac{1}{\alpha}} \boldsymbol \xi_{i}}^{\text{L\'evy motion}}
\label{eq:LangLevyNDdisc}
\end{equation}
\noindent which makes clear that next gaze position is obtained by shifting from current gaze position following a L\'evy displacement that is constrained by the external potential field. The external potential summaries the informative properties of the ``visual landscape'' of the forager.

For instance in \cite{bfpha04} Eq. (\ref{eq:LangLevyNDdisc}) was used as a \emph{generative model}  of   eye movements. In that case,  the external potential was taken as a function of the salience field and   $\boldsymbol \xi_i$ was sampled from a Cauchy distribution. Each sampled gaze shift was then accepted according to a Metropolis-like algorithm (cfr. Box \ref{tab:Sample}), governed  by a ``temperature'' parameter suitable to  tune the randomness of the visual exploration (i.e., the attitude of the forager to to frequently engage in longer relocations/saccade rather then  keeping with  local/fixational exploration). 
One example is provided in Fig. \ref{Fig:simul}\footnote{Matlab software for the simulation is freely downloadable at \url{http://www.mathworks.com/matlabcentral/fileexchange/38512}}.  

 As to the parameters of Eq. (\ref{eq:LangLevyNDdisc}), it is worth noting   that   established numerical techniques are available for fitting such parameters from real data (see, e.g. \cite{siegert2001}). Once the parameters have been learned, the generative capabilities of Eq. (\ref{eq:LangLevyNDdisc}) can be straightforwardly used to Monte Carlo simulate gaze shifts whose characteristics can then be   compare with  human data. For instance in a recent paper by Liberati \emph{et al.}~\cite{LiberatiASD2017}, it is shown that scan paths of children with typical development (TD) and Autism Spectrum Disorder (ASD) can be characterised by eye movements geometrically equivalent to L\'evy flights, but with a different degree of randomness, which can be captured by a temperature parameter as in \cite{bfpha04}.


 

As previously discussed, the heavy-tailed distributions of  gaze shift amplitudes are close to those characterising the foraging behaviour of many animal species.   L\'evy  flights  have been used to model optimal searches of foraging animals, namely their moment-to-moment relocations/flights used to sample the perceived habitat \cite{viswa08}. 
However, the general applicability of L\'evy flights in  ecology and biological sciences is still open to debate. In complex environments, optimal searches are likely to  result from  a \emph{composite strategy},  in which Brownian and Le\'vy motions can be adopted  depending 
on the structure  of the landscape in which  the organism moves \cite{plank2008}.  L\'evy flights are best  suited  for the location of randomly, sparsely
distributed patches and Brownian motion gives the best results for the location of densely but random distributed within-patch resources \cite{reynolds2008many}.

\begin{figure}[t]
\sidecaption[t]
\includegraphics[scale=0.19,keepaspectratio=true]{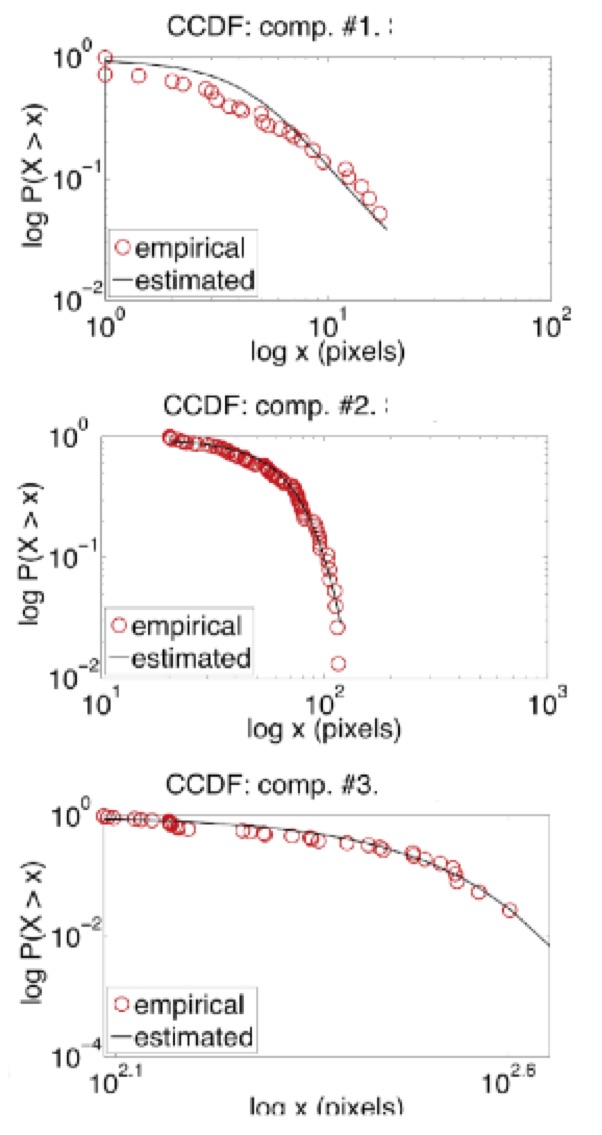}
\caption{Analysis of gaze shift dynamics from a video. The different components were automatically separated by using a clustering procedure based on the Variational Expectation-Maximization  algorithm (see \cite{BocFerSMCB2013} for details). Then, each component was fitted by an $\alpha$-stable distribution. 
Fitting results for one eye-tracked subject are shown in terms of  double log plot of the CCDF. From
top to bottom:  first component accounting for smooth-pursuit and FEMs motions; the medium saccade component; the long saccade  component}
\label{Fig:3comp}
\end{figure}

Also it is possible to compose the strategy as a hybrid strategy in which Brownian-like motion is adopted for local exploration (e.g., FEMs and small saccades) and L\'evy displacements are exploited for long relocations (medium / long saccades)
A preliminary attempt towards such a composite  strategy for modelling gaze shift mechanisms has been presented in \cite{BocFerIciap2011}. However, that approach only conjectured a  simple  binary switch between a Gaussian   and  a Cauchy-like  walk.  In \cite{BocFerSMCB2013} the approach was generalised to handle observers watching videos and thus accounting for multiple kinds of shifts: FEMs, saccade and smooth pursuit (cfr. Fig. \ref{Fig:3comp}).
To this end,  Eq. (\ref{eq:LangLevyNDdisc}) is  reformulated as a 2-dimensional dynamical system  in which   the stochastic part is driven by one-of-$K$ possible types of $\alpha$-stable motion $\boldsymbol \xi_{i}^{k}$.

In the Ecological Sampling model \cite{BocFerSMCB2013} the switch from one motion type to the other was bottom-up determined as a multinomial choice  biased by the complexity of the sampled visual landscape. More recently this idea was extended to a top-down  and task-dependent  probabilistic choice in the framework of Bayesian Decision theory \cite{BocCOGN2014}. Bayesian Decision theory has gained currency in modelling active sensing behaviour \cite{yang2016theoretical}: though minimal, from a theoretical standpoint, a gaze shift action can indeed be considered as the result of a decision-making process (either conscious or uncounscious).

%
%

\section{From patterns of movement to  patterns of the mind: unveiling observer's hidden states}
\label{sec:ML}

The last point we are addressing in this Chapter is: can we infer  target hidden states of observer's mind by analysing his eye movement trajectories? Or, in foraging terms, can we say something on forager's internal state by observing his foraging patterns?

Generally speaking, the hidden states that we may target could be, for example, the task the observer is accomplishing, his expertise, his emotional state (but, also, a certain pathology affecting a group of patients as opposed to a control group). 

More formally, if $\mathbf{T}$ denotes the target internal state (or a set of states) and $\mathbf{X}$ the visible eye movement behaviour, e.g., $\mathbf{X} = \{\mathbf{x}_{F}(1), \mathbf{x}_{F}(2),\cdots \} $, (or, alternatively, the sequence of gaze, amplitudes, directions and durations), one can assume a generative process $\mathbf{T} \rightarrow \mathbf{X}$, where the observer's hidden state shapes the kind of eye trajectories.

\begin{figure}[t]
\sidecaption[t]
\includegraphics[scale=0.15,keepaspectratio=true]{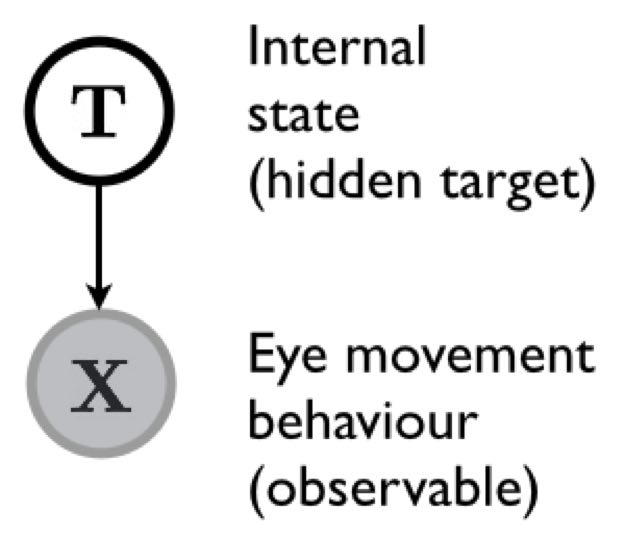}
\caption{The PGM specifying the generative process $\mathbf{T} \rightarrow \mathbf{X}$  through the joint pdf  factorisation: $P(\mathbf{T}, \mathbf{X}) = P(\mathbf{X} \mid \mathbf{T})P(\mathbf{T})$. The shaded node  denotes that RV $\mathbf{X}$ is observable}
\label{Fig:hidden}
\end{figure}

In probabilistic terms the generative process can be captured by the simple PGM sketched in Fig.~\ref{Fig:hidden}, which factorises the joint pdf $P(\mathbf{T}, \mathbf{X})$ as 
$P(\mathbf{T}, \mathbf{X}) = P(\mathbf{X} \mid \mathbf{T})P(\mathbf{T})$ (product rule).

This way, anything we can infer on the hidden state given the observable behaviour is obtained by ``inverting the arrow'', i.e., by applying Bayes' rule:

\begin{equation}
P(\mathbf{T} \mid \mathbf{X}) = \frac{P(\mathbf{X} \mid \mathbf{T}) P(\mathbf{T})} {P(\mathbf{X})}.
\label{eq:invYarbus}
\end{equation}

Once the posterior has been computed, then  we use decision theory to determine output $\mathbf{T}=t$  for each new input $\mathbf{X}=\mathbf{x}$

Note that this is a very general formulation of the problem, which actually may entail a large number of solutions, and the PGM shown in Figure~\ref{Fig:hidden} could be further specified/specialised in number of ways. Also, to keep the description simple, we have omitted the set of parameters involved by the actual specification of the pdfs in Eq.~(\ref{eq:invYarbus}). Clearly, before Eq.~(\ref{eq:invYarbus}) can be put into work, such parameters are to be specified and fitted, or, adopting a more modern term, learned. To this end, a huge amount of \textbf{Machine Learning} (ML) techniques are today available (see Box~\ref{tab:ML}, for ML basic terminology in a nutshell, and \cite{BishopPRML,murphy2012machine} for an in-depth presentation)

Keeping to such general level, from a methodological standpoint there are at least  three distinct approaches to cope with the inverse inference problem posed by Eq.(~\ref{eq:invYarbus}). In decreasing order of complexity:
\begin{enumerate}
\item First solve the inference problem of determining the likelihood function $P(\mathbf{X} \mid \mathbf{T})$  and the prior  probabilities $P(\mathbf{X})$. Then use Bayes' theorem in the form given in Eq. (\ref{eq:invYarbus}). Equivalently, we can model the joint distribution $P(\mathbf{T}, \mathbf{X})$ directly and then normalise to obtain the posterior probabilities $P(\mathbf{T}, \mathbf{X})$. Approaches that explicitly or implicitly model the distribution of inputs as well as outputs are known as \emph{generative models}, because by sampling from them it is possible to generate synthetic data points in the input space. The popular \textbf{Naive Bayes} and Linear Discriminant Analysis methods are  very simple instances (though effective in many practical cases) of a generative approach; \textbf{Hidden Markov Models} (HMM) \cite{BishopPRML,murphy2012machine} for modelling time series provide an appealing example of a generative approach  that has been often exploited  for eye movement analysis. 

\item Solve straightforwardly  the inference problem of determining the posterior  probabilities $P(\mathbf{T}, \mathbf{X})$, and then subsequently use decision theory to assign each new $\mathbf{X}=\mathbf{x}$ to an output target. Approaches that model the posterior probabilities directly are called \emph{discriminative models}. \textbf{Logistic regression} is one notable and classic example,  \textbf{Conditional Random Fields} (CRFs) for modelling time series are more sophisticated one \cite{BishopPRML,murphy2012machine}.

\item Find a function $f : \mathbf{X} \to \mathbf{T}$, called a \emph{discriminant function}, which maps each input $\mathbf{X}=\mathbf{x}$ directly onto a class label. For instance, in the case of two-class problems, e.g., distinguishing experts from novice observers, $\mathbf{T}$ might be binary valued and such that $f = 0$ represents class $\mathbf{T}=t_1$ and $f = 1$ represents class $\mathbf{T}=t_2$. In this case, probabilities play no role. Many popular artificial neural nets or modern methods such as the \textbf{Support Vector Machine} (SVM) for regression and classification (a baseline technique in ML) implement this approach \cite{BishopPRML,murphy2012machine}.
\end{enumerate}

Note that, since many applications require a posterior class probability,  methods based on discriminant functions can be ``transformed'' into discriminative ones in order to gain an output in probabilistic form. For instance, the output $f(\cdot)$ of a binary SVM classifier  can be fed into a sigmoid function, to approximate the posterior (e.g., $P(\mathbf{t} \mid \mathbf{X}) \approx \frac{1}{1+ \exp(Af +B)}$, where $A,B$ are parameters that can be determined via regularised maximum likelihood).

Clearly, the generative approach is in principle the most appealing one. However it should be recalled that  apart from simple cases such as Na{\"i}ve Bayes (see Box~\ref{tab:NB}), the normalisation of the joint pdf can be a hard task. Referring again to Eq. (\ref{eq:invYarbus}), calculating the normalisation factor $P(\mathbf{X})$ requires the marginalisation  $P(\mathbf{X})= \sum_{\mathbf{T}} P(\mathbf{T}, \mathbf{X}) $ ($P(\mathbf{X})= \int P(\mathbf{T}, \mathbf{X}) d\mathbf{T}$ when RVs are continuous), which, in real cases, is hardly computable. Thus, complex approximation techniques such as Monte Carlo or Variational Bayes are to be taken into account \cite{BishopPRML,murphy2012machine}.

\begin{figure}[t]
\includegraphics[scale=0.15,keepaspectratio=true]{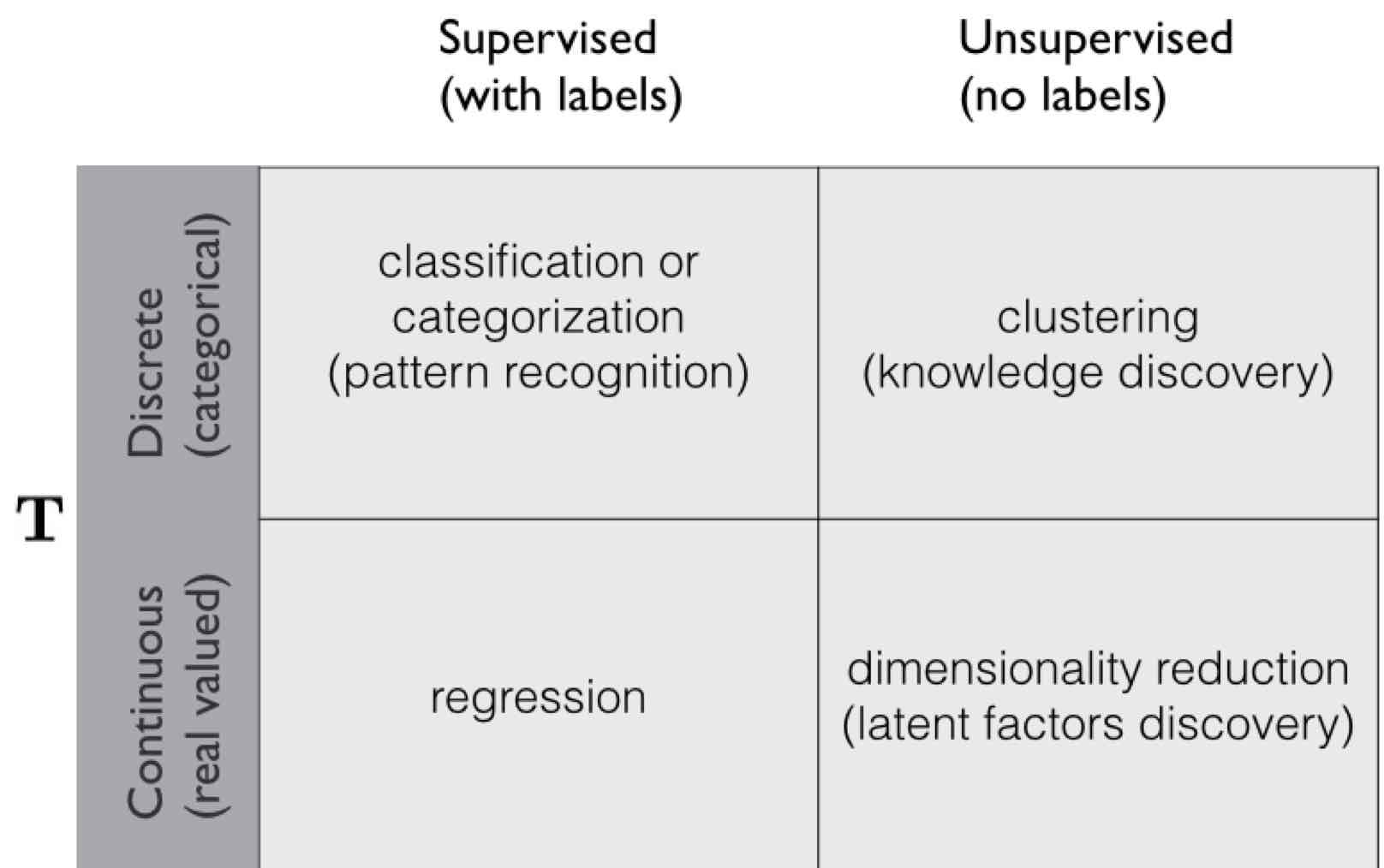}
\caption{The main problems  Machine Learning is addressing}
\label{Fig:ML}
\end{figure}

\begin{table}
\begin{svgraybox}
\caption{\textbf{Machine Learning in a nutshell}}
\label{tab:ML}

In the inferential process defined by Eq. (~\ref{eq:invYarbus}), $\mathbf{T}$ represents the output of the process and $\mathbf{X}$ the given input. In Statistical  Machine Learning (briefly, ML) terminology  $\mathbf{X}$ is usually shaped as a random vector of  features (or attributes, or covariates), $\mathbf{X}= \{ \mathbf{X}_i \}_{i=1}^{N}$, where $i$ is a suitable index. For example when $i$ is a time index, then $\{ \mathbf{x}_1, \mathbf{x}_2 \}$ is the realisation of a   stochastic process. ML does not relate to a specific problem thus $\mathbf{X}_i$ could be a complex structured object, such as an image, a sentence, an email message,  a graph, etc.

The form of the output or response variable $\mathbf{T}$ can be either discrete (categorical, nominal) or continuous (real-valued). One example of the first type, is when $\mathbf{T}$ can take the label of one of two tasks given to the observer, e.g., $\mathbf{T} = t_{k}$  where $t_{1} = \text{``look for people''}$ and $t_{2} = \text{``look for cars''}$. Another example, is $\mathbf{T}$  taking values over the discrete set of basic emotion (``fear'', ``disgust'', ``joy'', etc.).  As opposed to this latter example, we could try instead, to infer from eye movements a continuous affect state, so that $ \mathbf{T}$ is taking values  $t_{i}$ in the real valued space of valence and  arousal. 

From a practical standpoint, when using a ML approach to analyse our eye tracking data we can be in one of these two conditions: 
\begin{enumerate}
\item \textbf{supervised learning}: we know where input $\mathbf{x}_i$ comes  from (e.g., $\mathbf{x}_i$ was measured while the observer was scanning a happy face); more formally $\mathbf{x}_i$ is paired with target value or label $t_{i}$, thus we have a training set $\mathcal{D} = \{ (\mathbf{x}_i, t_{i})\}_{i=1}^{N}$;
\item \textbf{unsupervised learning}: we have no labels, and our dataset is represented by the bare input data $\mathcal{D} = \{ (\mathbf{x}_i)\}_{i=1}^{N}$
\end{enumerate}

Thus, in the supervised setting the goal is to learn a mapping from input $\mathbf{X}$ to output $\mathbf{T}$, given a labeled set of input-output pairs. When $\mathbf{T}$ is discrete the problem is known as \textbf{classification} or pattern recognition; when $\mathbf{T}$ is real-valued, we are performing \textbf{regression}.

In the unsupervised setting, we have no labels available, thus the goal is to find ``interesting patterns'' in the data. This is sometimes named knowledge discovery or data mining.  It is a much less well-defined problem, since we are not told what kinds of patterns to look for, and there is no obvious error metric to use (unlike supervised learning, where we can compare our prediction  for a given x to the observed value). When $\mathbf{T}$ is discrete, the problem is known as \textbf{clustering}. When $\mathbf{T}$ is real-valued we are typically in the case of \textbf{dimensionality reduction}. The latter is used  when dealing with high dimensional data: it is often useful to reduce the dimensionality by projecting the data to a lower dimensional subspace which captures the ``essence'' of the data. Indeed, although the input data may appear high dimensional, there may only be a small number of degrees of variability, corresponding to \textbf{latent factors} (Principal Component Analysis or Factor Analysis being well known examples).

Statistical Machine Learning is nowadays a broad and mathematically sophisticated field.   Two excellent and up-to-date textbooks are those by Bishop~\cite{BishopPRML} and Murphy~\cite{murphy2012machine}.

\end{svgraybox} 
\end{table}

\subsection{Inverting Yarbus to infer the task}
We now assume that the target internal state  $\mathbf{T}$ of the observer stems from 
a given visual task, and for simplicity we straightforwardly use $\mathbf{T}$ to denote the task.  In other terms,  the task $\mathbf{T}$ is the hidden state of interest.

The seminal experiment by Yarbus \cite{Yarbus}  studied the effect of the visual task    on the trajectories of eye movements $\mathbf{X}$. On the basis of our introductory discussion,   we know that in probabilistic terms the effect $\mathbf{T} \rightarrow \mathbf{X}$,  or forward mapping, is formally captured by the likelihood function  $P(\mathbf{X} \mid \mathbf{T})$.  Also, we   know  that Eq. (\ref{eq:invYarbus}) is an application of Bayes' rule, to compute the posterior probability of the task $\mathbf{T}$ after having observed eye movement trajectories  $\mathbf{X}$.  Summing up, the inference of the task requires the computation of the inverse mapping   $\mathbf{T} \leftarrow \mathbf{X}$, which is easily understood  as an ``inverse Yarbus'' process \cite{haji2013computational}.  


Clearly, as previously mentioned, there are several ways of ``inverting Yarbus''.

\subsubsection{Case study: Inverting Yarbus via Na{\"i}ve Bayes} 

In a recent study \cite{henderson2013predicting}, Henderson \emph{et al.} considered four tasks: scene search, scene memorisation, reading, and pseudo-reading. Task inference was achieved by classifying the observers'  task by implementing Eq.~(\ref{eq:invYarbus}) as the baseline Na{\"i}ve Bayes' (NB) classifier. Namely they addressed two problems: (i) whether the task associated with a trial could be identified using training from other trials within the same experimental session (within-session classification); (ii) whether the task performed in one session could be identified based on training from a session conducted on a different day (cross-session classification). Twelve members of the University of South Carolina community participated in the experiment. A dedicated  classifier was trained  for each observer, thus the baseline NB has proved to be sufficient. NB classifiers were trained on  a feature vector $\mathbf{X}$ of dimension $8$, i.e., eight eye movement features capturing eye movement patterns for each trial: the mean and standard deviation of fixation duration, the mean and standard deviation of saccade amplitude, the number of fixations per trial, and the three parameters $\mu$, $\sigma$, and $\tau$ quantifying the shape of the fixation duration distribution with an ex-Gaussian distribution, which is known to change for different eye-movement tasks (cfr., \cite{henderson2013predicting} for details.

\begin{table}
\begin{svgraybox}
\caption{\textbf{Na{\"i}ve Bayes}}
\label{tab:NB}
The Na{\"i}ve Bayes algorithm is a classification algorithm based on Bayes rule, that assumes the attributes $\mathbf{X} = \{ \mathbf{X}_1,  \mathbf{X}_2,  \cdots \}$ are all conditionally independent of one another given $\mathbf{T}$. Consider, for example, the two feature case, where $\mathbf{X} = \{ \mathbf{X}_1,  \mathbf{X}_2 \}$, then
\begin{equation}
P( \mathbf{X}   \mid   \mathbf{T}) = P(\mathbf{X}_1,  \mathbf{X}_2  \mid   \mathbf{T}) = 
P(\mathbf{X}_1 \mid \mathbf{X}_2,  \mathbf{T})P( \mathbf{X}_2  \mid   \mathbf{T}) =
P(\mathbf{X}_1 \mid   \mathbf{T})P( \mathbf{X}_2  \mid   \mathbf{T})
\end{equation}
Thus, if $\mathbf{X}$ contains $n$ attributes:  $P( \mathbf{X}   \mid   \mathbf{T}) = \prod_{i=1}^{n} P( \mathbf{X}_i   \mid   \mathbf{T})$. This way Eq.~\ref{eq:invYarbus} can be written as
\begin{equation}
P(\mathbf{T} = t_k \mid \mathbf{X}) = \frac{ P(\mathbf{T} =t_k) \prod_{i=1}^{n} P( \mathbf{X}_i   \mid   \mathbf{T} = t_k)} 
{\sum_j P(\mathbf{T} =t_j) \prod_{i=1}^{n} P( \mathbf{X}_i   \mid   \mathbf{T} = t_j)}.
\label{eq:invYarbusNB}
\end{equation}
If we are interested only in the most probable value of $\mathbf{T}$, then we have the Na{\"i}ve Bayes classification/decision rule: $\mathbf{T} \leftarrow \arg \max_{t_k} P(\mathbf{T} = t_k \mid \mathbf{X})$, which simplifies Eq.~(\ref{eq:invYarbusNB}) to the following (because the denominator does not depend on $t_k$ ):
\begin{equation}
\mathbf{T} \leftarrow \arg \max_{t_k}  P(\mathbf{T} =t_k) \prod_{i=1}^{n} P( \mathbf{X}_i   \mid   \mathbf{T} = t_k)
\label{eq:invYarbusNB2}
\end{equation}

\end{svgraybox} 
\end{table}

\subsubsection{Case study: Inverting Yarbus via HMM} 

\begin{figure}[t]
\sidecaption[t]
\includegraphics[scale=0.11,keepaspectratio=true]{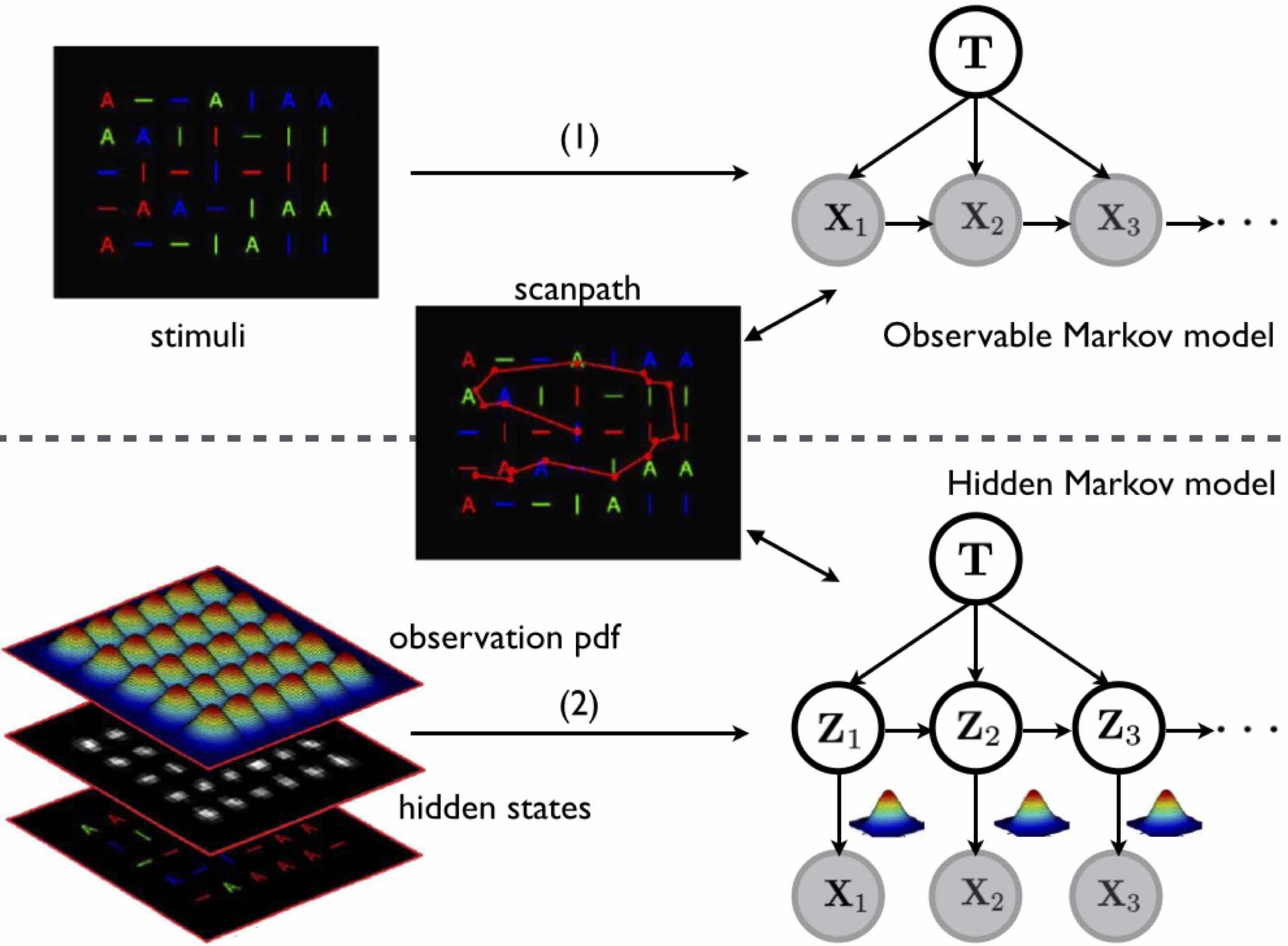}
\caption{Two solutions for the inverse Yarbus given a time series of observations: (1) the task-conditioned observable Markov model \cite{hacisalihzade1992visual,ellistark}  (top panel) and the task-conditioned hidden Markov model \cite{haji2013computational} (bottom panel)}
\label{Fig:TaskHMM}
\end{figure}

In \cite{haji2013computational} Haji-Abolhassani and Clark present a study in which Eq.~(\ref{eq:invYarbus}) is shaped to explicitly account for the gaze shift sequence as a stochastic process. They present different experiments and models, but here, for clarity sake, we will consider a basic condition and a baseline model so  to capture the  rationale behind their approach and, also, to compare with the Ellis and Stark model presented in Section \ref{sec:markovshift}. Recall that in that case the transition matrix $A_{kj}$ was directly estimated by ``counting'' the percentage of transitions from one point of interest to another (more formally, via Maximum Likelihood estimation). Thus, the model was an observable Markov model. In the inverse Yarbus setting, it can be represented as the Markov chain conditioned on task $\mathbf{T}$ that is depicted in Fig. ~\ref{Fig:TaskHMM} (top panel, solution $1$). Different tasks are likely to give rise to different transition matrices. 
In \cite{haji2013computational} visual tasks considered in the simplest experiment were counting red bars, green bars, blue bars, horizontal
bars, vertical bars, or characters. 

\begin{figure}[t]
\sidecaption[t]
\includegraphics[scale=0.17,keepaspectratio=true]{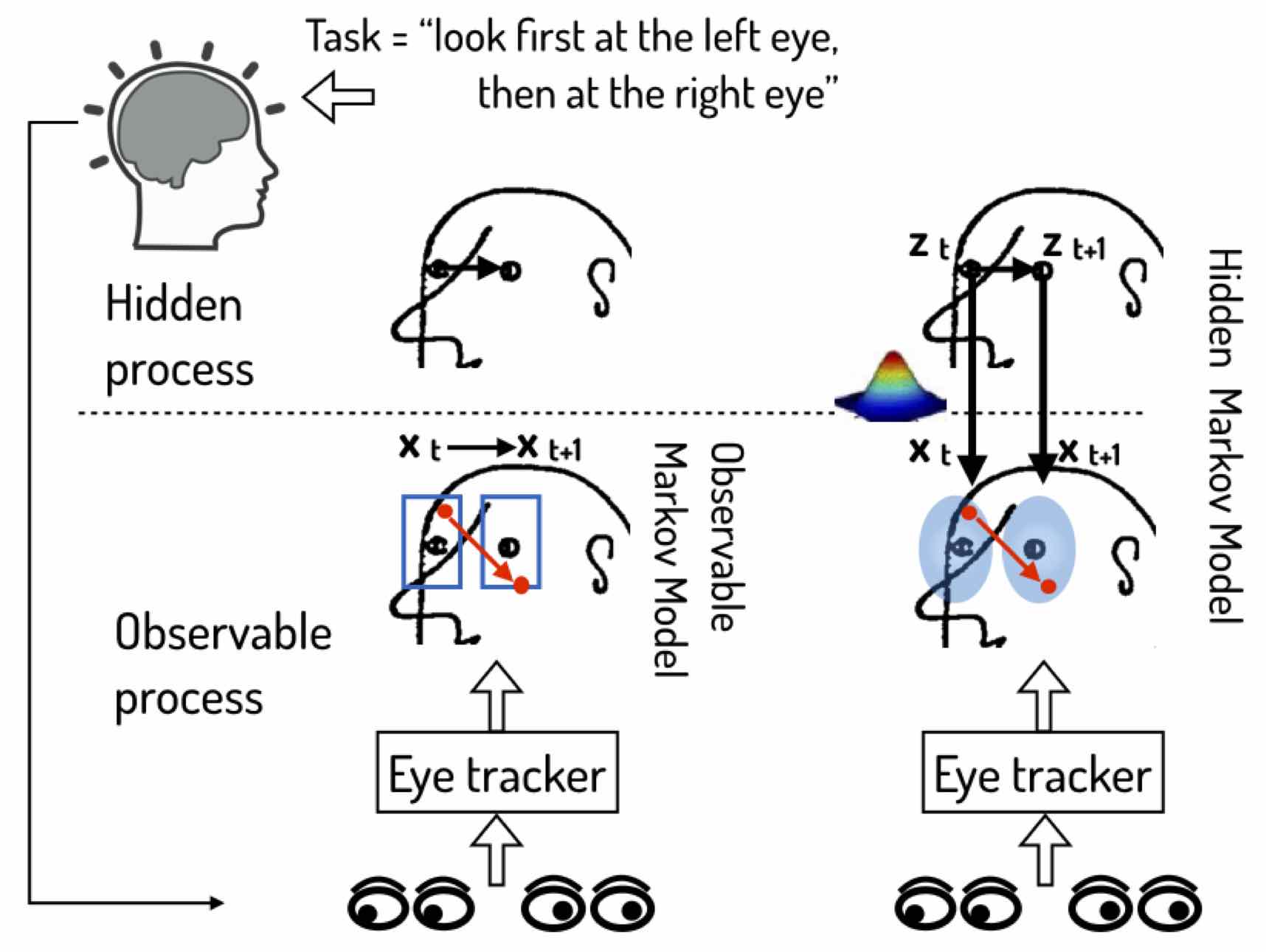}
\caption{Observable Markov Model vs. HMM. We assume the task ``\textsf{look first at the left eye, then at the right eye}'' of  Klee's drawing presented in Fig.~\ref{Fig:stark}.  The  noise-free state space is here   $S= \{ s_1=\text{``left eye''}, s_2=\text{``right eye''}\}$, where, ideally, $s_1, s_2$ are exactly centred on the eyes. Thus, the ``perfect shift'' of the mind's eye would be $\mathbf{z}_t = s_1 \rightarrow \mathbf{z}_{t+1} = s_2$. However, as in Plato's cave, we can only observe the noisy and variable eye-tracked shift $\mathbf{x}_t, \rightarrow \mathbf{x}_{t+1}$ as a surrogate.    The observable Markov Model simplifies the analysis by considering  the two ROIs as a coarse-grained representation of the hidden states $s_1, s_2$  and assumes  all  $\mathbf{x}_t$'s falling within the ROI as equivalent. On the contrary, the HMM allows the hidden shift $\mathbf{z}_t  \rightarrow \mathbf{z}_{t+1} $ to be part of the model, and the visible shift $\mathbf{x}_t  \rightarrow \mathbf{x}_{t+1} $  to be nothing but  its sampled noisy realisation (e.g., Gaussian). See text for  details.}
\label{Fig:HMM}
\end{figure}

Even if we leave apart subtle issues such as the dissociation between the centre of gaze and the covert focus of attention \cite{haji2013computational},  it is very unlikely for a saccade to land exactly on the chosen point of interest (objects or salient locations). The fixation locations may undershoot or overshoot the targets due to  oculomotor properties of human eyes or the noisiness of the eye tracker. To account for this problem, in terms of an observable Markov model, a ``practical'' viable solution is to relax the exact point of interest condition to a more flexible region of interest  surrounding the  point. This indeed was the solution adopted by Stark and colleagues~\cite{hacisalihzade1992visual,ellistark}, depicted in Fig.~\ref{Fig:stark}, and which we now recall in Fig.~\ref{Fig:HMM}.

As  a more principled alternative (see, Fig.~\ref{Fig:HMM}) one can assume that the exact points of interest,  correspond to ``hidden'' targets or states: when, under a given task,  one such target is chosen at time $t$, say  $\mathbf{z}_t$, the corresponding actual fixation   $\mathbf{x}_t$ will be generated by  adding some noise $\epsilon$ (e.g., distributed according to a zero mean Gaussian pdf), i.e. $\mathbf{x}_t = \mathbf{z}_t + \epsilon$. In other terms, we are assuming that $P(\mathbf{x}_t \mid \mathbf{z}_t) = \mathcal{N}(\mathbf{x}_t; \mathbf{z}_t, \Sigma)$. That is, when $\mathbf{z}_t$ is chosen, the actual observation is obtained by sampling from a Gaussian distribution $\mathcal{N}(\mathbf{x}_t; \mathbf{z}_t, \Sigma)$ centred on the true target $\mathbf{z}_t$, where the inverse of the covariance matrix  $\Sigma$ will define the precision of target shooting.

\begin{figure}[t]
\sidecaption[t]
\includegraphics[scale=0.22,keepaspectratio=true]{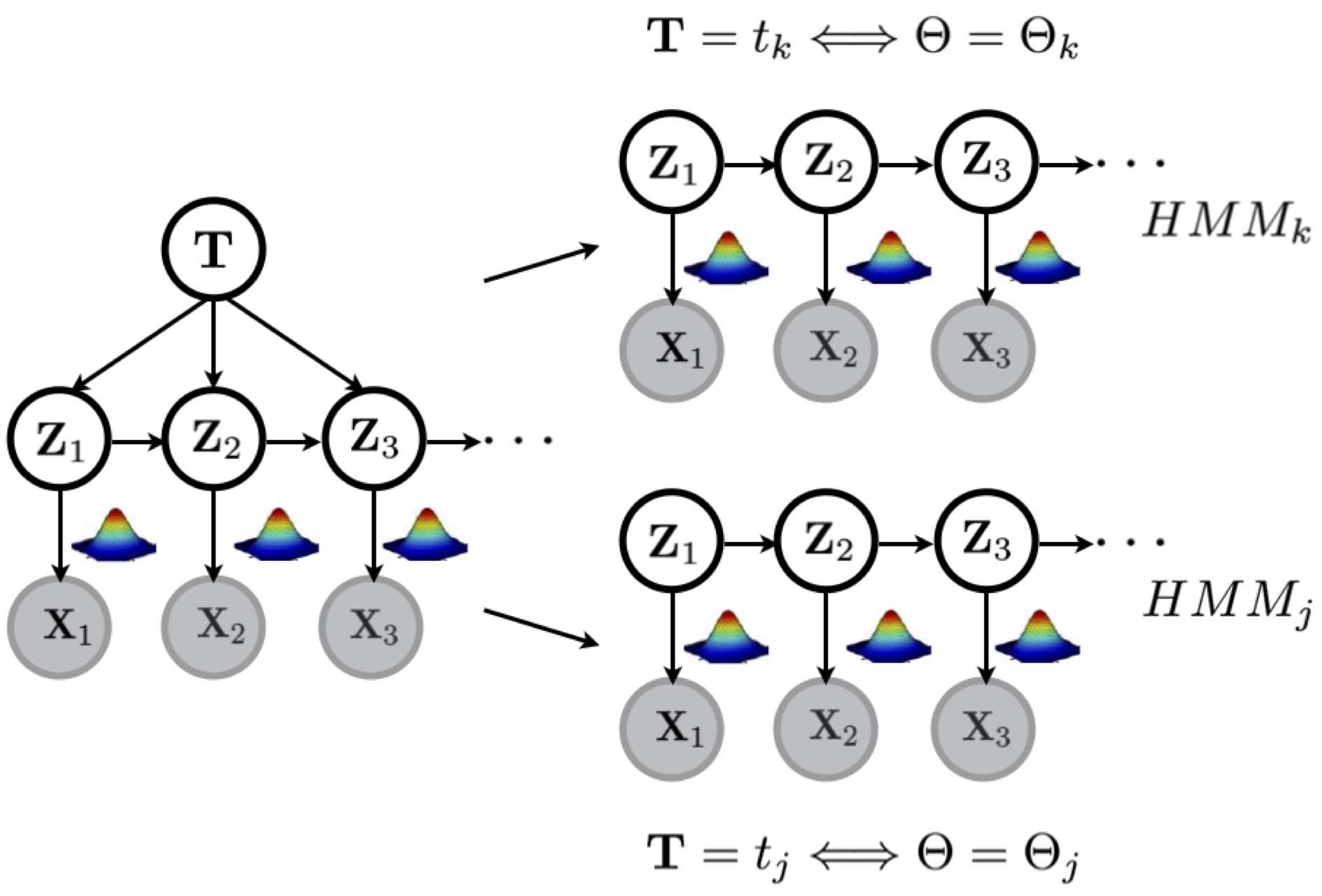}
\caption{For $K$ tasks, the problem of learning the parameters for the DBN on the left is simplified to learning $K$ simple HMM parameters. In the learning  stage, for each task $\mathbf{T} = t_k$ a specific set of parameter $\Theta_k$ is from observations, obtaining the $k$-th HMM. Task inference is performed by choosing the HMM providing the higher likelihood for the input observation $\mathbf{x}_{new}$}
\label{Fig:HMMsplit}
\end{figure}

An HMM can be explicitly conditioned on task $\mathbf{T}$ generalising to the the Dynamic Bayesian Network  depicted in Fig. ~\ref{Fig:TaskHMM} (bottom panel, solution $2$). The problem of learning such DBN can be further simplified by learning a separate HMM for a given  task $\mathbf{T} = t_k$. This way each task will be implicitly defined through the set of parameters defining the corresponding HMM, $\mathbf{T} = t_k \Longleftrightarrow \Theta = \Theta_k$ (cfr, Fig. \ref{Fig:HMMsplit}

Eventually, task inference is performed by choosing the HMM providing the higher likelihood for the input observation $\mathbf{x}_{new}$.

A nice study using HMM in the specific context of face exploration modelling has been presented by Coutrot \emph{et al.} \cite{coutrot2016face}. In this study, similarly to the example outlined in Fig. \ref{Fig:HMM},  each hidden state represents  specific parts  of the face that are likely to be fixated; the actual distribution of eye positions (emission density)  is modelled as a 2-D Gaussian distribution. Coutrot \emph{et al.}~\cite{Coutrot2017} have recently a Matlab toolbox freely available to the community to support scan path modeling and classification with HMMs\footnote{\url{http://antoinecoutrot.magix.net/public/code.html}}.
A more complex example of how to exploit more general DBNs - namely, an Input-Output Coupled HMM - for dynamically intertwining eye movements and hand actions in a drawing task is provided in Coen-Cagli \emph{et al.} \cite{coen2009visuomotor,cagli2008draughtsman}.

\subsection{Assessing cognitive impairments and expertise} 
Eq.~(\ref{eq:invYarbus}), can be used beyond the important issue of task classification. More generally, the value of $\mathbf{T}$  can represent a label $\ell$ to identify groups of observers that exhibit different eye movement behaviour with respect to a given task. In these circumstances, Eq.~(\ref{eq:invYarbus}) formalises the probability that one observer  belongs to one group. The posterior can then be used for classification (e.g., via the $\arg \max$ decision rule).

\subsubsection{Case study: Assessing cognitive impairments} 

On the rationale that patients with mild cognitive impairment (MCI) often progress to Alzheimer's disease (AD), Lagun \emph{et al.} \cite{lagun2011detecting} applied ML methods  to analyse and exploit the information contained in the characteristics of eye movement exhibited by healthy and impaired subjects during the viewing of stimuli in the Visual Paired Comparison (VPC) task for the detection of memory impairment associated with MCI. The VPC assessment proceeds in two steps. During the familiarisation phase, subjects are presented with two identical visual stimuli, side by side, on a computer screen. Eye tracked subjects are allowed to look at the pictures for a specified amount of time. During the test phase,  subjects are presented with pictures of the old stimulus and a novel stimulus, side by side. Control subjects typically spend $70\%$ of the time during the test phase looking at the novel stimulus, which  indicates that they have a memory for the repeated, and now less interesting, stimulus. In contrast, age-matched MCI patients did not spend more time looking at the novel stimulus than the repeated stimulus.

Data analysis was conducted via supervised classification (two class/label problem, $\mathbf{T}$ taking values in  $\ell = \{ \text{``impaired''}, \text{``control''}\}$),  by exploiting standard techniques, namely Na{\"i}ve Bayes, Logistic Regression, and the Support Vector Machine. They first trained the classification models on the multidimensional representation $\mathbf{X}$ of eye movements from a sample of the impaired and control subjects, $\mathcal{D} = \{\mathbf{x}_{train},\ell \}$ and then used the model to predict the status of new subjects based on their eye movement characteristics, i.e, $P(\mathbf{T} \mid \mathbf{X}= \mathbf{x}_{new})$. The results showed that eye movement characteristics including fixation duration, saccade length and direction, and re-fixation patterns (gaze position re-visits on previously seen parts of the stimuli) can be used to automatically distinguish impaired and normal subjects. In this study the SVM classifier  outperformed the other techniques.

Beyond the specific issue addressed by Lagun \emph{et al.}, it is worth looking at their paper \cite{lagun2011detecting} because it provides a gentle introduction to the  Na{\"i}ve Bayes, Logistic Regression, and  SVM algorithms.

\subsubsection{Case study: Classifying billiard player expertise}
The study presented in \cite{boccignone_jemr2014} analysed the oculomotor behaviour of individual observers engaged in a visual task, with the aim of classifying them as experts or novices (two class/label problem, $\mathbf{T}$ taking values in  
$\ell = \{ \text{``expert''}, \text{``novice''}\}$). To this end, various visual stimuli and tasks were administered to $42$ subjects, half novices and half expert billiard players. Stimuli were a portion of a real match, video-recorded from the top, containing several shots of variable length and complexity, as well as a number of ad-hoc individual shots, also video-recorded from the top in a real setting. The match stimulus was associated to a free-viewing observation condition, while for the individual shots, which were occluded in the final part of the trajectory, observers were asked to predict the outcome of the shot, which placed implicitly a significant constraint on the deployment of visuospatial attention, and, consequently, on the overt scan path.

The input $\mathbf{X}$ was obtained as follows. For each observer, given the sequence of fixations $\{\mathbf{x}_t\}_{t=1}^{N_T}$, where the vector $\mathbf{x}_t$ represents the fixation position (coordinates) at time $t$, the amplitude and direction of each gaze shift were computed:  $\{l_t,\theta_t \}_{t=1}$. Third feature was the fixation duration $\{f_t \}_{t=1}^{N_T}$.


The random sample $\{l_t,\theta_t, f_t\}_{t=1}^{N_T}$ was summarised through the empirical distribution functions (histograms), that is  the random vectors $\mathbf{x}^l = \left[x^{l}_{1} \cdots x^{l}_{D} \right]^T$, $\mathbf{ x}^{\theta}=\left[x^{\theta}_{1} \cdots x^{\theta}_{D} \right]^T$ and $\mathbf{x}^f=\left[ x^{f}_{1} \cdots x^{f}_{D} \right]^T$, respectively, where the vector dimension $D$ represents the number of bins of the histogram. 
The feature vector $\mathbf{x}^s$ is thus a summary of the behaviour of a single observer with respect to a particular feature space or source of information $s=1, \dots S$, here $S=3$. Thus, eventually, $\mathbf{X} = \{\mathbf{x}^s\}_{s=1}^{S}$.
\begin{figure}[t]
\sidecaption[t]
\centering
     \subfigure[Fixation duration]{\includegraphics[width=0.52\textwidth]{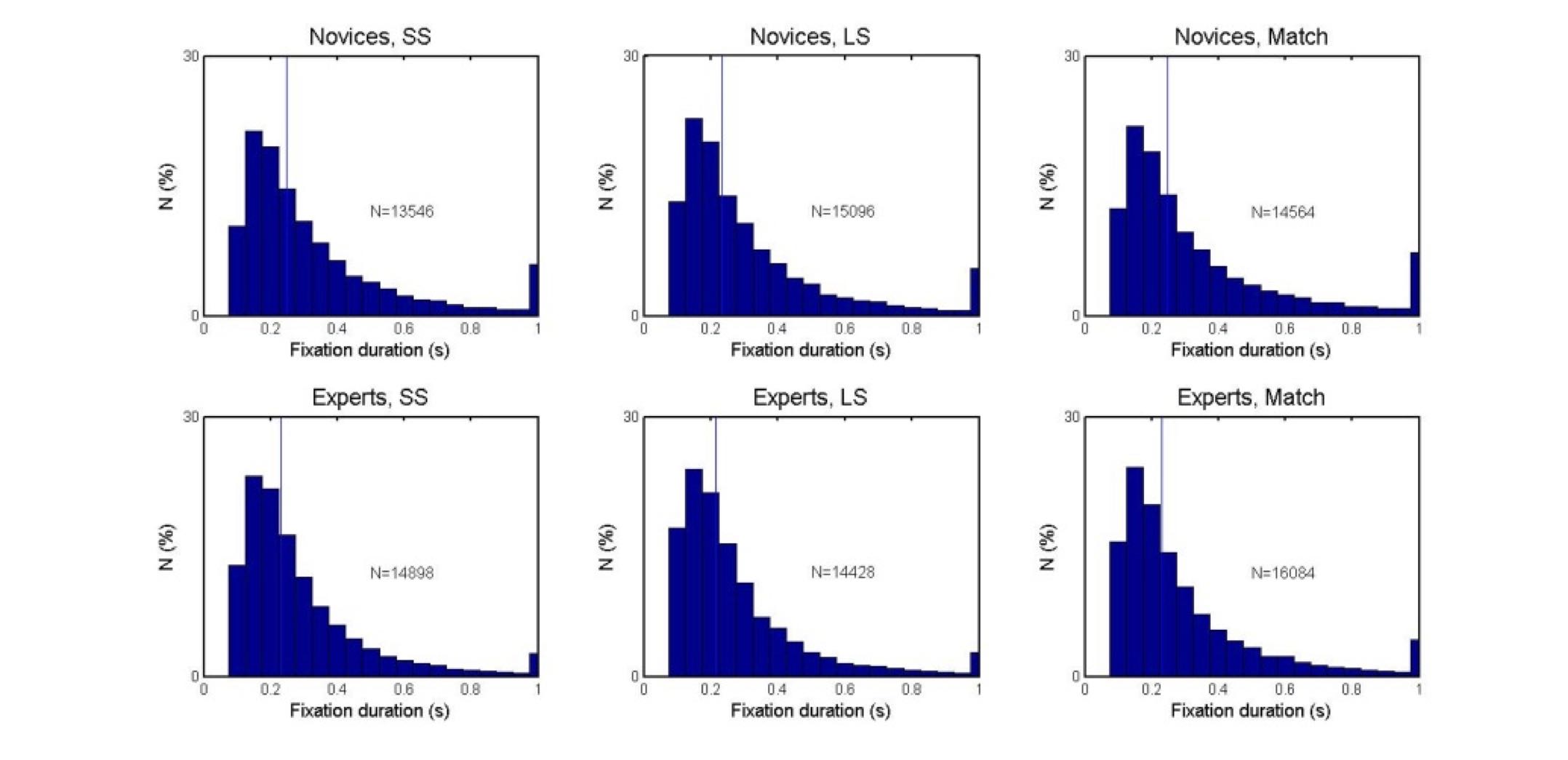}\label{gazefixtime}}
   \subfigure[Gaze shift amplitude]{\includegraphics[width=0.52\textwidth]{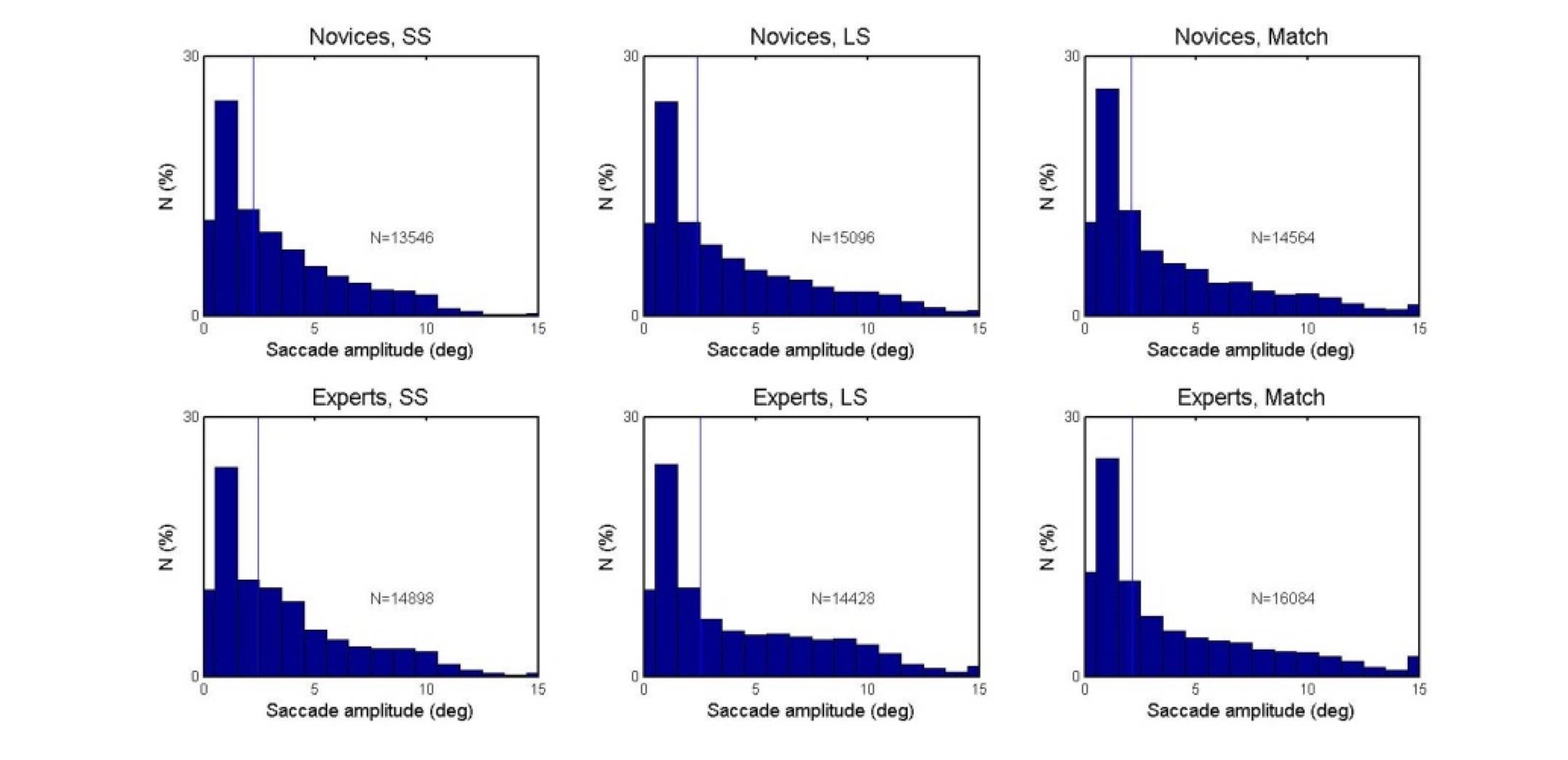}\label{gazeamp}}
    \subfigure[Gaze shift direction]{\includegraphics[width=0.54\textwidth]{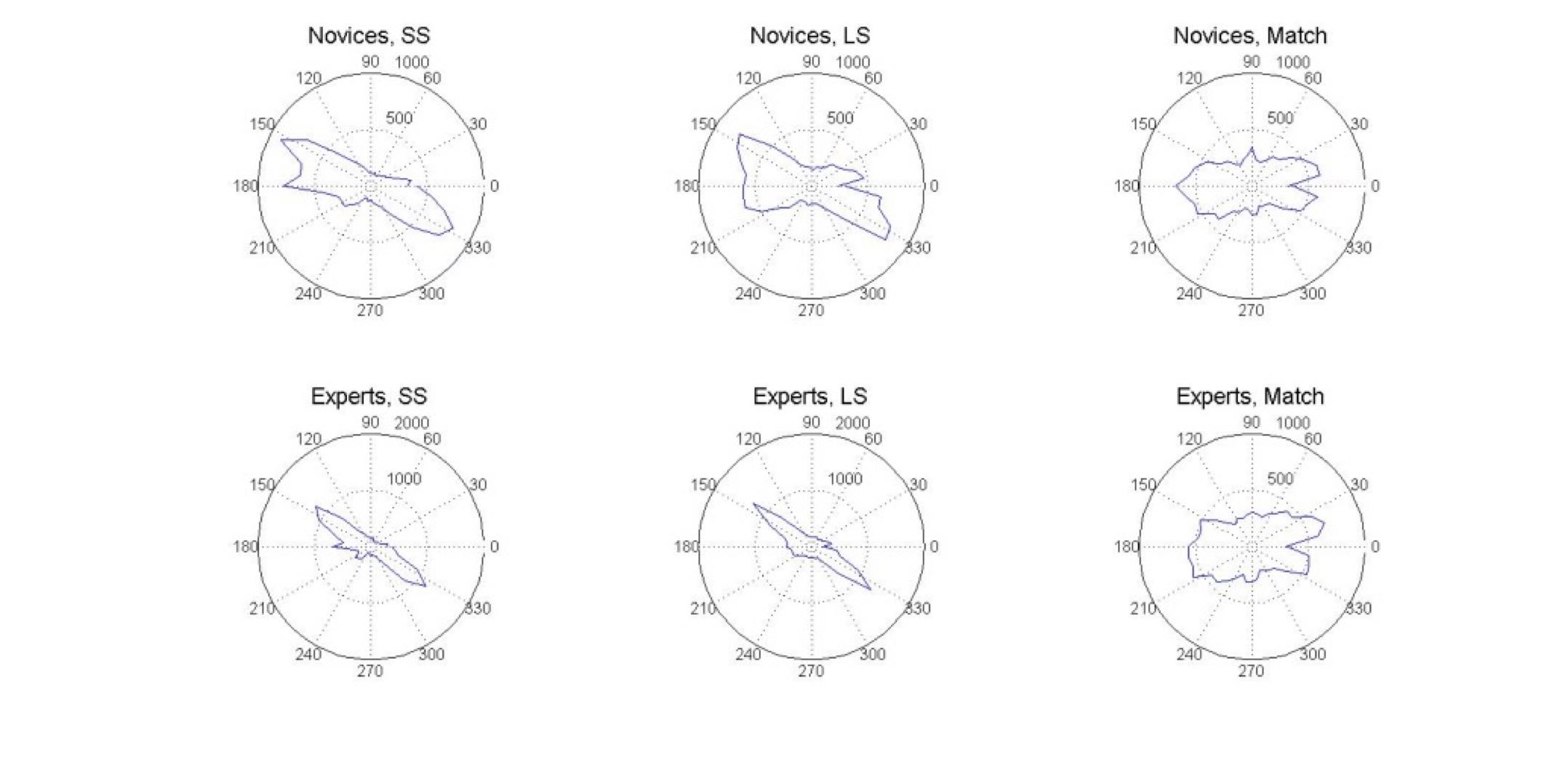}\label{gazedir}}
\caption{Empirical distributions (histograms) of $\{l_t,\theta_t, f_t\}_{t=1}^{N_T}$ used to classify expertise. Top panels (\ref{gazefixtime}), fixation duration; middle panels (\ref{gazeamp}), gaze shift amplitude; bottom panels (\ref{gazedir}), gaze shift direction. Vertical solid lines, median values. SS=Short Shots, LS=Long Shots. Modified after \cite{boccignone_jemr2014}}
\label{Fig:features}
\end{figure}

\begin{figure}[t]
\sidecaption[t]
\includegraphics[scale=0.1,keepaspectratio=true]{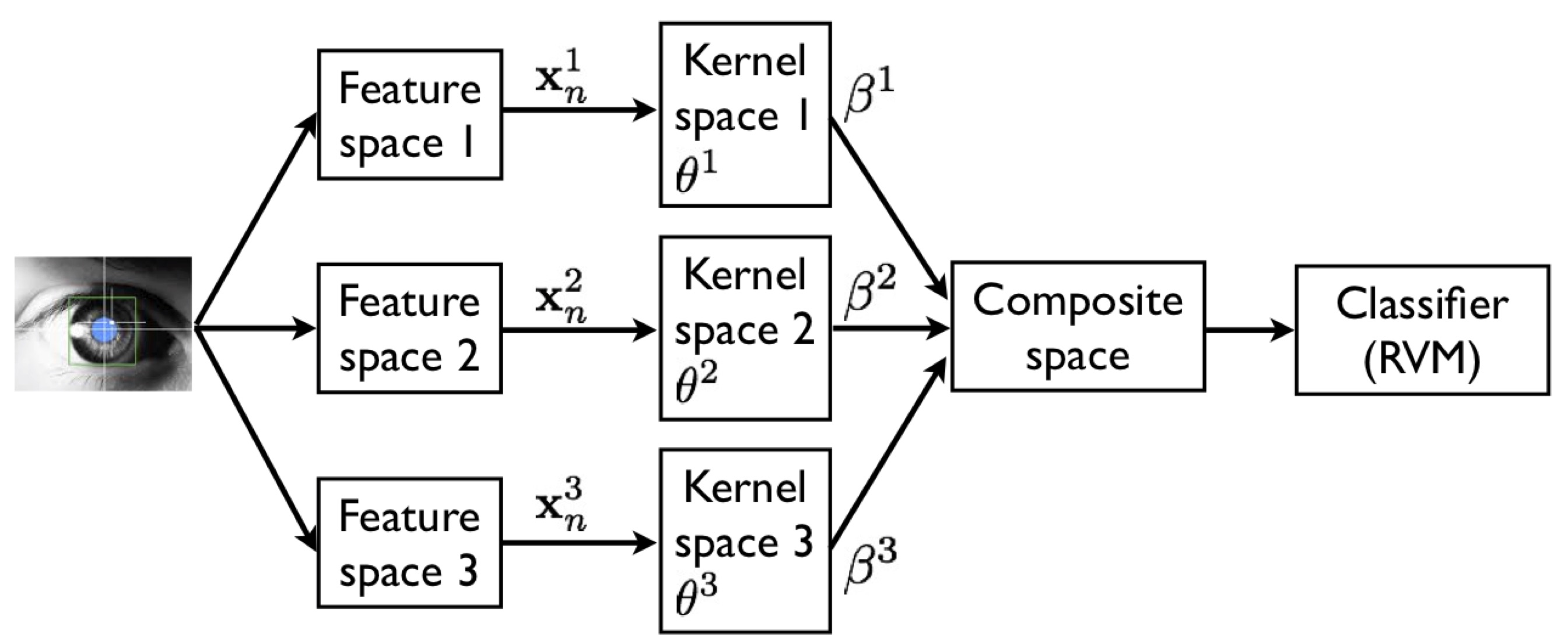}
\caption{Data analysis in multiple-kernel representation. The fixation sequence is represented in different feature spaces $s=1,\cdots,S$; each feature $\mathbf{x}^{s}$ is then separately mapped in a kernel space, each space being generated via kernel $K^s$ of parameters $\theta^s$. The separate kernel spaces are then combined in a composite kernel space, which is eventually used for classification. Modified after \cite{boccignone_jemr2014}}
\label{Fig:schema}
\end{figure}

From Fig. \ref{Fig:features},  note that differences between experts and novices are barely noticeable in terms of features.
Clearly, when addressing a scenario in which individual observers are classified as belonging to one or another population, and differences between features are so subtle, more sophisticated ML tools are needed. On this basis, each feature space $s$ was treated as independent and mapped  to a specific kernel space (either linear or Gaussian, \cite{murphy2012machine}). Then, the posterior $P(\mathbf{T} \mid  \mathbf{X})$ was rewritten as $P(t_n | \mathbf{x}_n^1,...,\mathbf{x}_n^S)= P(t_n  | \mathbf{W} , \mathbf{k}^{\beta}_{n})$, where the term on the r.h.s  is the Multinomial probit likelihood. Here, $\mathbf{W} \in \mathbb{R}^{N \times C}$ is the matrix of model parameters;  the variable $\mathbf{k}^{\beta}_{n}$ is a row of the kernel matrix $\mathbf{K}^{\beta} \in \mathbb{R}^{N \times N}$ - whose elements are the $K^{\beta}(\mathbf{x}_i, \mathbf{x}_j) $, i.e. the different kernels - and it  expresses how related, based on the selected kernel function, observation $\mathbf{x}_n$ is to the others of the training set.  This way (cfr. Fig. \ref{Fig:schema}), sources can be combined within a composite kernel space level and classified through a Relevance Vector Machine (RVM),  namely a multiple-kernel RVM  \cite{psorakis2010multiclass,damoulas2009combining} .

Discussing in detail this solution is out of the scope of this chapter (see \cite{boccignone_jemr2014} of a short presentation, and Bishop~\cite{BishopPRML} for more details). However, just to give some hints, RVMs can be considered the Bayesian counterpart of  SVMs.  They are Bayesian sparse machines, that is they employ sparse Bayesian learning via an appropriate prior formulation. Not only do they overcome some of the limitations affecting SVMs  but also they achieve sparser solutions (and hence they are faster at test time) than SVM.
Indeed, by combining only three basic parameters of visual exploration, the overall classification accuracy, expressed as percent correct and averaged across stimulus types and oculomotor features, scored a respectable $78\%$. More interesting is to consider the best performance for each stimulus type, which testifies the achievement of the classifier, and which depends on the features used. The best performance ranged between $81.90\% $ and $88.09\%$ - $1.852$ to $2.399$ in terms of $d^{'}$, which is an interesting  result, especially considering the naturalistic, unconstrained viewing condition.

\section{A final note on the use of Machine Learning in visual attention modelling}

For the sake of completeness, it is worth mentioning that modern Statistical Machine Learning techniques such as those described in the previous Section, are currently adopted even at the earliest stages of visual attention modelling, markedly in salience computation. 

Going back to Eqs. \ref{eq:models} and \ref{eq:modelItti} (Section~\ref{sec:tour}), as we already pointed out,
in the case  the  prior $P(\mathbf{L})$ is assumed to be uniform (no spatial bias, no preferred locations), then $ P(\mathbf{L} = 1 \mid \mathbf{F}) \simeq P(  \mathbf{F} \mid  \mathbf{L} = 1) $, where $\mathbf{L} = 1$ is now a binary RV ($1$ or $0$) simply denoting location $(x,y)$ as salient/non salient.  In such case, the likelihood function $P(  \mathbf{F} \mid  \mathbf{L} = 1)$ can be determined in many ways;  e.g.,  nonparametric kernel density estimation has been addressed~\cite{seo2009},  where  center / surround local regression kernels are exploited for computing  $\mathbf{F}$. 

More generally, taking into account the ratio 
$f(\mathbf{L})=\frac{P(\mathbf{L} = 1 \mid \mathbf{F})}{P(\mathbf{L} = 0 \mid \mathbf{F})}$ 
 (or, commonly, the log-ratio) casts the saliency detection problem in a classification problem, in particular a discriminative one,   for which a variety of learning techniques are readily available. Kienzle and colleagues \cite{kienzle2006nonparametric}  pioneered this approach by learning the saliency discriminant function $f(\mathbf{L})$  directly from human eye tracking data using an SVM. Their approach has paved the way to a relevant number of works: from \cite{judd2009learning} -- where a linear SVM is trained from human fixation data using a set of low, middle and high-level  features to define salient locations--, to most recent ones that wholeheartedly  endorse ML trends.  Henceforth, methods have been proposed  relying on  sparse representation of ``feature words'' (atoms)   encoded in salient and non-salient dictionaries; these are  either learned from local image patches \cite{yan2010visual,lang2012saliency}  or  from eye tracking data of training images \cite{jiang2015image}. Graph-based learning is one other trend,  from the seminal work of Harel \emph{et al.}~\cite{harel2007graph} to \cite{yu2014maximal} (see the latter, for a brief review of this field).  Crucially, for the research practice, data-driven learning methods allow to contend with large scale  dynamic datasets.  SVMs are used by  Mathe and Sminchisescu~\cite{mathe2015actions} in the vein of \cite{kienzle2006nonparametric} and \cite{judd2009learning}, but  they remarkably exploit state-of-the art computer vision datasets  annotated with human eye movements collected under the ecological constraints of a visual action recognition task.
 
As a general comment on (discriminative) ML-based methods, on the one hand it is embraceable the criticism  by Borji and Itti~\cite{BorItti2012}, who surmise that these techniques make  ``models data-dependent, thus influencing fair model comparison, slow, and to some extent, black-box.''  But on the other hand, one important lesson of these approaches lies in that they  provides a data-driven way of deriving the most relevant visual features as optimal predictors. The learned  patterns  can shape receptive fields (filters) that have equivalent or superior predictive power when compared against hand-crafted (and sometimes more complicated) models \cite{kienzle2009center}. Certainly, this lesson is at the base of the current  exponentially growth of methods based on \emph{deep learning techniques} \cite{lecun2015deep},  in particular Convolutional Neural Networks (CNN,  cfr. \cite{deepTaxonomy2016} for a focused review), where the computed features seem to outperform, at least from an engineering perspective, most of, if not all, the state-of-the art features conceived in computer vision. 

Again, CNNs,  as commonly exploited in the current practice, bring no significant conceptual novelty as to the use of Eq. \ref{eq:models}: fixation prediction is formulated  as a supervised  binary classification problem (in some case, regression  is addressed, \cite{wang2016deep}). For example,  in \cite{vig2014large}   a linear SVM is used for learning the saliency discriminant function $f(\mathbf{L})$ after a large-scale search for optimal features $\mathbf{F}$. Similarly, Shen \emph{et al.}~\cite{shenZhao2014learning} detect salient region  via linear SVM fed with features computed from multi-layer sparse network model. Work described in~\cite{lin2014saliency} uses the simple normalization step \cite{IttiKoch98} to  approximate $P(\mathbf{L} = 1 \mid \mathbf{F})$, where in  \cite{kruthiventi2015deepfix}  the last  $1 \times 1$ convolutional layer of a fully convolutional net is exploited. Cogent here  is the outstanding performance of CNN in learning and representing  features that correlate well with eye fixations, like objects, faces, context. 
 
Clearly, one problem is  the enormous amount of training data necessary to train these networks, and the  engineering expertise required, which makes them difficult to apply  for predicting saliency.  However, K\"ummerer \emph{et al.}~\cite{kummerer2014deep} by exploiting the well known  network from \cite{AlexNetNIPS2012} as starting point,  have given evidence that deep CNN  trained on computer vision tasks like object detection boost saliency prediction. The network  described in \cite{AlexNetNIPS2012} has been optimised for object recognition using a massive dataset consisting of more than one million images, and  results reported \cite{kummerer2014deep}  on static pictures are impressive when compared to state-of-the-art methods, even  to previous CNN-based proposals \cite{vig2014large}.   

\section{Suggested readings}

To explore beyond the contents of this Chapter, we recommend the following. A brief and clear introduction to stochastic processes (from a physicist's point of view) can be found in a few chapters of  Huang's ``\emph{Introduction to statistical physics}'' \cite{huang2001introduction}.  A comprehensive  treatment of stochastic processes is given in Gardiner's ``\emph{Stochastic Methods: A Handbook for the Natural and Social Sciences}'' \cite{gardiner2009stochastic}; it is a great starting point if you are looking for specific information about a specific stochastic process. One of the finest books on stochastic processes is  van Kampen's classic  ``\emph{Stochastic processes in physics and chemistry}'' \cite{vankampen}; difficult reading, but well--worth the effort.  A  modern treatment of the subject is provided in Paul and Baschnagel ``\emph{Stochastic Processes -- From Physics to Finance}'' \cite{paul2013stochastic}, with a clear discussion of what happens beyond the Central Limit Theorem.

A beautiful bridge between stochastic processes and foraging is outlined in M{\'e}ndez,  Campos,  and Bartumeus,  ``\emph{Stochastic Foundations in Movement Ecology: Anomalous Diffusion, Front Propagation and Random Searches}'' \cite{mendez2014stochastic}. However, if one wants to skip more technical details, an affordable, easy to read introduction to foraging and L\'evy flights is  ``\emph{The physics of foraging}'' by Viswanathan \emph{et al.} \cite{viswanathan2011physics}.

Eventually, for what concerns Statistical Machine Learning, which is nowadays a vast field, a thorough and simple introduction is provided by Rogers and Girolami \cite{rogers2011first}. A deeper insight can be gained by reading Bishop's textbook~\cite{BishopPRML}. The most comprehensive and up-to-date textbook is that by Kevin Murphy~\cite{murphy2012machine}. If some of the readers are daring to surf the big wave of deep learning, then the book by Goodfellow, Bengio and Courville~\cite{Goodfellow-et-al-2016} provides the vital outfit.

\section{Questions students should be able to answer}
\begin{enumerate}
\item What are the main reasons for considering the sequence of gaze shifts to be the observable outcome of a stochastic process?
\item If you were to set up a probabilistic model of gaze shifts, which factors would you consider to design a prior distribution $P(\mathbf{x})$ on gaze position  $\mathbf{x}$?

\item What kind of information is provided by a Probabilistic Graphical Model?

\item You have conducted an eye tracking experiment where you  recorded the first ten fixations for each subject, say $\mathbf{x}_i$, $i=1,2,\cdots,10$ at times $t_i$, $i=1,2,\cdots,10$. You devise two possible models:
\begin{eqnarray}
P( \mathbf{x}_{1}, t_{1};   \mathbf{x}_{2}, t_{2}; \cdots  ) = \prod_{i=1}^{10} P( \mathbf{x}_{i}, t_{i}), \nonumber \\
P( \mathbf{x}_{1}, t_{1};   \mathbf{x}_{2}, t_{2}; \cdots  ) = P(\mathbf{x}_{1}, t_{1}) \prod_{i=2}^{10} P(\mathbf{x}_{i}, t_{i} \mid \mathbf{x}_{i-1}, t_{i-1}).\nonumber
\end{eqnarray}
What are the assumptions behind the two models? What are the pros and cons of each model?

\item You set up an eye tracking experiment where a control group and a patient group observe pictures displaying a number of objects of interest. At the end of the experiment you define objects as Areas of Interest (AOIs) and compute the observation transition frequency between AOIs in each picture for each subject of the two groups. What is the  probabilistic model behind such statistics? 

\item You are simulating a 2-dimensional stochastic process according to the following equations:
\begin{eqnarray}
x_{t} =  \xi_{x,t } \nonumber \\ 
y_{t} =  \xi_{y,t }, \nonumber
\end{eqnarray}
where $x_{t}, y_{t}$ denote spatial coordinates and the random variable $\xi_{x,t }$ is sampled from a $1$-dimensional Gaussian distribution $\mathcal{N}(\mu_x,\sigma^2)$, i.e. 
$\xi_{x,t } \sim \mathcal{N}(\mu,\sigma^2)$, with $\mu_x = 10, \sigma = 1$,  while $\xi_{y,t } \sim \mathcal{N}(\mu_y,\sigma^2)$, with $\mu_y = 10$. What kind of $2$-dimensional pattern is likely to be drawn if you run the simulation for $t=1,2,\cdots 100$ iterations?
\item You are studying the central fixation bias in scene viewing (i.e., the marked tendency to fixate the center of the screen when viewing scenes on computer monitors). The resolution of the CRT monitor where stimuli are presented is $1280 \times 1020$ pixels. What could be a possible microscopic level description of such bias? What a possible macroscopic description? (\emph{Hint}: Reconsider  Question 5)
 
\item Assume that you are able to fit the empirical distribution of some experimental data with the law $P(x) \approx x^{-\mu}$. Which kind of information cold you infer from such result?

\item Repeat the simulation proposed in Question 5, but now assume to sample $\xi_{x,t }$ and $ \xi_{y,t }$ from a Cauchy distribution $\frac{1}{\pi \gamma} \left[ \frac{\gamma ^2}{(\xi- \delta)^2 + \gamma^2} \right]$ with $\delta = 10$, $\gamma =1$. What kind of $2$-dimensional pattern do you expect? Would the histogram of the sequence of step amplitudes have a bell shape? Could we still consider this kind of random walk a Markovian walk?

\item You plan an eye tracking experiment with $n$ subjects to distinguish experts from non experts  when viewing five paintings. After recording the data, you define a number $N_k$ of AOIs on paintings (for simplicity, assume an equal number for all paintings). Then, for each subject and each painting, you measure $l_k$ and $t_k$, namely   the number of fixations and  the average fixation time, respectively,  in $k$-th AOI, $k=1,2,\cdots, N_k$. How would you proceed to make the desired discrimination?
\end{enumerate}



\bibliographystyle{spmpsci}   
\bibliography{levyeye,billiard}

\end{document}